\documentclass[a4paper,fleqn]{cas-sc}
\usepackage{footnote}
\makesavenoteenv{tabular}
\usepackage[square,numbers,compress]{natbib}
\usepackage{csquotes}
\usepackage{xcolor}
\usepackage{longtable}
\usepackage{tabu}
\usepackage{tablefootnote}
\usepackage{url}
\usepackage[colorinlistoftodos,prependcaption]{todonotes}
\usepackage[inline]{enumitem}
\usepackage{xspace}
\usepackage{subcaption}
\usepackage{float}
\usepackage{amssymb}
\usepackage{pifont}
\newcommand{\cmark}{\ding{51}}%
\newcommand{\xmark}{\text{\ding{55}}}
\usepackage{enumitem}

\usepackage{hyperref}

\def\tsc#1{\csdef{#1}{\textsc{\lowercase{#1}}\xspace}}
\tsc{WGM}
\tsc{QE}
\tsc{EP}
\tsc{PMS}
\tsc{BEC}
\tsc{DE}

\usepackage{mathtools}
\DeclarePairedDelimiter\norm{\lVert}{\rVert}

\begin{document}
\let\WriteBookmarks\relax
\def\floatpagepagefraction{1}
\def\textpagefraction{.001}

\shorttitle{Deep Transfer Learning for Automatic Speech Recognition: Towards
Better Generalization}

\shortauthors{H Kheddar et~al.}

\title [mode = title]{Deep Transfer Learning for Automatic Speech Recognition: Towards Better Generalization }                      

\vskip2mm

\author[1]{Hamza Kheddar\corref{cor1}}
\ead{kheddar.hamza@univ-medea.dz}

\author[2,3]{Yassine Himeur}
\ead{yhimeur@ud.ac.ae}

\author[3]{Somaya Al-Maadeed}
\ead{s_alali@qu.edu.qa}

\author[4,5]{Abbes Amira}
\ead{aamira@sharjah.ac.ae}

\author[6]{Faycal Bensaali}
\ead{f.bensaali@qu.edu.qa}


\address[1]{LSEA Laboratory, Electrical Engineering Department, University of Medea, Algeria}
\address[2]{College of Engineering and Information Technology, University of Dubai, Dubai, UAE}
\address[3]{Computer Science and Engineering, Qatar University, Qatar}
\address[4]{Department of Computer Science, University of Sharjah, UAE}
\address[5]{Institute of Artificial Intelligence, De Montfort University, Leicester, United Kingdom}
\address[6]{Department of Electrical Engineering, Qatar University, Doha, Qatar}


%
%
%
%
%
%
%
%
%
%
%

\begin{abstract}
Automatic speech recognition (ASR) has recently become an important challenge when using deep learning (DL).
It requires large-scale training datasets and high computational and storage resources.
Moreover, DL techniques and machine learning (ML) approaches in general, hypothesize that training and testing data come from the same domain, with the same input feature space and data distribution characteristics. This assumption, however, is not applicable in some real-world artificial intelligence (AI) applications. Moreover, there are situations where gathering real data is challenging, expensive, or rarely occurring, which can not meet the data requirements of DL models. deep transfer learning (DTL) has been introduced to overcome these issues, which helps develop high-performing models using real datasets that are small or slightly different but related to the training data.
This paper presents a comprehensive survey of DTL-based ASR frameworks to shed light on the latest developments and helps academics and professionals understand current challenges. Specifically, after presenting the DTL background, a well-designed taxonomy is adopted to inform the state-of-the-art. A critical analysis is then conducted to identify the limitations and advantages of each framework. Moving on, a comparative study is introduced to highlight the current challenges before deriving opportunities for future research.
\end{abstract}




\maketitle

\section{Introduction}   \label{sec1}
\subsection{Preliminary}
The human-machine interaction (HMI) has become increasingly ubiquitous with the development of AI techniques that can reproduce speech ready for transmission to a system that executes actions. Automatic speech recognition is considered as a cutting-edge communication technology in HMI \cite{nedjah2023automatic}. Large companies and even service providers widely use ASR-based systems, where orders or transactions can be completed by communicating with some AI servers, such as a chat robot or virtual assistant \cite{anoop2023suitability}. Spoken language is the basis of these communications, which is a critical component to properly consider when building an AI-based system dedicated to ASR. 
ASR takes into account (i) acoustic, lexical, and syntactic information;  and (ii) semantic knowledge.
Acoustic model (AM) processing includes speech coding ~\cite{haneche2021compressed}, speech enhancement \cite{michelsanti2021overview}, source separation \cite{michelsanti2021overview,luo2021group}, speech security (e.g, steganography \cite{kheddar2019pitch,kheddar2022speech,kheddar2018fourier} and watermarking \cite{yassine2012secure,yamni2022efficient,chen2020specmark}), and other technologies that can all be used in audio analysis. 
On the other hand, semantic model (SM), commonly known in the literature as language model (LM) processing, includes all techniques of natural language processing (NLP). This branch of AI is concerned with teaching computers to comprehend and interpret human language. It is the foundation of music information retrieval \cite{olivieri2021audio}, collecting sound files based on similar content \cite{wold1996content}, audio tagging and sound event detection \cite{boes2021audiovisual}, converting speech to text and vise versa \cite{tang2021general}, hate speech detection \cite{plaza2021comparing}, etc.  When NLP is employed as a tool in various domains, AI models can understand humans and respond to them appropriately, revealing immense research possibilities in a variety of sectors.

ASR has significantly benefited from the latest advances made possible by deep learning (DL) algorithms, where a plethora of DL models have been proposed in the literature, offering promising performance and outperforming actual state-of-the-art techniques  \cite{meghraoui2021novel,lin2021speech}.
However, using DL in ASR is a challenging task that plays a crucial role in natural HMI. Despite all its advantages, its suffers from different problems.
The complexity of DL models is enormous due to the huge amounts of training data required for their training to achieve excellent performance. Thus, DL models require high computational and storage resources \cite{kumar2022novel}.
Moreover, data scarcity is among the challenges of ASR, which refers to the case of having insufficient quantities of training data to develop and completely explore complex DL algorithms \cite{padi2021improved}. 
Additionally, the lack of annotated data is another issue that impedes building supervised DL-based ASR models.  
On the other hand, generally, DL models (and ML tools) assume that training and testing data come from the same domain, with the same input feature space and data distribution characteristics. This assumption, however, is not applicable in some real-world applications \cite{himeur2022next}.
Consequently, DL models can not perform well if (i) small training datasets are utilized and; (ii) there is a discrepancy or data distribution inconsistency between training and test data  \cite{niu2020decade}.



\begin{figure}[ht!]
\centering
\includegraphics[scale=0.9]{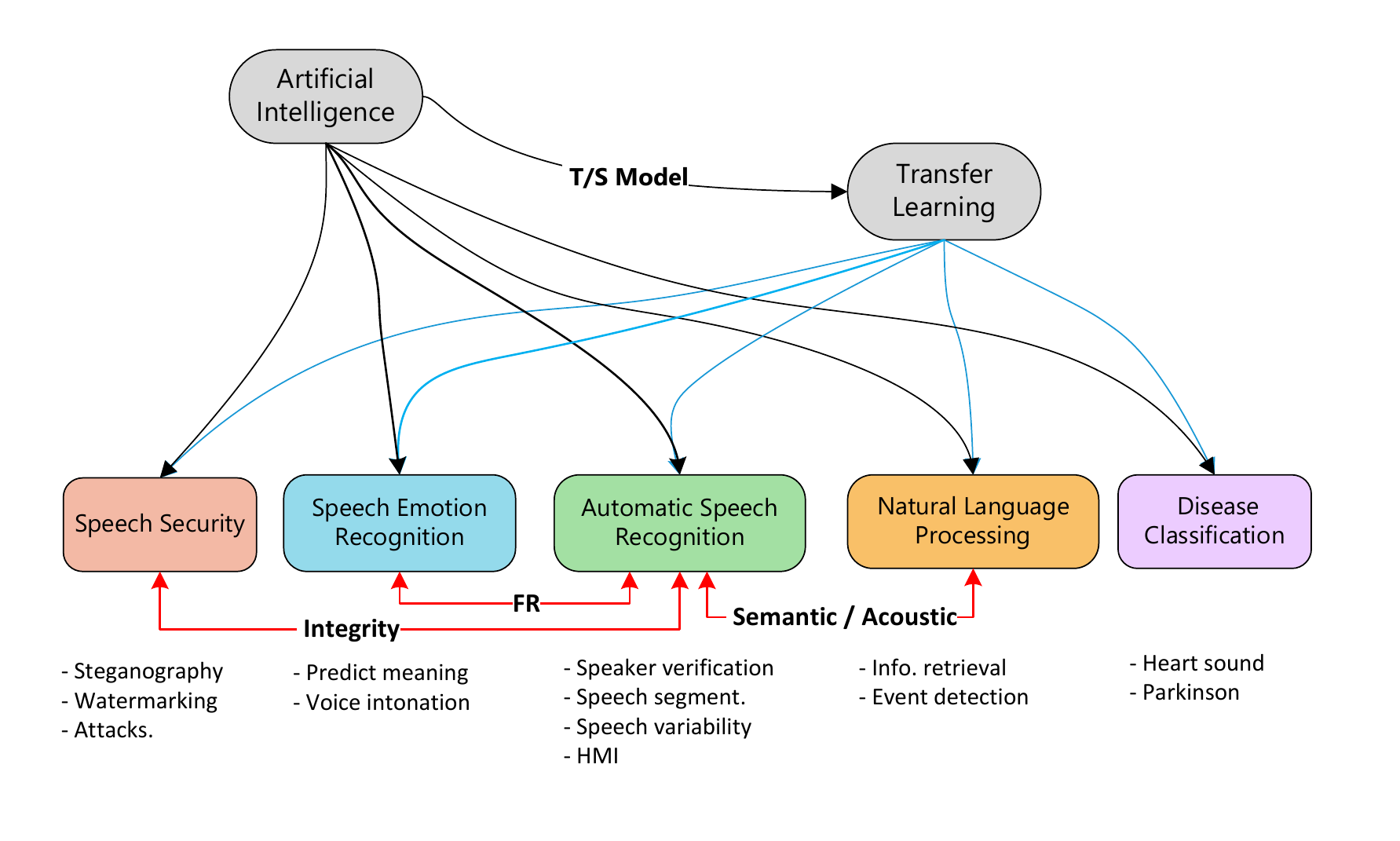}
\caption{ Summary of the main speech processing disciplines where DTL can be applied.}
\label{fig:01}
\end{figure}

Deep transfer learning (DTL) targets the mentioned problems by pooling knowledge from the source domain (SD)/task and transferring it to a target domain (TD)/task or multiple domains/tasks. This concept can also be named the teacher-to-student (T/S) knowledge transfer \cite{sayed2023time}.
Put differently, in DTL, a pre-trained model is reused as the starting point for a DTL model on a new task, which helps (i) optimizing and rapidly progressing the second task, (ii) achieving higher performance on small datasets, and (iii) reducing the effect of overfitting.
Furthermore, DTL can merge pre-trained knowledge from different domains and tasks to deal with data distribution inconsistency. 
In DL, some trainable neurons and hyperparameters can be frozen to better preserve the knowledge learned from the original datasets \cite{himeur2023face}.
Fig. \ref{fig:01} points out some critical areas in speech processing that can apply DL and DTL. 
Accordingly, ASR, speech emotion recognition (SER), NLP, and speech security (SS) domains are significantly related. 
ASR provides the acoustic parameters to NLP, which provides the semantic details to ASR. Frequently, SER is similar to ASR but with a form recognition (FR) module. Additionally, ASR can be used in the SS domain, as a steganalytic process, to check the integrity of the speech \cite{kheddar2019pitch,kheddar2022high}.

To that end, the importance of DTL to resolve the above-mentioned problems encountered with the use of DL tools in ASR has motivated the scientists to propose numerous DTL-based ASR solutions for different applications. Typically, DTL methodologies have been widely applied in speech processing, such as speaker verification (SV) \cite{jia2018transfer}, ASR \cite{malik2021automatic}, SER \cite{padi2021improved,hazarika2021conversational,vryzas2021web} and NLP field \cite{malhotra2021bidirectional}. Additionally, DTL has been widely used in the speech-based medical disease diagnosis, such as the heart sound classification \cite{hettiarachchi2021novel} and early diagnosis of Parkinson's disease (PD)  \cite{karaman2021robust}, Speech-based depression prediction \cite{harati2021speech}.



\subsection{Contribution of the paper}
This paper overviews existing DTL-based ASR frameworks published in the 2015-2023 period. Accordingly, a well-defined taxonomy is introduced to classify them into various categories based on different aspects, including the nature of transferred knowledge, availability of labels in the source and TDs and adopted strategy. To the best of the authors' knowledge, to date, there is no research paper that goes into the details and reviews critically DTL-based ASR contributions.  
Moreover, this review is conducted by focusing on different aspects: i) presenting the background of DTL, introducing the issue of ASR and expounding the importance of DTL for it; (ii) analyzing the DTL frameworks and identifying their limitations and comparing their performance; (iii) discussing the importance applications of DTL-based ASR; (iv) clarifying the methodological merits and elucidating the main DTL challenges and issues; and (v) suggesting future directions to further improve the performance of DTL-based ASR solutions and predicting the prospective development of DTL for ASR applications in the near future. 


On the other hand, despite that, there were some attempts to survey the generic applications of transfer learning (TL), most of them did not investigate recent DTL advances for ASR tasks. Table \ref{tab:2} summarizes some of the main contributions of the proposed study compared to other TL surveys.
It is clearly seen that this survey article has numerous major enhancements and additions since it merge works that employ both ASR and DTL as compared to previous TL surveys. It also provides a quantitative analysis of existing ASR-based DTL solutions, discusses the application of DTL-based ASR for medical diagnosis and describes DTL-based ASR attacks and security. Moreover, actual current challenges  such as the negative transfer (NT), knowledge gain measurement and unification of DTL have been covered. The main contributions of this article can be summarized as follows:

\begin{itemize}
\item Describing the evaluation metrics and datasets used for validation of DTL-based ASR schemes. 
\item Presenting a well-defined taxonomy of DTL-based ASR methodologies with regard to AM and LM domains.
\item Investigating for the first time adversarial DTL-based ASR methods and summarizing DA techniques used in ASR applications.
\item Explaining the relationship between DTL-based ASR and other application fields, such as speech translation (ST), speech evaluation, medical diagnosis, etc.  
\item Identifying DTL-based ASR challenges and gaps, such as NT, knowledge gain measurement and unification of DTL.
\item Suggesting future directions to further improve the performance of DTL-based ASR solutions and predicting the prospective development of DTL for ASR 
\end{itemize}

\begin{table}[ht!]
\centering
\caption{{Contribution comparison of the proposed study against other TL surveys. The tick mark (\cmark) indicates that the specific field has been addressed, whereas the cross mark (\xmark
) means addressing the specific fields has been missed. When ($\sim$) is presented, it indicates that most critical concerns/fields have not been  addressed. }}
\label{tab:2}

\begin{tabular}{llllllllll}
\hline
{\scriptsize Survey} & {\scriptsize Description} & {\scriptsize TL} & 
{\scriptsize Quantitative } & {\scriptsize TL-ASR } & {\scriptsize ASR
attacks } & \multicolumn{3}{c}{\scriptsize Current challenges} & 
{\scriptsize Future} \\ \cline{7-9}
&  & {\scriptsize Background} & {\scriptsize analysis} & {\scriptsize in
medical } & {\scriptsize and security} & {\scriptsize Negative } & 
{\scriptsize Knowledge } & {\scriptsize Unification} & {\scriptsize %
Directions} \\ 
&  &  &  & {\scriptsize diagnosis} &  & {\scriptsize transfer} & 
{\scriptsize gain} & {\scriptsize of TL} &  {\scriptsize for ASR} \\ \hline
{\scriptsize \cite{lu2015transfer}} & {\scriptsize TL for computational
intelligence} & {\scriptsize \xmark} & {\scriptsize \xmark} & {\scriptsize %
\xmark} & {\scriptsize \xmark} & {\scriptsize \xmark} & {\scriptsize \xmark}
& {\scriptsize \xmark} & {\scriptsize \xmark} \\ 
{\scriptsize \cite{weiss2016survey} } & {\scriptsize General information on
TL} & {\scriptsize \cmark} & {\scriptsize \xmark} & {\scriptsize \xmark} & 
{\scriptsize \xmark} & {\scriptsize \xmark} & {\scriptsize \xmark} & 
{\scriptsize \xmark} & {\scriptsize \xmark} \\ 
{\scriptsize \cite{niu2020decade}} & {\scriptsize Generic TL contributions}
& {\scriptsize \xmark} & {\scriptsize \xmark} & {\scriptsize \xmark} & 
{\scriptsize \xmark} & {\scriptsize \xmark} & {\scriptsize \xmark} & 
{\scriptsize \xmark} & {\scriptsize \xmark} \\ 
{\scriptsize \cite{himeur2023video}} & {\scriptsize DTL for video surveillance} & {\scriptsize \cmark} & {\scriptsize \cmark} & {\scriptsize \xmark} & {\scriptsize \xmark} & {\scriptsize \cmark}
& {\scriptsize \xmark} &  {\scriptsize \xmark} & {\scriptsize $\sim$} \\ 
{\scriptsize \cite{zhuang2020comprehensive}} & {\scriptsize Focus on
homogeneous TL} & {\scriptsize \cmark} & {\scriptsize \xmark} & {\scriptsize %
\xmark} & {\scriptsize \xmark} & {\scriptsize \xmark} & {\scriptsize \xmark}
& {\scriptsize \xmark} & {\scriptsize \xmark} \\ 
{\scriptsize \cite{durrani2021transfer} } & {\scriptsize TL for NLP} & 
{\scriptsize \cmark} & {\scriptsize \xmark} & {\scriptsize \xmark} & 
{\scriptsize \xmark} & {\scriptsize \xmark} & {\scriptsize \xmark} & 
{\scriptsize \xmark} & {\scriptsize \xmark} \\ 
{\scriptsize \cite{wan2021review} } & {\scriptsize TL in EEG} & {\scriptsize %
\cmark} & {\scriptsize \xmark} & {\scriptsize \cmark} & {\scriptsize \xmark}
& {\scriptsize \xmark} & {\scriptsize \xmark} & {\scriptsize \xmark} & 
{\scriptsize \xmark} \\ 
{\scriptsize \cite{bashath2022data}} & {\scriptsize TL for text data} & 
{\scriptsize \cmark} & {\scriptsize \xmark} & {\scriptsize \xmark} & 
{\scriptsize \xmark} & {\scriptsize \xmark} & {\scriptsize \xmark} & 
{\scriptsize \xmark} & {\scriptsize \xmark} \\ 
{\scriptsize \cite{kheddar2023deep}} & {\scriptsize DTL for intrusion detection} & {\scriptsize \cmark}
& {\scriptsize \cmark} & {\scriptsize \xmark} & {\scriptsize \xmark} & 
{\scriptsize \xmark} & {\scriptsize \xmark} & {\scriptsize \xmark} & 
{\scriptsize $\sim$} \\ 
{\scriptsize Ours } & {\scriptsize DTL for ASR applications} & {\scriptsize %
\cmark} & {\scriptsize \cmark} & {\scriptsize \cmark} & {\scriptsize \cmark}
& {\scriptsize \cmark} & {\scriptsize \cmark} & {\scriptsize \cmark} & 
{\scriptsize \cmark} \\ 
&  &  &  &  &  &  &  &  &  \\ \hline
\end{tabular}

\end{table}


The remainder of this survey article is structured as follows. Section \ref{sec2} explains the review methodology of our survey. Moving on, existing DTL-based ASR are reviewed in Section \ref{sec3} using a well-defined taxonomy. Next, Section \ref{sec4} describes the main contributions made in different DTL-based ASR-based applications, including acoustic models, language models, cross-domain ASR, medical diagnosis, and attacks and security. After that, Section \ref{sec5} provides a critical discussion and outlines the open challenges before identifying future research directions in Section \ref{sec6}. Lastly, concluding remarks are derived in Section \ref{sec7}.

\section{Review methodology} \label{sec2}
The methodology of conducting this survey is outlined in this section, where the search strategy is discussed first, followed by a section on study selection. The inclusion criteria is then described, including the keyword match, creativity and effect, and uniqueness. All these procedures contribute to the development of our paper quality assessment protocol.

\subsection{Literature search strategy}
To identify and determine existing DTL-based ASR studies, a thorough search has been conducted on the popular publication databases, which are considered as the main source of high-quality scientific research articles. Thus, the search has been done in Scopus, Web of Science, Elsevier, IEEE, ACM Digital Library, Wiley and IET Digital Library.

\subsection{Selection study}
The following three criteria have been considered to search and select the studies included in this review.

\begin{enumerate}
\item {Keyword match:} the preliminary references' keywords were manually extracted and grouped. These publications were grouped using "theme clustering" based on keywords that construct to the following query: 
\begin{center} References=SELECT\big( Papers WHERE keywords=("\textbf{Transfer  learning}" OR "\textbf{Knowledge transfer}" OR "\textbf{Model adaptation}" OR "\textbf{domain adaptation}" "\textbf{Model combination}" OR "\textbf{Fine-tuning}") AND ("\textbf{Automatic speech recognition}" OR "\textbf{Speech processing}" OR "\textbf{Natural language processing}" OR  "\textbf{Spoken language}" OR "\textbf{Applications}")\big).
\end{center}

\item {Innovation and impact:} the publications have been filtered according to ASR-based innovation, the quality of the study and presented contributions and results, and the nature of publication (i.e., journal papers, conference proceedings articles and book chapters). The articles that present repetitive contributions (or highly similar content) or not written in English, have been eliminated. 

\item {Novelty:} only the research contributions published during the 2015-2023 period have been included.
\end{enumerate}

Table \ref{tab:3} summarises the number of included papers per database after applying the selection protocol.

\begin{table}[t!]
\caption{Literature acquisition databases.}
\label{tab:3}
\begin{tabular}{ p{3cm}p{3cm}p{3cm}p{3cm}p{3cm}}
\toprule
\textbf{Database}& \textbf{Research Articles} & \textbf{Conference Papers} &\textbf{Book Chapter} & \textbf{Total}  \\
\midrule
 ACM& {7} & {3} & -- & {10}  \\
 Elsevier& {32} & -- & -- & {32}  \\
 Springer& {29} & -- & {13} &{42} \\
 IEEE&  {46} & {40} & -- & {86} \\
Others & {65} & {29} & -- &{94}  \\
\bottomrule
\end{tabular}
\end{table}

\subsection{Quantitative analysis}
\begin{figure}[t!]
\begin{center}
\includegraphics[scale=0.9]{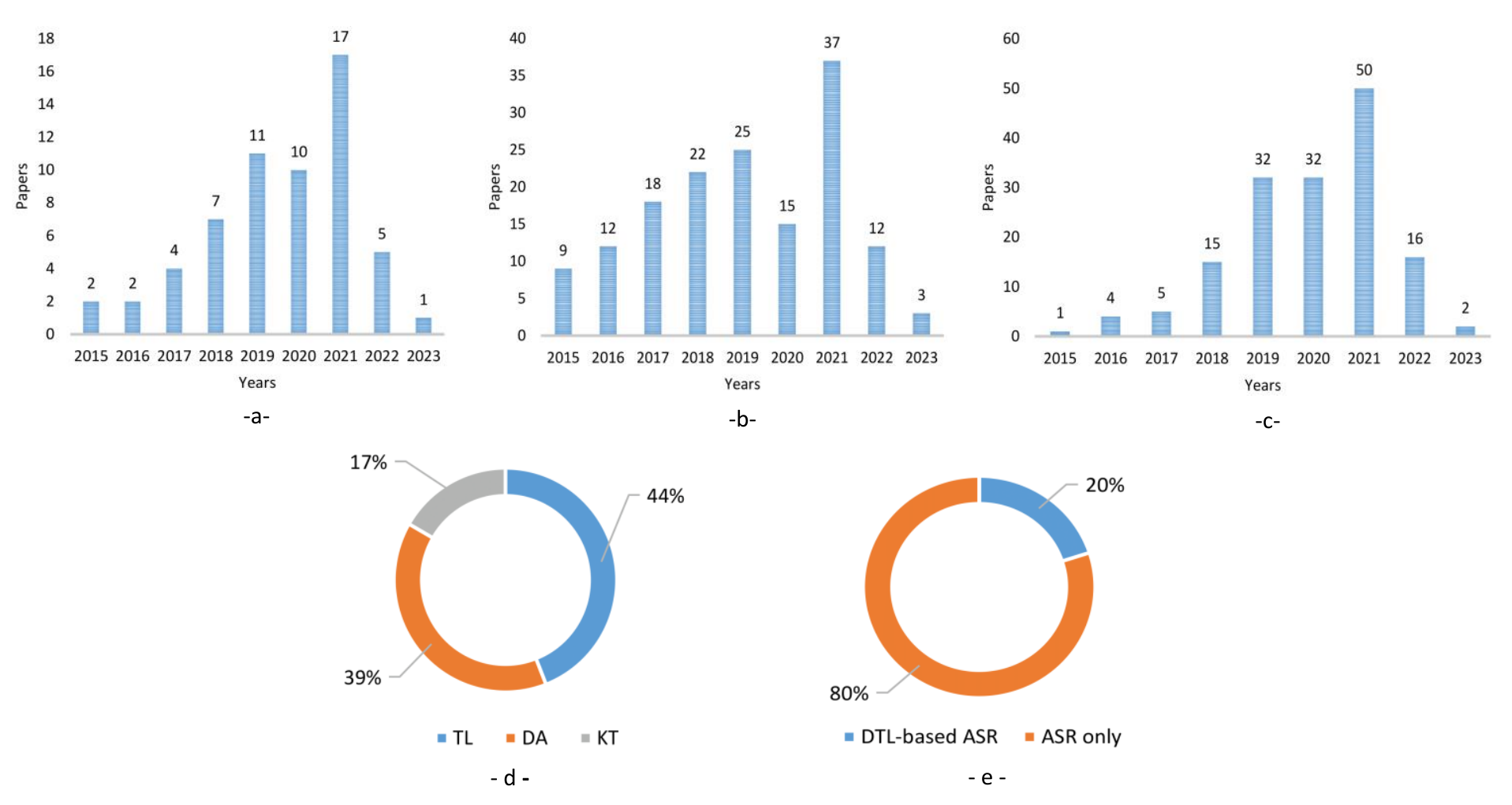}\\
\end{center}
\caption{{The number of publications on ASR and DTL-based ASR (source "Scopus database") from 2015 to 2023, where: (a) papers searched using ASR and Knowledge transfer (KT) keywords, (b) papers searched using ASR and DA keywords, (c) papers searched using ASR and TL keywords, (d) comparison-based Keywords search, and (e) pure ASR vs. TL-based ASR papers search.}}
\label{fig2:statistics}
\end{figure}

With the advance of DL, a large number of studies have been introduced to improve ASR, in which a significant part has been reserved for treating DTL-based contributions. 
Fig. \ref{fig2:statistics} presents the statistics extracted from the Scopus database, with reference to the (i) yearly published papers using adopted keywords (including \enquote{ASR \& knowledge transfer}, \enquote{ASR \& domain adaptation} and \enquote{ASR \& transfer learning}), (ii) included papers per keyword, and (iii) published DTL-based ASR articles compared to the overall ASR papers.
In general, the number of published DTL-based ASR studies increases yearly, and this was clear even by using different keywords as depicted in Figs. \ref{fig2:statistics} (a), (b) and (c).
In addition, most included contributions have been found using the \enquote{transfer learning \& ASR}, \enquote{domain adaptation \& ASR} and \enquote{knowledge transfer \& ASR} keywords, as depicted in Fig. \ref{fig2:statistics} (d). 
Besides, it can be seen from Fig. \ref{fig2:statistics} (e) that DTL-based ASR contributions stand for 17\% of all existing ASR studies. Moreover, it is worth noting that there is an overlapping between these keywords, where some articles search by one keyword could be found by the others.

\section{Overview of DTL techniques for speech recognition}
\label{sec3}

\subsection{DTL conceptual background}   \label{sub12}
Overall, DTL consists of training a DL model on a specific domain (or task) and then transferring the acquired knowledge to a new, similar domain (or task). In what follows, we present some of the definitions that are essential to understand the principle of DTL for ASR applications.

\noindent \textbf{\textit{Def. 1 - Domain:}} Let us consider a specific dataset $X=\left\{
x_{1},\cdots ,x_{n}\right\} \in \chi $, in which $\chi $ represents the feature space, and $P(X)$ refers to the marginal probability distribution of $X$. A domain is defined as $\mathbb{D}=\left\{ X,~P(X)\right\}$. In DTL, the domain that contains the initial knowledge is defined as the SD, where it is represented by $\mathbb{D}_{S}$. By contrast, the domain including the unknown knowledge to be learnt is named the TD, it is corresponding to $\mathbb{D}_{T}$~\cite{lu2021general}.

\vskip2mm
\noindent \textbf{\textit{Def. 2 - Task:}} Considering the previously defined dataset $%
X=\left\{ x_{1},\cdots ,x_{n}\right\} \in \chi $, which corresponds to a set
of labels $Y=\left\{ y1,\cdots ,yn\right\} \in \gamma $, where $\gamma $
represents the label space. A task can be defined as $\mathbb{T}=\left\{
Y,F(X)\right\} $, where $F$ denotes the learning objective predictive
function that could be represented as well as a conditional distribution $%
P(Y|X)$. Following the definition of task, the label spaces of the source
and TDs are represented as $\gamma _{s}$ and $\gamma _{T}$,
respectively \cite{ramirez2019learning}.

\vskip2mm
\noindent \textbf{\textit{Def. 3 - DTL:}} 
A learned function $\mathbb{F}_{S}$ may be viewed as the knowledge gained in $\mathbb{D}_{S}$ using $\mathbb{T}_{S}$ if we consider a SD $\mathbb{D}_{S}$ and its related task $\mathbb{T}_{S}$. When domains or tasks differ, the purpose of DTL is to find the target predictive function $\mathbb{F}_{T}$ for the target task $\mathbb{T}_{T}$ with the TD $\mathbb{D}_{T}$. To put it another way, DTL tries to increase the performance of $\mathbb{F}_{T}$ by leveraging the knowledge $\mathbb{F}_{S}$, where $\mathbb{D}_{S}\neq \mathbb{D}_{T}$ and $
\mathbb{T}_{S}\neq \mathbb{T}_{T}$ are used. As a result, DTL may be written as follows \cite{lu2015transfer}:

\begin{equation}
\mathbb{D}_{S}=\{X_{S},P(X_{S})\},~\mathbb{T}_{S}=%
\{Y_{S},P(Y_{S}/X_{S})\}\rightarrow \mathbb{D}_{T}=\{X_{T},P(X_{T})\},%
\mathbb{T}_{T}=\{Y_{T},P(Y_{T}/ X_{T})\}
\end{equation}

\vskip2mm
\noindent \textbf{\textit{Def. 4 - Domain adaptation (DA):}} 
Considering the SD $\mathbb{D}_{S}$ for the task $\mathbb{T}_{S}$ and the TD $\mathbb{D}_{T}$ for the task $\mathbb{T}_{T}$, where $\mathbb{D}_{S}\neq \mathbb{D}_{T}$. DA tries to learn a prediction function $\mathbb{F}_{T}$ that can be utilized to improve $\mathbb{F}_{T}$ using the knowledge gained from $\mathbb{D}_{S}$ and $\mathbb{T}_{S}$. In other words, in $\mathbb{F}_{T}$, the domain divergence is adapted \cite{li2020transfer}.

\begin{figure}[t!]
\begin{center}
\includegraphics[width=0.9\textwidth]{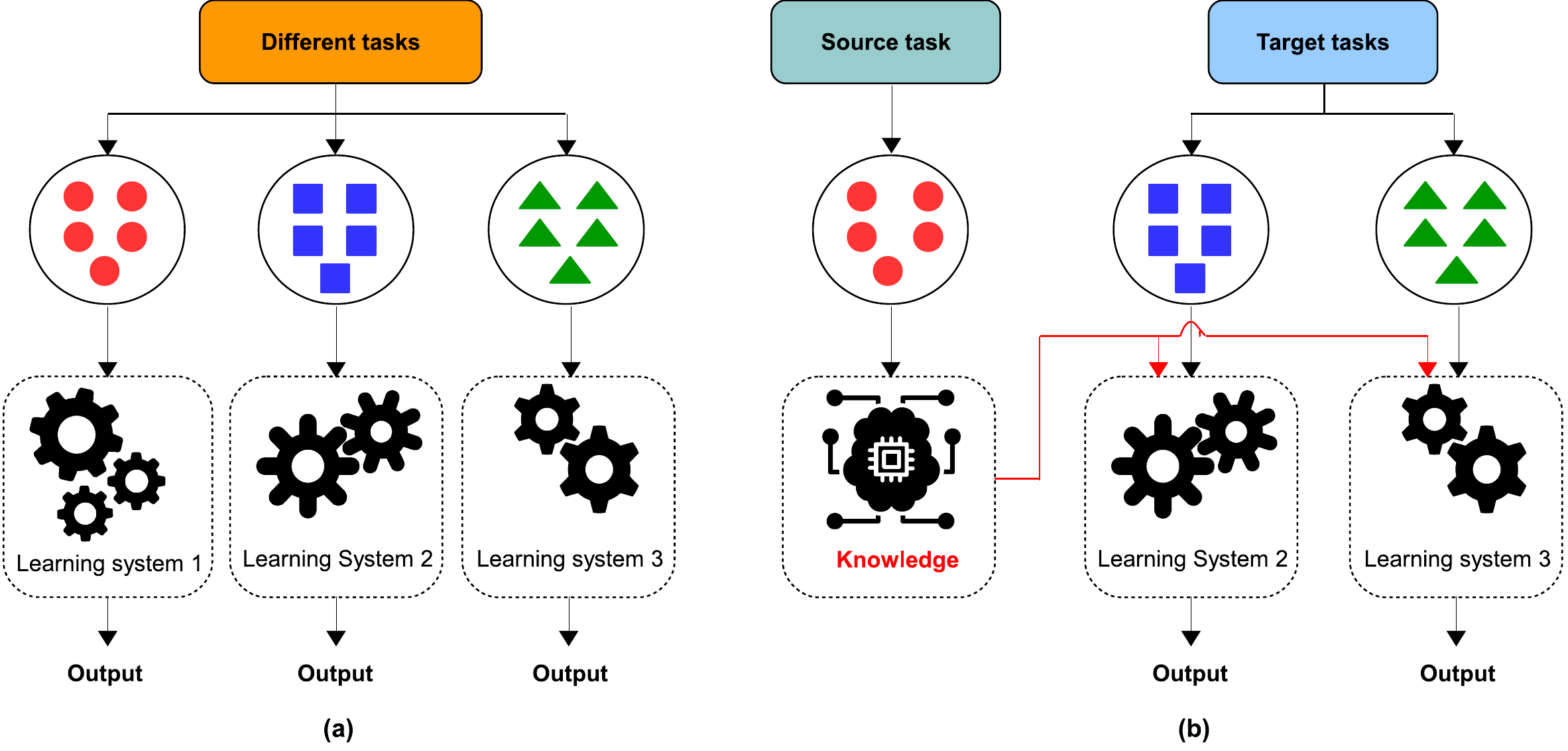}\\
\end{center}
\caption{Difference between conventional DL and DTL techniques for multiple tasks: (a) conventional DL and (b) DTL.}
\label{domain_adapt}
\end{figure}

All in all, classifying data where $\mathbb{D}_{S}\neq \mathbb{D}_{T}$ or $%
\mathbb{T}_{S}\neq \mathbb{T}_{T}$ is the main challenge that DTL algorithms
attempt to meet. One popular idea to do so is by reducing the difference
between domains or tasks, which ensures certain similarity between the
corresponding feature or label spaces~\cite{tuia2016domain}. Fig. \ref{domain_adapt} explains the difference between conventional DL and DTL techniques.

\subsection{Taxonomy of existing DTL techniques}

{DTL and classical TL differ in terms of the models used and the level of abstraction in feature representation. Classical TL employs traditional ML models with shallow architectures and relies on handcrafted features. In contrast, DTL utilizes DNNs with millions of parameters to automatically learn features from raw data. Classical TL directly applies a trained model to a related task, while DTL involves pre-training on a large dataset and fine-tuning on the target task. DTL generally exhibits higher generalization capability, as DNNs can capture complex patterns and transfer knowledge across domains.}





To date, there is no standardized and comprehensive technique for classifying DTL into categories. However, DTL algorithms could be classified into several types depending on \textit{what}, \textit{when}, and \textit{how} knowledge is transferred. Moreover, some studies have attempted to introduce a taxonomy of DTL-based ASR techniques. For example, Niu et al. \cite{niu2020decade} divide the DTL methods into two levels. The first one is divided into four sub-groups based on (i) the availability of labeled data, and (ii) the data modality in the source and TDs. Typically, this has resulted in inductive DTL, transductive DTL, cross-modality DTL, and unsupervised DTL \cite{kheddar2023deep}.
Table \ref{tab:4} summarises these possibilities. Going deeper, each sub-group in the first level can be further subdivided into four distinct learning types, including learning on instances, learning on features, learning on parameters, and learning on relations.

\begin{table}[ht!]
\caption{DTL possibilities. whereas the mark ($\varsubsetneq$) indicates that the domains/tasks are different but related, ($\exists!$) indicates that there exists one and only one domain/task, and ($\cong$) indicates that domains, tasks, or spaces are not always equals.}
\label{tab:4}
\begin{tabular}{{lllm{2.5cm}m{6cm}}}
\hline\noalign{\smallskip}
 &     Domains & Tasks &Math. propriety  &Sub-categories / Usage \\
 \hline
Traditional ML/DL & $\mathbb{D}_{S}= \mathbb{D}_{T}$  & $\mathbb{T}_{S}= \mathbb{T}_{T}$  & $X_{S}\neq X_{T}$, \newline $ Y_{S}= Y_{T}$& ASR model trained with $X_{S}$ database and used to recognise $X_{T}$ database.
\\ 
Inductive DTL & $\mathbb{D}_{S}\cong \mathbb{D}_{T}$  & $\mathbb{T}_{S}\neq \mathbb{T}_{T}$  & $X_{S}\neq X_{T}$, \newline $ Y_{S}\exists,  Y_{T}\exists$& If  $Y_{S}\exists$, DTL is multitask learning. If $ Y_{S}\nexists$, DTL is self-taught
learning, thus $\chi_{S}\cong\chi_{T}$. \\
Transductive DTL   &  $\mathbb{D}_{S}\neq \mathbb{D}_{T}$   & $\mathbb{T}_{S}= \mathbb{T}_{T}$ & $P(X_S)\neq P(X_T)$, \newline $ Y_{S}\exists,  Y_{T} \nexists$ , \newline $\chi_{S}=\chi_{T}$& When $\chi_{S}=\chi_{T}$, DTL is  is related to DA. If $   \mathbb{D}_{T}\exists!$ and $  \mathbb{T}_{T}\exists!$, DTL is used for sample selection bias or covariate shift.\\
Cross-modality DTL  & $\mathbb{D}_{S}\neq\mathbb{D}_{T}$   & $\mathbb{T}_{S}\neq \mathbb{T}_{T}$ & $P(Y_{S}/X_{S})\neq P(Y_{T}/X_{T})$,\newline$Y_{S}\neq Y_{T}$,  $\chi_{S}\neq\chi_{T}$ &
i,e. the dataset $X_{S}$ of $\mathbb{D}_{S}$ is speech data, and the dataset $X_{T}$ of $\mathbb{D}_{T}$ is text data. \\
Unsupervised DTL  & $\mathbb{D}_{S}\varsubsetneq\mathbb{D}_{T}$   & $\mathbb{T}_{S}\varsubsetneq \mathbb{T}_{T}$ & $Y_{S}\nexists, Y_{T}\nexists $ & DTL used for clustering, dimensionality reduction,
and density estimation, etc.\\
 \hline
\end{tabular}
\end{table}

\subsubsection{Inductive DTL}
In comparison to classical ML, which may be used as a reference for DTL comparison, and given that the target tasks $\mathbb{T}_{T}$ are distinct from the source tasks $\mathbb{T}_{S}$, the goal of inductive DTL is to enhance the target prediction function $\mathbb{F}_T$ in the TD, mentioned above in subsection \ref{sub12}. However, the SD $\mathbb{D}_{S}$ and TD $\mathbb{D}_{T}$ may not always be the same (Table \ref{tab:4}). The inductive DTL can be stated similarly to the following two cases, depending on whether labeled or unlabeled data is available: \\
\noindent \textbf{(a) Multi-task DTL:} The SD has a huge labeled database ($X_{S}$ labeled with $Y_{S}$), which is a distinctive form of multi-task learning. However, with multi-task approaches, many tasks $(T_1, T_2,\dots, T_n)$ are learned at the same time (in parallel), including both source and target activities (tasks). 

\begin{figure}[t!]
\begin{center}
\includegraphics[width=1.0\textwidth]{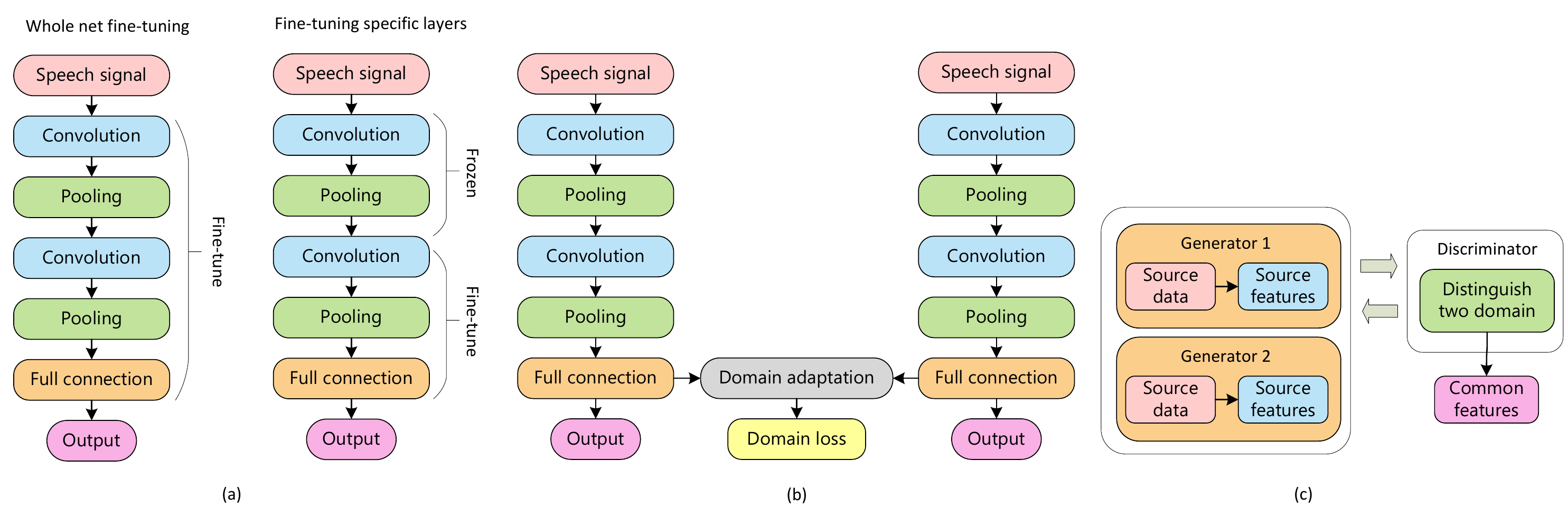}\\
\end{center}
\caption{Structures of: (a) Fine-tuning, (b) DA, and (c) DTL-based GAN.}
\label{fig:4}
\end{figure}

\noindent \textbf{(b) Sequential DTL:} ( Commonly known as  \textit{self-taught learning}) Dataset is not labeled in the SD ($X_{S}$ is not labeled with $Y_{S}$) but the labels are available in the destination domain ($X_{T}$ is labeled with $Y_{T}$). Sequential learning is a DL system that can be realized in two steps for classification purposes. The first step is the feature representation transfer, which is learned from a large collection of unlabeled datasets, and the second stage is when this learned representation is applied to labeled data to accomplish classification tasks. Hence, sequential DTL is a method of sequentially learning a number of activities (Tasks). The spaces between the source and destination domains may differ. For example, let us suppose we have a pre-trained model $M$ and consider applying DTL to a number of tasks $(T_1, T_2,\dots, T_n)$. We learn a specific task $\mathbb{T}_{T}$ at each time step $t$, which is slower than multi-task learning. However, when not all the tasks are present during training time, it might be beneficial. Sequential DTL is additionally classified into several types \cite{alyafeai2020survey}:
\begin{itemize}
    \item [1-] \textbf{Fine-tuning:} The principle is to learn a new function $\mathbb{F}_T$ that translates the parameters $\mathbb{F}_T(W_S) = W_T$ by using $M$, given a pre-trained model $M_S$ having $W_S$ as weights and target task $\mathbb{T}_{T}$ having $W_T$ as weights. The settings can be adjusted across all layers or just some of (Fig. \ref{fig:4} (a)). The learning rate for each layer could be distinct (discriminative fine-tuning). A new set of parameters $K$ could be added to most of the tasks so that $\mathbb{F}_T(W_T, K) = W_S\circ K$.
\item [2-] \textbf{Adapter modules:} Given an $M_S$ model  that has been pre-trained and output $W_S$, for a target task $\mathbb{T}_{T}$. The adapter module aims to lunch a different set of parameters $K$ that is too much less than $W_S$, i.e, $K\ll W_S$. $K$ and $W_S$ must have the ability to be decomposed into more compact modules such that,  $W_S=\{w\}_n$ and $K=\{k\}_n$. The adapter module permit learning the following new function $\mathbb{F}_T$:
\begin{equation}
\label{adapt}
 \mathbb{F}_T(K, W_S)= k_{1}'\circ w_{1}\circ \dots  k_n'\circ w_n.    
\end{equation}

According to equation \ref{adapt}, during the adaptation procedure, the set of original weights $W_S=\{w\}_n$ is left unaltered, but the set of weights K is changed to $K'=\{k'\}_n$. The principle of the adaptation domain is illustrated in Fig. \ref{fig:4} (b).

\item [3-] \textbf{Feature-based:} Interested only in learning concepts and representations, at various levels, such as word, character, phrase, or paragraph embedding $E$. The collection of $E$ based on a model $M$ remains unaltered, i.e., $\mathbb{F}_T(W_S, E) = E \circ W'$, in the way that $W'$ is fine-tuned.  For example, researchers have applied the generative adversarial network (GAN) principle to DTL where, the generators send features from the SD and the TD to a discriminator, which determines the source of the features and feeds the result back to the generators until they can no longer be distinguished. In this procedure, GAN obtains the common properties of two domains, as shown in Fig. \ref{fig:4} (c).
\item [4-] \textbf{Zero-shot:} Is the easiest method among all of the others. Making the assumption that the parameters $W_S$ can't be modified or add $K$ as a new parameter to a pre-trained model $M_S$ using $W_S$. To put this into context, in zero-shot there is no training technique to optimize or learn new parameters.
\end{itemize}

\subsubsection{Transductive DTL}
Compared to the traditional ML, which can be considered as a reference for DTL comparison, and given that the TDs $\mathbb{D}_{T}$ are distinct from the SDs $\mathbb{D}_{S}$. The SD has a labeled dataset ($X_{S}$ labeled with $Y_{S}$), whereas the TD has no labeled dataset, the source and target tasks are equal ( Table \ref{tab:4}). The goal of transductive DTL is to build the target prediction function $\mathbb{F}_T$ in the $\mathbb{D}_{T}$ by knowledge of the $\mathbb{D}_{S}$ and $\mathbb{T}_{T}$. Furthermore, the transductive DTL environment may be further classified into two categories depending on different conditions between the source and destination domains:\\
\noindent \textbf{(a) Domain adaptation (DA):} The feature spaces across domains, $\chi_{S}$ and $\chi_{T}$, are the identical, but the marginal probability distributions of the input dataset are not, $P(Y_{S}/ X_{S})\neq P(Y_{T}/ X_{T})$. For example, an assessment may be done on the topic of the resort in the $\mathbb{D}_{S}$ and it will be used to train a model for restaurants in the $\mathbb{D}_{T}$. DA is mostly effective when the $\mathbb{T}_{T}$ has a distinct distribution or there is a scarcity of labeled data. 

\noindent \textbf{(b) Cross-modality DTL:} Also known as cross-lingual DTL in the spoken language field, most DTL methods, more or less, a connection in feature spaces or label spaces is required between $\mathbb{D}_{S}$ and $\mathbb{D}_{T}$. In other words, DTL can only occur when the source and destination data are both in the same modality, like video, speech, or text. cross-lingual DTL, in contrast to all other DTL approaches, is one of the most complicated issues in DTL. It is assumed that the feature spaces of the source and destination domains are completely distinct ($\chi_{S}\neq\chi_{T}$), as in speech-to-image, image-to-text, and text-to-speech. Furthermore, the label spaces of source $Y_S$ and destination $Y_S$ domains might differ ($Y_S\neq Y_T$).

\noindent \textbf{(c) Unsupervised DTL:}  Intends to enhance the learning of the target predictive function $\mathbb{F}_T$ in $\mathbb{D}_{T}$ using the knowledge in $\mathbb{D}_{S}$ and $\mathbb{T}_{S}$, where $\mathbb{T}_{S}$ different from $\mathbb{T}_{T}$ but related, and $Y_S$ and $Y_T$ are not visible, given a SD $\mathbb{D}_{S}$ with a learning task $\mathbb{T}_{S}$, a TD $\mathbb{D}_{T}$ and a matching learning task $\mathbb{T}_{T}$ ($\mathbb{D}_{S}$ different from $\mathbb{D}_{T}$, but related).

\subsubsection{Adversarial DTL} 
In contrast to the methods described above for DTL, adversarial learning \cite{wang2020transfer} aids in the learning of more transferable and discriminative representations. The work in \cite{ganin2016domain}, was the first that introduced the domain-adversarial neural network (DANN). Instead of using a predefined distance function like maximum mean discrepancy (MMD), the core idea is to use a domain-adversarial loss in the network. This has greatly aided the network's ability to learn more discriminative data. Many studies have used domain-adversarial training as a result of DANN's idea \cite{bousmalis2016domain,chen2019joint,long2017deep,zhang2018collaborative}. All of the previous work ignores the different effects of marginal and conditional distributions in adversarial TL, whereas in \cite{wang2020transfer}, the proposed scheme, named dynamic distribution alignment (DDA), can dynamically evaluate the importance of each distribution.

\subsection{ASR conceptual background}
\subsubsection{Structure of ASR systems} 
\begin{figure}[ht!]
\centering
\includegraphics[scale=0.65]{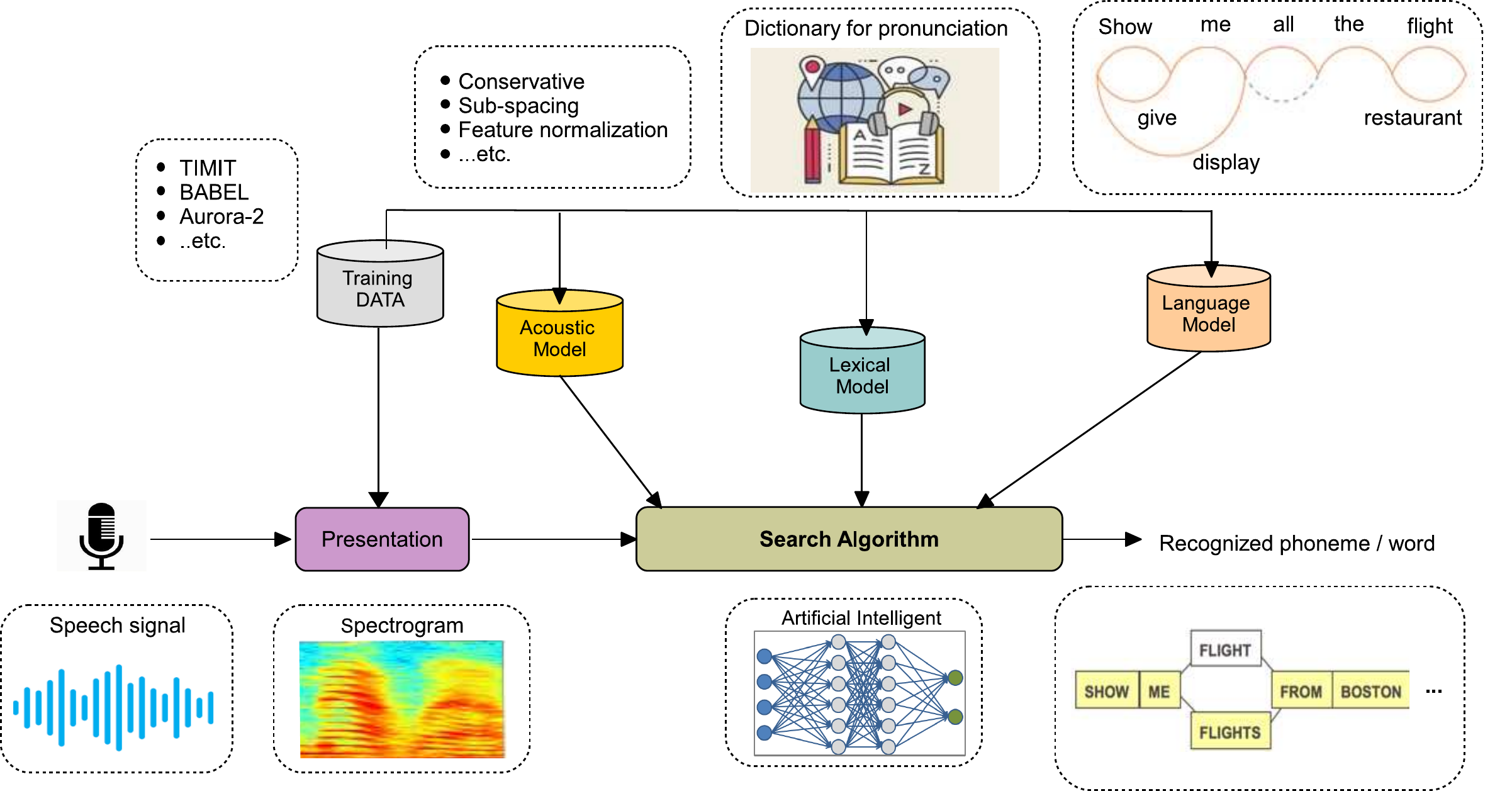}
\caption{ Flowchart of end-to-end ASR framework.}
\label{fig:ASR}
\end{figure}

The speech signal is embedded in an ASR module, which in turn, converts the speech to a list of words (text format). A list of candidate texts is generated throughout the ASR process, and the most suitable text for the original sound signal is eventually chosen. An acoustic front-end processes the speech data to extract usable characteristics in a conventional ASR system before creating a feature vector. 
In doing so, different kinds of features can be extracted using the principle component analysis (PCA) characteristics, Cepstral mean subtraction (CMS) data, linear predictive coding (LPC), independent component analysis (ICA), linear discriminant analysis (LDA), Cepstral analysis, filter-bank analysis, Mel-frequency cepstral coefficients (MFCC), kernal-based feature extraction, wavelet analysis dynamic feature extraction, spectral subtraction \cite{filippidou2020alpha}. 
Moving on, a decoder (search algorithm) uses the acoustic lexicon and LM to construct the hypothesized word or phoneme in the processing stage, as demonstrated in Fig. \ref{fig:ASR}.

The AM comprises acoustic characteristics for each of the different phonetic units. It often refers to the operation of generating statistical measures for characteristic vector sequences of the audio waveform. Typically, HMM is one of the most frequently utilized AMs. Segmental and super segmental models, neural networks, maximum entropy models, and conditional random fields are some of the other auditory models. A file containing statistical measures of various speeches that make up a word is known as an AM. The lexicon comprises terms from the current application's vocabulary. The limits connected with the word sequence that is acceptable in a specific language form the LM \cite{filippidou2020alpha}. The Stanford research institute language modeling and statistical language modeling (SLM) are two widely used toolkits for language modeling. Using appropriate models, the decoder attempts to identify the most likely word sequences that match the audio stream. The decoding algorithms then generate the n-best list \cite{filippidou2020alpha}.

\subsubsection{Evaluation criteria in ASR}
ASR techniques, including voice search, games, and interactive systems in the context of ASR, have substantially improved human-machine communication in recent years. For this purpose, several methods have been adopted by the ASR research community to assess the quality and generalizability of ASR techniques, which are as follows:

\begin{itemize}
\item \textbf{WER:} the proportion of wrong words in terms of the total number of words processed represents the word error rate (WER), which is the most often used statistic for ASR assessment. This has been used in a large number of studies, such as in \cite{jiang2021gdpr,kumar2021exploration,chan2016listen,yu2021enhancing}. It is described as:
\begin{equation}
\mathrm{ WER}=\mathrm{\frac{S+D+I}{N}=\frac{S+D+I}{H+S+D}}. 
\end{equation}
Where $\mathrm{I, D, S, H}$, and $\mathrm{N}$ represent the number of insertions, deletions, substitutions, hits, and input words, respectively. The greater the performance of an ASR system, the lower the WER (i.e., better AM or LM). There are schemes that process character-by-character instead of word-by-word in the ASR evaluation, such as in \cite{yu2021enhancing,lin2021speech,bai2021fast}. To that end, the character error rate (CER) has been used instead of WER, although the evaluation principle remains the same. Besides, WER is also called phoneme error rate (PER) in schemes that adopt phoneme as a unit of measure rather than a word  \cite{zhang2017towards,qin2018towards,zhu2020end}. The word recognition rate (WRR) is a version of WER that may be used to assess  ASR performance such that $\mathrm{WRR=1-WER}$ and $\mathrm{N - (S + D)}$ is the total number of successfully predicted words \cite{malik2021automatic}.  

Despite being the most widely utilized, WER has different pitfalls or weaknesses. For example, it does not indicate how excellent a system is because there is no scale for comparison. Furthermore, it may reach 100\% under noisy situations since it gives more weight to $\mathrm{I}$ than $\mathrm{D}$ because WER is not D/I symmetric.

\item \textbf{RIL:} relative information lost (RIL) is determined using the Shannon entropy $H$ and is based on mutual information $I$ \cite{errattahi2018automatic}, which evaluates the statistical dependency between the input words $X$ and output words $Y$ (Equation
\ref{eqRIL}).

\begin{equation}
\label{eqRIL}
\mathrm{RIL=\frac{H(Y/ X)}{H(Y)}},
\end{equation}
where  
\begin{equation}
\mathrm{H(Y)=-\sum_{i=1}^{n}P(y_i)log P(y_i)},
\end{equation}
and
\begin{equation}
\mathrm{H(X/ Y)=-\sum_{i=1}^{n}P(x_i,y_i)log P(x_i,y_i)}.
\end{equation}

However, the RIL is still so far from becoming a good performance metric because it is challenging to assess zero error for any one-to-one mapping of input/output words, which does not meet the criteria for an ideal ASR assessment metric. The alternative is called word information lost (WIL), and it's a rough approximation of RIL. WIL, on the other hand, is easy to use since it is based only on WER parameters and is presented as \cite{errattahi2018automatic}:
\begin{equation}
\mathrm{WIL=1-\frac{H^2}{(H+S+D)(H+S+I)}}.
\end{equation}

\item \textbf{Accuracy, recall, precision, and specificity:} to test the efficiency of a suggested method in ASR field, the classification \textit{accuracy}, \textit{recall} (sensitivity) or commonly known as true positive rate (TPR), \textit{precision}  (known also as positive predictive
value), and  \textit{specificity} (commonly known as true negative rate (TNR)) are often used as assessment criteria for experiment results, such as in \cite{li2021insight,karaman2021robust,ramadan2021detecting}. The accuracy rate is defined as the proportion of properly assessed samples to the total number of samples. Three often used metrics to illustrate the accuracy of ASR tests are recall, precision, and specificity. The above-mentioned indications could well be written as follows:
\begin{equation}
 \mathrm{Accuracy (\%)= \frac{TP+TN}{TP+FP+TN+FN}\cdot 100},   
\end{equation}
\begin{equation}
\mathrm{Recall(\%)= \frac{TP}{TP+FN}\cdot 100}, 
\end{equation}
\begin{equation}
\mathrm{Precision(\%)= \frac{TP}{TP+FP} \cdot 100},
\end{equation}
\begin{equation}
\mathrm{Specificity(\%)= \frac{TN}{TN+FP}\cdot 100}. 
\end{equation}
Where TP, TN, FP, and FN refer to true positive, true negative, false positive, and false negative, respectively. Equal error rate (EER) is a metric used to predetermine the threshold values for its false acceptance rate and false rejection rate. The EER has been adopted as a metric in many ASR frameworks, such as in \cite{zhang2021multi,hong2017transfer}. It is calculated using the following formula:  
\begin{equation}
\mathrm{EER=1-\frac{1}{2}(specificity+sensitivity)}.
\end{equation}

Where, $\mathrm{\frac{1}{2}(specificity+sensitivity)}$ is called the area under curve metric (AUC). The lower the EER value, the higher the accuracy of the ASR algorithm. If the measure of recall and precision at a pre-specified
global threshold $Th$, the metric is called term weighted value (TWV) \cite{yusuf2019low}, such that:  
\begin{equation}
\mathrm{TWV(\theta, Th)=1-\frac{1}{|Th|}\sum_{t\in Th}  \big(P_{mis}(t,\theta)+\beta . P_{fa}(t,\theta)\big)}.
\end{equation}
Where, $\mathrm{P_{mis}(t,\theta)}$ and $\mathrm{P_{fa}(t,\theta)}$ are the probabilities of misses and false alarms respectively, and $\beta$ is a parameter that controls the relative costs of false alarms and misses. 

Another famous metric inspired by measures of recall and precision metrics is F-measure (or F1-score), which is calculated as follows:

\begin{equation}
    \mathrm{F-measure (\%)=\frac{2\cdot Precision\cdot Recall}{Precision + Recall}\cdot 100}.
\end{equation}

The F1 score replaces the accuracy metric when the FN and FP are crucial; besides, when imbalanced class distributions exist, the F1 score is a much better metric to evaluate ASR models, as explained in \cite{lu2021detection,arora2017study,kumar2021development}.

\item \textbf{Unweighted average recall (UAR):} it is also called the balanced accuracy. It is calculated by averaging the recall of all the classes regardless of how many samples exist in each class. UAR gives the correct expectation on class predictions despite the correlation between UAR and accuracy. UAR is defined as follows \cite{koike2020audio,markitantov2020transfer,durrani2021transfer}:

\begin{equation}
\mathrm{UAR=\frac{\sum_{i=1}^{N_c} Recall_i}{N_c}},
\end{equation}
\noindent where $\mathrm{N_c}$ is the number of classes.

\item \textbf{Mean opinion score (MOS):} is widely used as \textit{subjective} metric for assessing the naturalness of audio and audiovisual quality. 
The MOS is usually a single rational number between 1 and 5, with 1 being the worst and 5 being the best-perceived quality. 
In \cite{fahmy2020transfer}, MOS has been used as a subjective means to assess text-to-speech (TTS) conversion using DTL. 
In \cite{oord2016wavenet}, MOS has been used to subjectively evaluate the performance of the proposed WaveNet-based speech synthesizers, LSTM-RNN-based statistical parametric, and HMM-driven unit selection concatenative schemes. 

\item \textbf{Perceptual evaluation of speech quality (PESQ):} is an \textit{objective} voice automated evaluation of speech quality as perceived by a telecommunication system user \cite{recommendation2001perceptual}. It provides raw scores in the range –0.5 to 4.5. For example, in \cite{siddiqui2020using}, PESQ is used to assess speech enhancement concerning ASR. Besides, PESQ is used to evaluate DTL-based ASR scheme dedicated for PD patients in \cite{yu2021enhancing}. 
On the other hand, mapping between MOS and PESQ results in a new evaluation metric called mean opinion score-listening quality objective (MOS-LQO), also known as PESQ Rec.862.1, that allows a linear comparison of PESQ with MOS \cite{peng2019security}. The mapping function is expressed as follows:
\begin{equation}
\mathrm{MOS-LQO}=0.999 +\frac{4.999-0.999}{1+e^{-1.4945.PESQ +   4.6607}}.
\end{equation}

Moving on, different frameworks, including \cite{kheddar2019pitch,kheddar2022high, peng2019security} have utilized ASR principles as a steganalysis process and MOS-LQO as an assessment tool to check the integrity and/or the steganography quality-loss (SQ-Loss) between the original unprocessed speech and degraded speech version that has been passed through the steganography distorting system.

\item \textbf{Real time metrics:} 
the real-time factor (RTF) and the average processing time (APT) are frequently used to evaluate speed \cite{bai2021fast}.
RTF is a standard metric for assessing an ASR system's processing time cost. It is the average processing time for a one-second speech, which is defined as follows:
\begin{equation}
 \mathrm{RTF=\frac{\text{Total Processing Time}}{\text{Total Duration}}}
\end{equation}
Computing the APT helps account for the impact of utterance duration on processing time. Moreover, using APT helps demonstrate how quickly one utterance can be processed.
It is estimated as:
\begin{equation}
\mathrm{APT=\frac{\text{Total Processing Time}}{\text{Total Number of Utterance}}}   
\end{equation}

\item \textbf{Quantitative measure of generalizability in DTL}
In DTL, quantitatively measuring the generalizability can be challenging since it depends on various factors such as the specific task, the source and target domains, and the chosen DTL approach. However, there exist a few common techniques and metrics that can be used to assess generalizability:

\begin{itemize}
\item \textit{DTL Performance:} compare the model's performance with a baseline model that is trained from scratch on the target task alone. Metrics such as accuracy, precision, recall, F1-score, or AUC can be used to quantify the model's performance.
\item \textit{Domain similarity metrics:} Domain similarity metrics, such as domain adaptation or domain discrepancy measures, can be used to quantify the differences between the feature distributions of the source and target domains. Some popular metrics include maximum mean discrepancy (MMD) as described in Equation \ref{mmd}, and Kullback-Leibler divergence (KLD) as described in Equation \ref{kld}.

 \begin{equation}
\label{mmd}
    MMD(\mathbb{D}_{S}, \mathbb{D}_{T})=\norm[\bigg]{ \frac{1}{n_s} \sum_{i=1}^{n_s} x_s^i.\beta_s-\frac{1}{n_t} \sum_{i=1}^{n_t} x_t^i.\beta_t}_\mathcal{H}
\end{equation}
where $n_s$ and $n_t$  are the numbers of samples of the source and target domain, respectively. $\beta_s$ and $\beta_t$ denote the representation of the source and target datasets (i.e., $x_s^i$ $x_t^i$), respectively. $\Vert. \Vert_\mathcal{H} $
 represents the 2-norm operation
in reproducing kernel Hilbert space \cite{vu2020deep}.

\item \textit{Fine-tuning Analysis:}
When the pre-trained model is fine-tuned on the target task, the convergence behavior can be analyzed by observing the rate of convergence of the training and validation curves for the target task. Faster convergence and better generalization to the target task indicate higher generalizability \cite{tendle2021study}.

\item \textit{DTL stability:} generalizability can also be assessed by evaluating the stability of the DTL process. The steps are as follows: randomly split the source and target domains into multiple train-test splits, perform DTL on each split, and measure the variance in performance across different splits. Higher stability indicates better generalizability \cite{liu2022improved}.

\item \textit{Ablation studies:} gradually remove or modify elements such as pre-training, specific layers, or specific data augmentation techniques, and observe the impact on the model's performance. This helps to understand which factors contribute the most to generalizability \cite{liu2023dropout}.

\item \textit{Cross-validation:} split the target data into multiple folds, train and evaluate the model on each fold while ensuring that the training and validation data remain independent. Calculate the average performance across all folds to obtain a more robust estimate of generalizability \cite{bu2023achieving}.

\end{itemize}

\end{itemize}

\subsubsection{ASR datasets} 
Many datasets have been used in the literature for different ASR tasks. Table \ref{tab:dataset} lists some of the datasets used for DTL-based ASR applications and their characteristics. Typically, only publicly available repositories are reported in this table. It is also worth noting that some of these datasets have been updated many times and have become more developed over time.

\begin{table}[H]
\caption{List of publicly available datasets used for DTL-based ASR applications}. 
\label{tab:dataset}
\small
\resizebox{!}{160pt}{
\begin{tabular}{lp{1.5cm}p{4cm}p{7cm}}
\hline
Dataset & Used by & Default ASR task & Characteristics \\
 \hline
TIMIT \cite{garofolo1993darpa}& \cite{qin2018towards,li2021insight,wang2020cross} & Acoustic-phonetic knowledge and ASR evaluation & 630 speakers speak 10 phonetically-rich phrases, reflecting eight major dialect divisions of American English. \\
LibriSpeech \cite{panayotov2015librispeech}& \cite{luo2021group,tang2021general,jia2018transfer,yu2022enhancing} & Train and evaluate speech recognition systems  & corpus comprises 1000 hours of speech sampled at 16 kHz and is derived from audiobooks that are part of the LibriVox project.\\
MuST-C \cite{di2019must}&\cite{tang2021general}&Facilitate the training of end-to-end systems from English into 8 languages & Contains at least 385 hours of audio recordings from English Talks that are automatically matched with their hand transcriptions and translations at the sentence level.\\

VCTK \cite{veaux2017superseded}  &\cite{jia2018transfer}&English multi-speaker corpus for voice cloning & Speech data from 110 English speakers with diverse accents is included. Approximately 400 sentences are read by each speaker.\\
GlobalPhone \cite{schultz2013globalphone} &\cite{sahraeian2018cross}& Multilingual speech processing dictionary resources& Contains more than 400 hours of transcribed audio record from over than 2000 native participants in 20 different languages.\\
IARPA Babel \cite{hartmann2017analysis}&\cite{yusuf2019low,yi2018language}&Analysis of cross-language  & Consists of 25 languages with approximately 40 hours of transcribed speech. \\
CinC \cite{liu2016open}&\cite{koike2020audio}&Classification of heart sound &It contains 2,435 heart sound recordings from 1,297 healthy and sick people, including those with heart valve dysfunction and coronary artery disease.\\
UASpeech \cite{kim2008dysarthric} &\cite{xiong2020source,shahamiri2021speech}&Dysarthric speakers classification&There are 15 dysarthric speakers with cerebral palsy and 13 normal speakers. Each speaker has three blocks of speech.\\
DCASE \cite{mesaros2016tut} &\cite{boes2021audiovisual,arora2017study,chen2018transfer}&Acoustic scene classification
and sound event detection&Binaural recordings from 15 distinct sound settings.\\
\hline
\end{tabular}}
\end{table}

\section{DTL-based ASR applications} \label{sec4}
As detailed in Section \ref{sec1}, the ASR field is split into two main axes, LM and AM. The LM in the ASR algorithm guides applicant searches and evaluates decoding output quality. Conventional statistical LMs, like the backoff n-gram LM, have been used in this field for decades because of their simplicity and reliability \cite{jiang2021gdpr}. The work in \cite{devlin2019bert} proposes the bidirectional encoder representations from transformers (BERT), which uses the popular attention model for an LM. It has been demonstrated that bidirectionally trained LMs have a better meaning of language context and flow than single-direction LMs. 
In terms of AM, DL-based AMs like the deep neural network-hidden Markov model (DNN-HMM) have made significant progress in the ASR research field, such as in \cite{wang2019overview,novoa2018uncertainty,fahad2021dnn}. The connectionist temporal classification (CTC), as suggested in \cite{wang2019overview,nakatani2019improving,salazar2019self}, is a fully end-to-end AM training method that does not need data pre-alignment and only requires one input and output sequence to train. The challenge of ASR is simply a direct conversion problem between two variable-length sequences. The excellent model structure and outstanding performance of the sequence-to-sequence (Seq2Seq) \cite{chiu2018state} model allow the voice recognition problem to be solved without using an LM or a pronunciation dictionary. 

On the other hand, ASR systems often offer "one-model-fits-all" to all users. Due to the disparity between trained and tested data, "one-model-fits-all" ASR systems invariably experience substantial performance loss in certain circumstances. Meanwhile, because the quantity and variety of utterances used to train the ASR algorithm are crucial for AM rendering, the speech data kept on users' devices (due to security and property protection reasons) is an excellent resource for ASR researchers looking to improve their ASR framework. 
With tight data privacy protection, ASR researchers have confronted problems in getting speech data generated in real-life circumstances. As a result, developing new frameworks that effectively incorporate DTL is becoming increasingly important to evade the above issues and achieve excellent performances.
To date, numerous research studies have focused on investigating how to enhance the performance of existing ASR schemes, and many frameworks have suggested applying DTL and DA. 
For example, the work in \cite{sukhadia2023domain} describes techniques to improve the performance of target-domain models in the field of ASR. The authors explore the use of encoder layers and embeddings from a well-trained ASR model, along with the application of spectral augmentation (SpecAug), to enhance the performance of low-resource target-domain models using DA along with model-based CTC that employs multi-task learning as DTL techniques. The proposed method shows a significant average relative improvement of approximately 40\% over the baseline. Always in the field of DA, \cite{fan2022towards} explores the possibility of improving DA for self-supervised models, with a specific focus on child ASR. The researchers introduce a novel framework called DRAFT, which aims to reduce domain shifting in pre-trained speech models. Through extensive experiments, they evaluate the effectiveness of DRAFT on different transformer architectures such as autoregressive predictive coding (APC), Wav2vec2.0, and HuBERT. Tests indicate that HuBERT when fine-tuned with the DRAFT framework, demonstrates the highest performance in terms of ASR accuracy on the OGI dataset. Similarly, Bethan et al. \cite{thomas2022efficient} discuss the efficient transfer of self-supervised speech models for ASR using adapters. The paper explains the concept of self-supervised learning and demonstrates how adapters can effectively reduce the number of parameters in pre-trained upstream models, enabling task-specific adaptation for target downstream ASR tasks.

\begin{figure}[ht!]
\centering
\includegraphics[height=7.5cm, width=18cm]{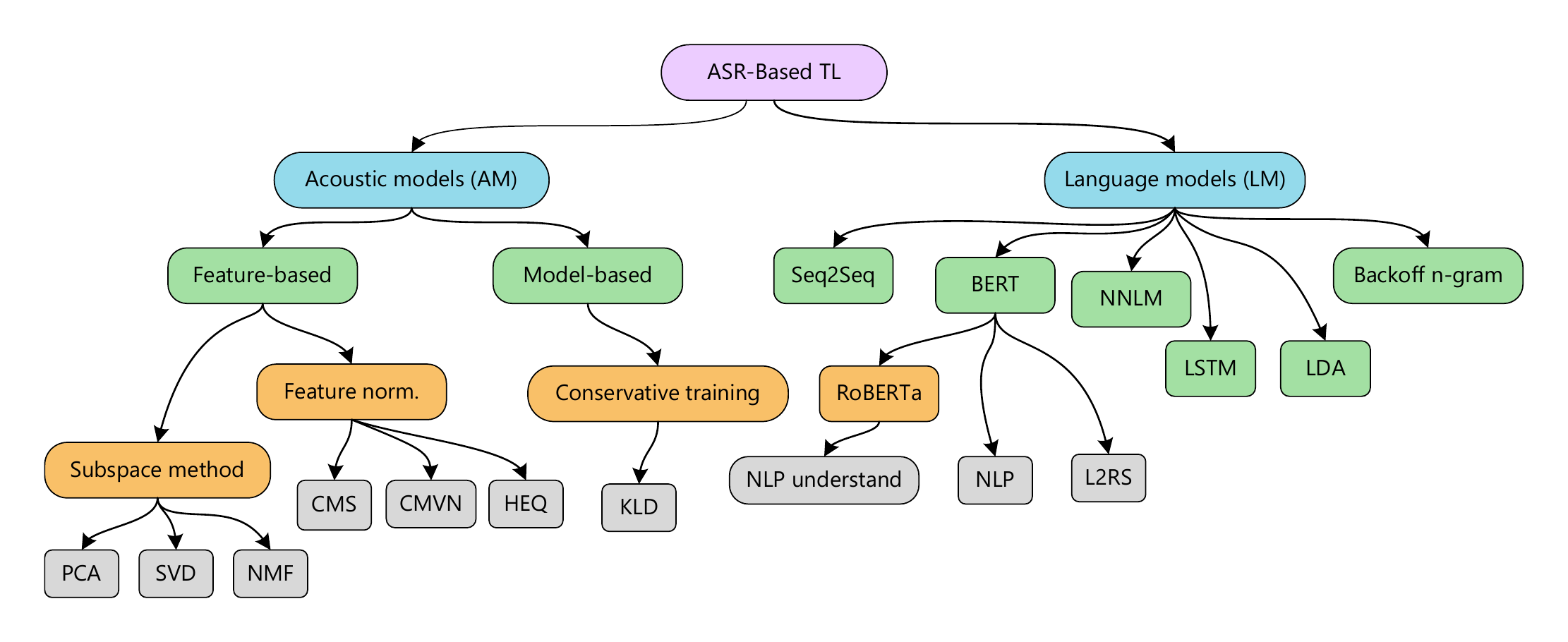}
\caption{Summary of DTL-based ASR subcategories and their most used models.}
\label{fig:04}
\end{figure}

Fig. \ref{fig:04} summarises the existing state-of-the-art DTL-based ASR subcategories.

\subsection{DTL for acoustic models (AMs)} 
The end-to-end and layered DNN-HMM models are two types of current DL–based AMs. In AM, DNN extracts high-level features from acoustic signals, e.g., MFCCs and HMM lexical sequences required for decoding into transcripts. The DNN takes acoustic characteristics as input and outputs context-dependent lexical units (tri-phonemes) that correspond to the input of the HMM component downstream. On the other hand, the end-to-end model is purely a DNN technique that receives acoustic characteristics (features) as input and immediately outputs the recognition rate. A typical end-to-end framework for voice recognition is shown in Fig. \ref{fig:e2e}. 
The neural network creates embeddings from input features, passing to a stack of recurrent layers. The recurrent layers create a final output by looking for patterns based on prior and current input information. Backpropagation is used to train the network with the CTC loss \cite{mridha2021study}. Generally, three main DTL strategies have been used with AMs:

\begin{figure}[t!]
     \centering
      \includegraphics[scale=1]{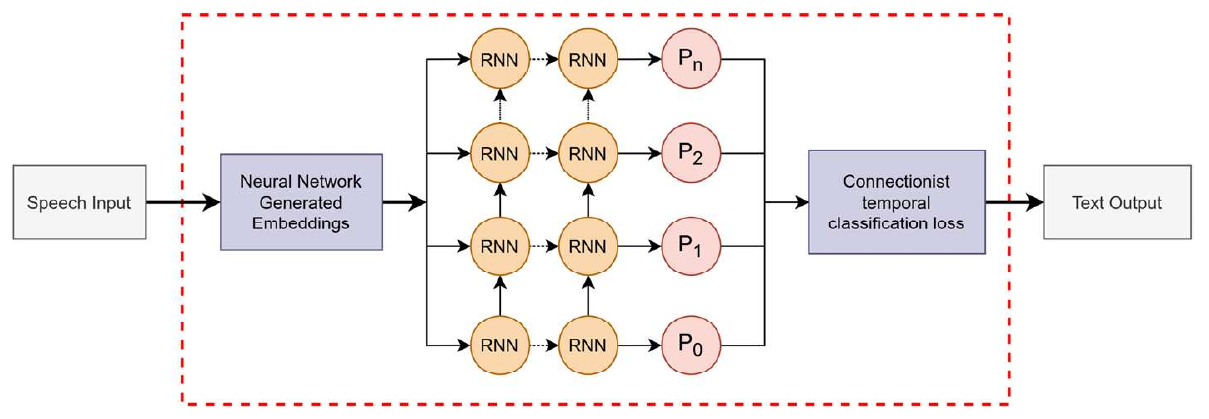}
      \caption{ An example of end-to-end source model for DTL-based ASR \cite{mridha2021study}. }
     \label{fig:e2e}
     \end{figure}

\subsubsection{Feature normalisation based-DTL} 
\begin{figure}[ht!]
     \centering
      \includegraphics[scale=0.7]{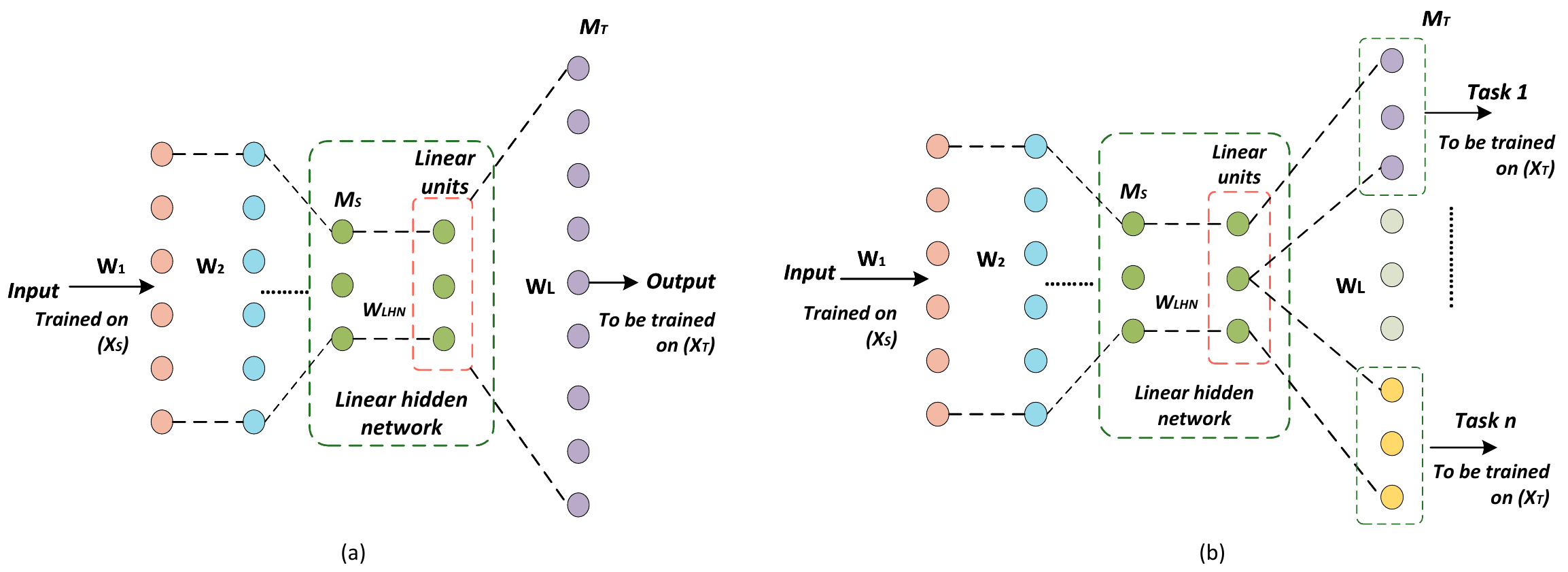}
      \caption{The fundamentals of transformation in DTL. The weights connected with the links in the dashed rectangles are estimated, while the rest of the weights are left unaltered: (a) feature normalization for mono-task DTL, and (b) feature normalization for multi-task DTL.}
     \label{fig:05}
     \end{figure}

The idea behind the linear transformation is that speech characteristics can be normalized by linear mapping. To perform linear mapping, a transformation network (or transformation layer) is simply added to an existing network. It is a prominent neural network adaption approach. The last hidden layer is generally designed to be a bottleneck in order to limit the number of parameters to adjust (less neurons), as described in Fig. \ref{fig:05}. The transformation can be a linear input network or a linear output network. 

If the last hidden network (LHN) is considered as a feature extractor and the output layer as a discriminative source model ($M_S$). The weights of the output layer's linear transform matrix, $W_L$, represent the target model, $M_T$, parameters. $M_T$ may now be written as follows \cite{huang2015maximum}:
    \begin{equation}
       \mathrm{M_T=softmax(W_L M_S)}.
    \end{equation}
Where the activation at the last hidden layer of $M_S$ can be used
as the new feature representation extracted by the hidden layers in $M_T$. Applying a transformation matrix $W_{LHN}$ on the model parameters to generate an adapted model parameter set is equal to adding an extended LHN after the last hidden layer \cite{huang2015maximum}:
\begin{equation}
        \mathrm{M_T=softmax(W_{LHN} W_L M_S)}.
\end{equation}

Many existing ASR research studies adopt the linear transformation strategy. For instance, the authors in \cite{elaraby2016deep} attempt to improve current models of a computer-aided language learner (CAPL) system that teaches Arabic pronunciation for Quran recitation regulations. They implemented four significant improvements: first, they used speaker adaptive training (SAT) to reduce inter-speaker variability; second, they integrated a hybrid DNN-HMM model to improve the AM and reduce the PER. Third, they combined the hybrid DNN with minimum phone error (MPE). Finally, they employed a grammar-based decoding graph in the testing phase to narrow the search area to the most common sorts of mistakes. 
In \cite{mimura2016joint}, a multi-target learning approach is used to present a combined optimization technique for the output of denoising auto-encoder (DAE) and the input of DNN. The output of DAE is trained in the first stage to reduce the mistakes propagated by the input of DNN. After that, the DAE and DNN unified network is fine-tuned for the phone state ASR objective, with an additional target of input voice augmentation placed on the DAE portion. 
Besides, in \cite{ma2017approaches}, the authors included an adaptation layer for fine-tuning and employed non-linearity to learn a function that is considerably more complicated than the linear transformation in the softmax layer. In the pre-trained model, they adjusted the cluster softmax and NNadapt layers during the fine-tuning phase while leaving the settings of the other layers unchanged. {The authors in \cite{kadyan2022transfer} explores data augmentation techniques to address data scarcity challenges and improve neural network consistency. Elastic spectral distortion and TL, particularly spectrogram augmentation, are proposed as methods to create synthetic datasets. Perturbation-based domain data augmentation is performed. The proposed ASR system utilizes DNN with concatenated feature analysis (MFCC, pitch features) and vocal tract length normalization on pooled datasets, resulting in a considerable improvement over the baseline system.} More schemes that employ linear transformation techniques and their performance details are summarised in Table \ref{tab:5}.


\subsubsection{Conservative training for DTL} 
Since trained models and only a limited quantity of spoken data can be used to accomplish performance adaptation, conservative training has become a widely used method of accent adaption. It is efficient and does not require excessive data to provide a reasonable outcome. However, it requires too many parameters, which may disrupt the model's structure. KL-Divergence (KLD) is a widely used conservative DNN-based DTL algorithm for ASR. KLD regularization has a neat and distinct mathematical model, essential for DL training. By minimizing the loss, KLD-regularized adaptation aims to make the output distributions of the source model $M_S$ and target model $M_T$ more similar. The KL-divergence prevents overfitting and keeps the adapted $M_T$ closed to the $M_S$ domain \cite{jiang2021gdpr,weninger2019listen}. 
Assuming that the loss functions in the source and TDs for training DNNs are $\mathbb{D}_{S}$ and $\mathbb{D}_{T}$, respectively, the conservative approach can be mathematically summarized in terms of model-based DTL with KLD-regularization as follows \cite{jiang2021gdpr,weninger2019listen}:
\begin{equation}
\label{kld}
 \mathbb{D}_{T}^{KLD}=(1-\rho) \mathbb{D}_{S} +\frac{\rho}{N} \sum P_S(x_T/ x_T)log P_T(y_T/ x_T),
\end{equation}
where ($x_T,y_T$) is the speech sample collected from the $\mathbb{D}_{T}$, $N$ is the number of speech samples in   $\mathbb{D}_{T}$, and $\rho$ is a hyper-parameter that controls the transfer ratio from $\mathbb{D}_{S}$. For example, DTL-based ASR using conservative training and a KLD-regularization algorithm have been employed in \cite{jiang2021gdpr} to build target AM. 
In \cite{weninger2019listen}, both KLD and LHN have been employed for speaker adaptation using a pre-trained seq2seq ASR model. Other techniques deploying conservative techniques along with their performance details are summarised in Table \ref{tab:5}.  


\subsubsection{Subspace-based DTL} 
It aims to identify a subspace of every model parameter or transformation and creates them as a point in the subspace. PCA, singular value decomposition (SVD), and nonnegative matrix factorization (NMF) are unsupervised approaches for data dimensionality reduction. They are widely used in subspace-based DTL for ASR applications. PCA is an orthogonal transformation that maps higher-dimensional data to smaller-dimensional subspaces while keeping the original variables' correlation and the original data's maximum variance on the lower-dimensional representation. 
When data comes to dimensionality reduction, SVD and PCA are practically similar. Only a specific number of singular values are chosen while using SVD to condense networks. The target weight matrix has several nonlinearities in the left-singular and right-singular matrices. 
As a result, neglecting some matrix components to build a linear project layer might result in harmful losses in some cases. In contrast to SVD, practically all NMF algorithms require at least one nonnegative matrix. As a result, the target matrix might be defined as the weighted sum of the base matrix's columns. 
This is an important limitation since it reduces the importance of the coefficient matrix when it simply needs to maintain the fundamental component of the original matrix. This can explain why NMF is more interpretable than SVD, and PCA \cite{qin2018towards}. 
Convex nonnegative matrix factorization (CNMF), a variant of NMF, has been used in \cite{qin2018towards} to extract high-level features. Then, a DTL is applied for both high and low-level features through multilingual training and multi-task learning. These techniques produce an incredibly relative performance compared to state-of-the-art speech ASR studies. Other studies that employ subspace techniques and their performance details are summarised in Table \ref{tab:5}.

\subsection{DTL for language models (LMs)} 
The backoff n-gram model is widely used as an LM since it is the most prevalent option in ASR systems. The backoff n-gram LM may be defined as a collection of tuples containing an n-gram and its related logarithm probability. 
In \cite{jiang2021gdpr}, the LM of the source $M_S$, which is considered the teacher model, is used to perform a DTL by linearly interpolating the LM trained on student speech ($X_T$) with the teacher LM already trained on teacher speech ($X_S$) to get target model $M_T$ (or student LM), the process can be expressed as follows:
\begin{equation}
    \mathrm{P(w)=\lambda P_{LM}^{S}(w)+(1-\lambda)P_{LM}^{T}(w)},
\end{equation}
where $\mathrm{P(w)}$ and $\mathrm{P^T_{LM}(w)}$ represent the probability of the n-gram w, provided by the source (teacher) LM, and the probability provided by the LM trained on the student data, respectively. $\mathrm{\lambda}$ is a hyper-parameter that adapts the teacher and student semantic model's weights. This type of DTL works well to increase the probability of n-grams and simultaneously maintain a broad coverage of general-purpose n-grams.

Besides, it is persistent in the field of data mining to transfer information from pre-trained models to new tasks \cite{devlin2018bert}. According to \cite{deena2018recurrent}, n-gram LMs are based on relative frequencies of n-gram events. Therefore, the adaptation techniques can further be broadly classified (in addition to the model interpolation) into (i) the \textit{constraint specification}, which entails the combination of multiple sources of data in the form of characteristics using methods, such as exponential models and the maximum entropy criterion; and (ii) the \textit{mixture language models} that extract topic information from the implicit text data to calculate the sub-models weights. Besides n-gram LM, LM-based ASR is built based on other state-of-the-art pillar models:


\begin{center}
\begin{longtable}[!t]{p{0.5cm}p{2cm}p{3cm}p{2cm}lp{2.5cm}p{2.8cm}p{1.6cm}}
\caption{A summary of the recent state-of-the-art frameworks organized by the pre-trained models employed, the tasks addressed, the sort of DTL technique utilized, and the performance metric used. Whereas the marks  ($\nearrow$) and ($\searrow$) indicate improvement and reduction, respectively. If many scenarios have been conducted in one metric, only the best result is mentioned.}
\label{tab:5}
\small
\label{tab:metrics} \\
\hline
Ref. &Model-based   & ASR Tasks ($\mathbb{T}_{T}$) $\nearrow$& DTL & LM/AM & Adaptation & Metric& {Code available?}   \\ 
\hline
\endfirsthead
\multicolumn{3}{c}{Table \thetable\ (Continue)} \\
\hline
Ref. &Model-based   & ASR Tasks ($\mathbb{T}_{T}$) $\nearrow$& DTL & LM/AM & Adaptation & Metric& {Code available?}\\ 
\hline 
\endhead
\hline
\endfoot
\hline \hline
\endlastfoot 
\cite{durrani2021transfer}&ResNet34 \newline (English)&Speech affect\newline (Urdu)&Cross-lingual &AM&Fine-tuning& 12.05\% UAR$\nearrow$ 
 & No\\
 
 \cite{jiang2021gdpr}&TFE&Vendor-client \newline ecosystem &Inductive&Both&KLD and \newline Linear norm.&$\simeq$ 50\% WER $\searrow$ & {Yes\footnote{\url{https://gitee.com/WeBank/FATE}}}\\
 \cite{bai2021fast}&BERT \newline (Text model)& LASO\newline (speech model)&Cross-modality&Both&Linear transform.& APT 50$\times \searrow$, \newline 12\% CER$\searrow$ &{Yes\footnote{\url{https://github.com/kaldi-asr/kaldi/blob/master/egs/aishell/s5/}}}\\
 \cite{qin2018towards}&CTC-attention &learning-based \newline end-to-end &Multilingual \newline Multi-task &LM&NMF& 03.90\% PER$\searrow$ & No\\
\cite{zhang2021multi}& ResNet-34 \newline (Near-Field) &Speaker verification\newline ( Far-field) &Inductive  &AM&Feature norm. \newline and space-level& 38.6\% EER$\searrow$ & No\\
\cite{hong2017transfer}&PLDA&Speaker verification&Transductive&AM&KLD&6.6\% EER$\searrow$ & No \\
\cite{lu2021detection}&AlexNet &Marine mammal sounds classification&Inductive& AM& Fine-tuning& Accuracy=99.96\%  &  No\\

\cite{koike2020audio}&PANNs&Heart sound&Transductive&AM&Fine-tuning&89.7\% UAR & No\\

\cite{shahamiri2021speech}&S-CNN&Dysarthric speech & Inductive &AM&Fine-tuning & 67\% Accuracy $\nearrow$ & No   \\

\cite{weninger2019listen}  & Seq2Seq & Speaker adapt. & Transductive & Both & KLD and LHN & 25.0\% WER$\searrow$ & No\\

\cite{deena2018recurrent}&RNNLM& Multi-genre broadcast&  Fine-tuning & Both & LHN with\newline K-Component  &16.0\% WER$\searrow$, \newline 02.0\% F1–score$\nearrow$ & No\\

\cite{song2019topic}&LDA&DSTM&Inductive&LM&Fine-tuning&1.48 \% WER$\searrow$ & No\\
\cite{hentschel2018feature}& RNNLM &LDA&Feature norm. &LM&LHUC& 16.0\% WER$\searrow$ & \\
\cite{ng2020cuhk}&DNN-HMM &Children’s speech&Transductive&Both&LDA, STC, and  \newline fMLLR &15\% CER$\searrow$ & No\\
\cite{chen2020darts} & DARTS-ASR & Multilingual adapt. & Transductive  & LM &Sub-spacing&  10.2\% CER $\searrow$ & No\\
\cite{sun2017unsupervised} & Clean speech training & Noisy speech testing &  Unsupervised& AM & DDA &37.8\% WER$\searrow$ & No\\
\cite{ghahremani2017investigation}&LF-MMI & LF-MMI & Multi-task & LM &TDNN with \newline i-vectors & 02.0\% WER$\searrow$ & No\\
\cite{huang2016unified}&CD-DNN-HMM&Speaker adapt.&Sequential \newline and Multi-task &AM&Feature norm.&02.16\% WER$\searrow$ & No\\
\cite{turan2021improving}&CNN/HMM&Throat microphones &Cross-domain&LM&SDA &12\% PER$\searrow$ & No\\
\cite{shivakumar2020transfer}&DNN adult’s speech& Children’s speech &Cross-domain &AM&SAT \newline VTLN &14.66\% WER$\searrow$ \newline  03.25\% WER$\searrow$ & No\\
\cite{sayed2021bimodal}&VGGFace2&Audiovisual &Cross-modality&AM&BiVAE&  $\simeq 2.79\%$ Accuracy $\nearrow$ & No\\
\cite{chen2017progressive}&PIT&Speech separation&Self-transfer & AM&KLD and T/S &30\% WER$\searrow$ & No\\
\cite{cho2018multilingual}&Seq2Seq&BABEL
speech&Multilingual&LM&Fine-tuning&6\% WER, 4\%CER  $\searrow$ & No\\

\cite{he2020multi}&AllenNLP &Low-resource \newline languages&Multi-level \newline cross-lingual&LM&Neural adapters &12.21\% F1–score$\nearrow$ & No\\

\cite{lin2021improving}&MCNN& Air traffic control&Cross-modality&AM&Fine-tuning&250\% CER  $\searrow$ & No\\

\cite{schneider2019wav2vec}&Wav2vec&WSJ data speech &Unsupervised&AM&Affine transform.& 36\% WER$\searrow$ & {Yes\footnote{\url{https://github.com/facebookresearch/fairseq}}}\\

\cite{manohar2017jhu}&TDNN-LSTM&Arabic MGB-3 &Multi-task&LM& Weights transfer& 32.78\% WER$\searrow$ & {Yes\footnote{ \url{https://github.com/kaldi-asr/kaldi/blob/master/
egs/wsj/s5/steps/cleanup/segment\_long_utterances.sh}}}\\

\cite{kim2017cross}&BLSTM&Part of speech tagging&Cross Lingual&LM&Fine-Tuning&Accuracy= 93.26\%  & No\\

\cite{wang2022arobert}&ARoBERT& Spoken language understanding&Self-supervised&LM&Fine-tuning&F1-score=92.56\%  & {Yes\footnote{\url{https://github.com/alibaba/EasyTransfer/tree/master/scripts/arobert}}}\\ 

\cite{song2019speech}&Speech-XLNet&Speech represent. learning &Unsupervised&AM&Fine-tuning&68\% WER$\searrow$ & No\\

\cite{Tian20225438} &{Wav2vec} &{MOS assessment} & {Multi-task}& {AM}& {Fine-tuning}& {MSE= 5.2}& {Yes\footnote{\url{ https://github.com/s3prl/s3prl}}}\\ 

\cite{jain2022text} &{MOSNet} &{Text-to-child-speech synthesis} & {Inductive}& {AM}& {Fine-tuning}& {MOS= 3.96}& {Yes\footnote{\url{https://github.com/C3Imaging/ChildTTS}}}\\

\cite{sancinetti2022transfer} &{DNN} &{Pronunciation scoring} & {Inductive}& {AM}& {Fine-tuning}& {20\% GOP-FT$\nearrow$}& {Yes\footnote{\url{https://github.com/MarceloSancinetti/epa-gop-pykaldi}}}\\

\cite{monica2022comparison} &{BERT} &{Cognitive impairment recog.} & {Inductive}& {LM}& {fine-tuning}& {Accuracy=81\%}& {Yes\footnote{\url{https://github.com/monicagoma/masters_thesis_dementia}}}\\

\cite{9900378} &{CNN-MLP-LiGRU} &{Dysarthric speech recognition} & {Inductive}& {AM}& {Parameters fusion\newline (VT and Exc)}& {WER= 30.3\%}& {Yes\footnote{\url{https://github.com/zhengjunyue/bntg}}}\\

\cite{9747374} &{Wav2vec} &{Speech representations} & {Sequential}& {LM}& {Fine-tuning}& {33.33\% Training time$\searrow$}& No\\

\cite{10023147} &{RoBERTa} &{Semantic alignment} & {Multi-task}& {LM}& {Cross-modal}& {27\% Accuracy$\nearrow$}& No\\

\cite{qin2022improving} &{DNN-HMM} &{Improve low resource Lhasa dialect} & {Multilingual}& {LM}& {Feature adapt. \newline (Self-fusion)}& {14.2\% CER$\searrow$}& {Yes\tablefootnote{\url{https://github.com/tensorflow/tensor2tensor}}}\\

\cite{schlotterbeck2022teacher}&{Wav2vec} &{Classroom environ.} & {Inductive}& {LM}& {Fine-Tuning}& {59\% CER$\searrow$}& No\\

\cite{medeiros2023domain} &{QuartzNet15} &{Speech-to-Text} & {Transductive}& {LM}& {DA}& { WER=5.03\%}& No \\

\hline
\end{longtable}
\end{center}

\subsubsection{BERT-based DTL} 
It is based on using the BERT model \cite{devlin2018bert}, which pretrains LMs and shows that they perform better on a variety of downstream tasks. 
DTL approaches for LM, utilized in the voice recognition sector, are known as the LM adaptation. They attempt to narrow the gap between the $\mathbb{D}_{S}$ and the $\mathbb{D}_{T}$. 
Song et al. \cite{song2019l2rs} present a unique learning-to-rescore (L2RS) process, which relies on (i) using various textual data from the state-of-the-art NLP models, including the BERT model, and (ii) automatically selecting their weights to rescore the N-best lists for ASR algorithms. {Kubo et al. in \cite{kubo2022knowledge} propose a method to improve the performance of end-to-end speech recognition systems by transferring knowledge from large-scale pre-trained language models. The proposed method involves multi-task learning with an auxiliary regression to the word embeddings. The experiments were conducted on the LibriSpeech dataset using a BERT language model pre-trained with BooksCorpus and Wikipedia text data. The results show that the proposed method can further reduce the WER even from a strong baseline with a pre-trained encoder. However, the advantage is relatively small, and future work highlights the possibility to improve the configuration for this type of combination.}

\subsubsection{LDA-based DTL} 
For discrete data collection, generative probabilistic models, such as the latent Dirichlet allocation (LDA), are used. Typically, LDA is a three-level hierarchical Bayesian model in which each item in a collection is modeled as a finite mixture over a set of underlying topics. Every topic is then modeled as an infinite mixture of topic probabilities. To catch the connection between words and construct LMs of a specific document, topic model-based techniques, e.g., LDA, have been applied in \cite{song2019topic}. 
In \cite{hentschel2018feature}, LDA features are transformed by a linear layer with the weight matrix and bias vector; then, they are used as features in the LHN input during the network training and evaluation.

\subsubsection{NNLM-based DTL} 
\begin{figure}[ht!]
     \centering
      \includegraphics[scale=0.65]{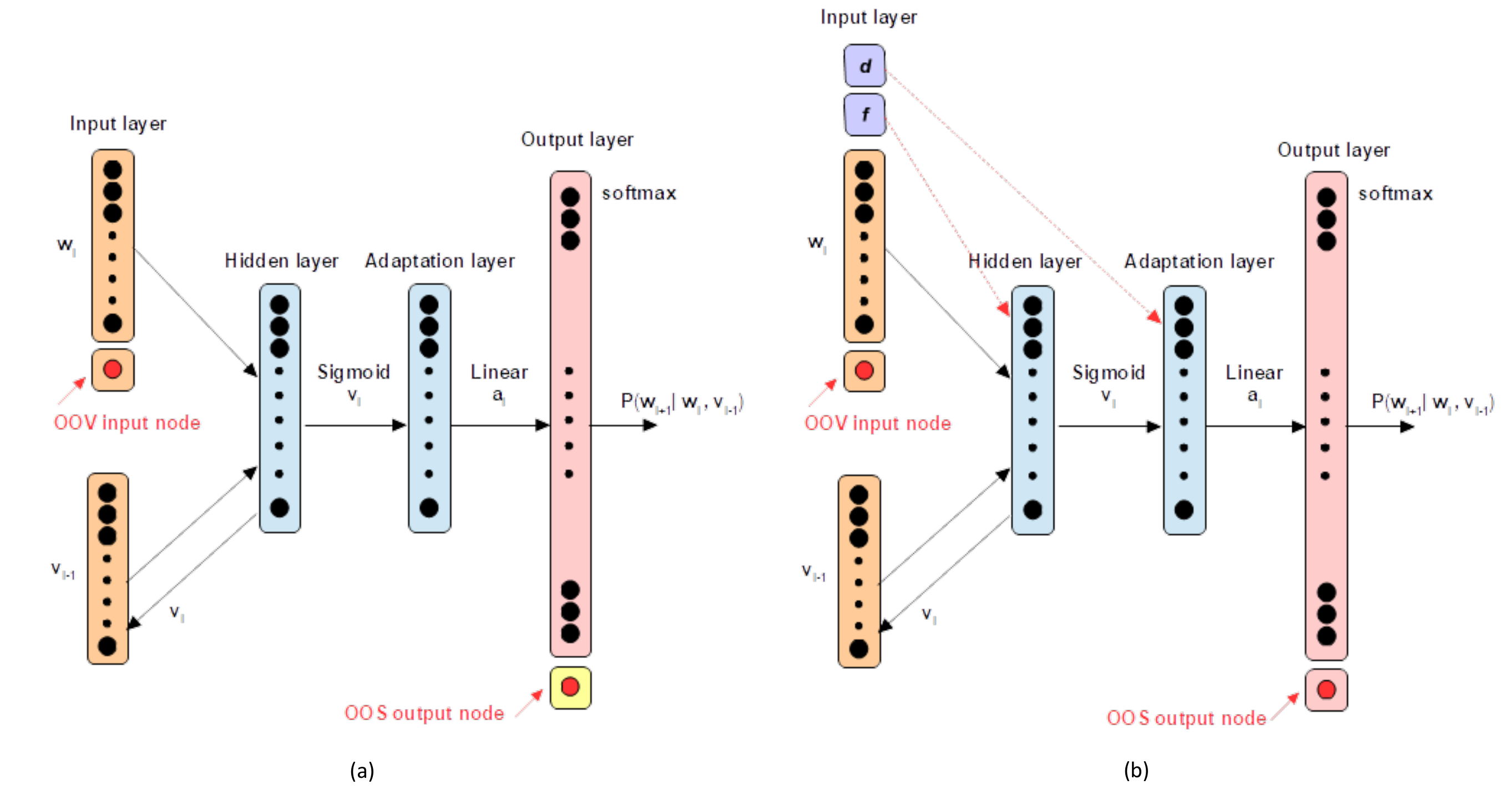}
      \caption{DTL-based RNNLM with different adaptation techniques, where the out-of-vocabulary (OOV) node represents an input word that does not belong to the specified vocabulary but can be included in the input. Similarly, out-of-shortlist (OOS) nodes can also be included in the output: (a) RNNLM with LHN adaptation layer. (b) RNNLM with feature-based adaptation layer.}
     \label{rnnLm}
     \end{figure}

In many tasks, neural network LMs (NN-LMs) outperform count-based LM models in ASR. Specifically, when applied to N-best rescoring, NN-LMs achieve a lower WER \cite{hentschel2018feature}. In this context, adapting NN-LM to new domains is a research challenge, and current approaches can be classified as model-based or feature-based. 
The input of an NN-LM is augmented with auxiliary features in feature-based adaptation, whereas the model-based adaptation includes fine-tuning and network layers adaptation. 
The authors in \cite{deena2018recurrent} develop a recurrent-neural-network-based LM (RNN-LM) approach, where both types of adaptation are investigated. 
Considering this study as an example, Figs. \ref{rnnLm} (a) and (b) explain in details the adopted RNN-based DTL. Moving forward, the authors in \cite{ng2020cuhk} propose a DNN-based model to modify the LM for ASR, where a factorized time-delay neural network (TDNN-F) has been considered. Concretely, it is trained using a combination of cross-entropy and lattice-free maximum mutual information objective functions (LF-MMI). The TDNN-F is shown to be effective in recognizing English child speech \cite{ng2020cuhk}.

\subsubsection{LSTM-based DTL} 
Generally, the NNLM models used in ASR are still trained on a sentence-level corpus, despite the attempts to train them at the document level. This is due to various factors; for example, a more extended context may not be relevant for enhancing next-word prediction in conventional ASR systems. 
It is also challenging to gather training data representing extended session contexts in many conversational circumstances. 
Long-span models are becoming more common in scenarios where they are beneficial. 
Long-span models will likely help scenarios, such as transcriptions of conversations and meetings, human-to-human communication, and document production by voice \cite{parthasarathy2019long}. 
LSTM models are widely employed, and their architectures are well-suited to variable-length sequences. Therefore, they can exploit extreme long-range dependencies without using n-gram approximation.
For instance, by employing equal context, the authors in \cite{tuske2018investigation} demonstrate that the deep 4-gram LSTM outperforms big interpolated count models by performing considerably better backing off and smoothing. 
In another example, the central part of a shared encoder is constructed using BLSTM \cite{cho2018multilingual}.

\subsubsection{DTL-based sequence-to-sequence model} 
\begin{figure}[ht!]
     \centering
      \includegraphics[scale=0.9]{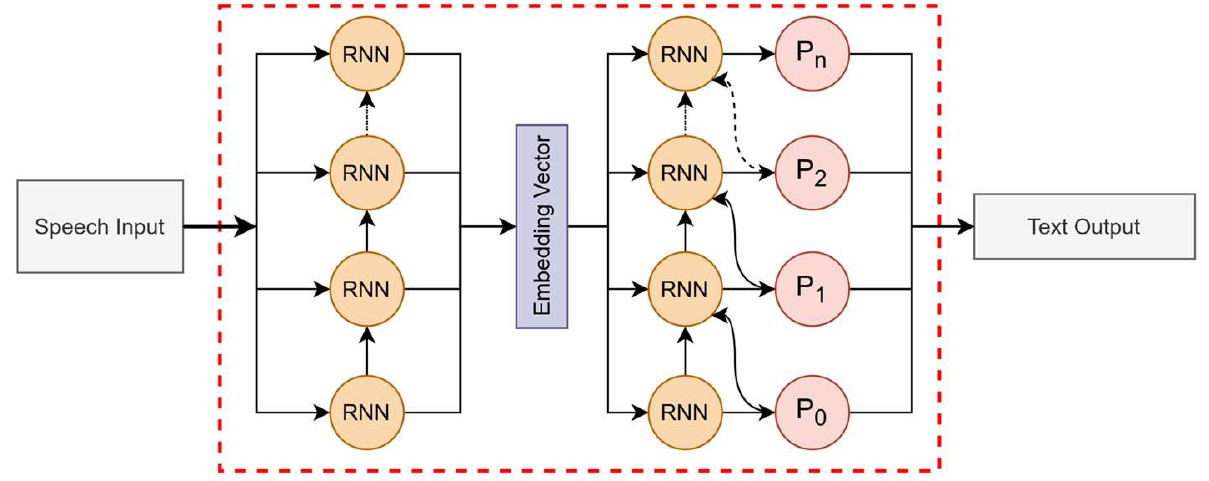}
      \caption{ An example of sequence-to-sequence source model for DTL-based ASR \cite{mridha2021study}. }
     \label{fig:s2s}
     \end{figure}

Recent advances in seq2seq models have demonstrated their promising outcomes for training monolingual ASR systems. 
The CTC \cite{nakatani2019improving} and encoder-decoder models \cite{winata2019code,dong2018speech,winata2020lightweight} are two popular architectures for end-to-end speech recognition. 
Moreover, end-to-end architectures have been further investigated by jointly training these designs in a multi-task hybrid approach \cite{kim2018towards,milde2017multitask}, where it has been discovered that they can increase the overall model's performance. 
For example, the architecture portrayed in Fig. \ref{fig:s2s}, aims to build a seq2seq model \cite{mridha2021study}. The seq2seq's encoder network consists of a stack of RNNs that create embedding vectors. 
The RNN decoder collects the embedding vectors and generates final results. The RNN, on the other hand, has access to the prior prediction ($P_i,  i=0,\dots, n$). As a result, the succeeding prediction has a better chance of being accurate.

Moving on, combining AM and LM techniques can improve or build an efficient DTL-based ASR model, such as in \cite{bai2021fast,jiang2021gdpr,weninger2019listen,deena2018recurrent,ng2020cuhk}.  Table \ref{tab:5} summarises the most recent DTL-based ASR techniques used in the LM domain.

\subsection{Cross-domain ASR} 

\subsubsection{ DTL-based ASR for emotion recognition} When language information is combined with acoustic data, multiple investigations have found that SER accuracy improves. As a result, it would be beneficial to combine both systems in order to improve ASR systems' ability to cope with emotional speech while also giving linguistic input to SER systems.  For example, this can be realised when a spectrogram is input into shared convolutional layers, which are then followed by a number of specialized layers that share one or more levels in order to interact.

The scheme in \cite{tits2018asr} used the internal representation of a speech-to-text system to investigate the relationship between valence/arousal and different modalities by the means of DTL. A speech-to-text system or ASR system learns a mapping between two modalities: an audio speech signal and its transcription. DTL is used to estimate valence and arousal using features learned for an ASR task, the proposed method has the advantage of allowing large datasets of speech with transcriptions to be combined with smaller datasets annotated in emotional dimensions. Another work in \cite{ananthram2020multi} fine-tune the speaker recognition  TDNN-based model on the task of emotion detection using the multi-modal emotion dataset crema-D with the canonical label clustering. The work in \cite{boateng2020speech} aims to extract features from the audio segments with the most extreme positive and negative ratings, and the audio's ending, they used the peak-end rule and DTL approach to extract the acoustic features. They used a pre-trained CNN speech model named YAMNet and a linear SVM to perform binary classification of partner valence.

\subsubsection{Cross-language DTL} 
Cross-language DTL is one of the application of cross-modality DTL. It is one of the most common methods for constructing ASR models for low resource languages from a model trained for another language, and it is based on the assumption that phoneme features can be shared across languages. A generic ASR model can also be adapted to another narrow domain using DTL. With the help of high-resource languages, several knowledge transfer methods are investigated in \cite{liu2019investigation} to overcome the data sparsity problem. The first is the DTL and fine-tuning techniques, which uses a well-trained neural network to initialize the LHN parameters. Second, progressive neural networks (Prognets) are examined. The latter is immune to the forgetting effect and excel at knowledge transfer owing to lateral connections in the network architecture. Finally, using cross-lingual DNN, bottleneck features are extracted as an enhanced feature to boost the effectiveness of the ASR system. Tables \ref{tab:5} and \ref{tab:6} summarise most recent schemes related to ASR using cross-language DTL and their performances.



\begin{table}[h!]
\caption{A summary of the recent \textit{ASR-based cross-language DTL} technique. Whereas the marks  ($\nearrow$) and ($\searrow$) indicate improvement and reduction, respectively. If many scenarios has been conducted in one metric, only the best result is mentioned.}
\label{tab:6}
\small
\begin{tabular}{p{0.5cm}p{1.8cm}p{7.5cm}p{2cm}p{2.2cm}p{1.6cm}}
\hline
Ref. &Model-based   & ASR Tasks ($\mathbb{T}_{T}$) & Characteristic & Performance & {Code available?}\\
 \hline
\cite{yusuf2019low}& EDML & Framework of performing
the DTL that reduces the impact of the prevalence of out-of vocabulary terms.& query-by-example task &  74\% TWV$\nearrow$ & {Yes\tablefootnote{\url{https://github.com/ysfb/crosslingual_exemplars}}}\\ [0.6cm]

\cite{sahraeian2018cross}&DNN  &W Schemes are used to combine the ensemble's constituents, with the combination weights being trained to minimize the cross-entropy objective function.&Weights interpolation  & 7.7\% WER$\searrow$ & No\\ [0.6cm]

\cite{liu2019investigation} & Prognets & Improving ASR scheme quality by overcoming the data sparsity problems by means of high-resource languages.  & Fine-tuning \newline LHN adapt. & 38.6\% WER$\searrow$ & No\\ [0.6cm]
\cite{feng2019low}&FNN and CNN & Indo-European
speech samples used to improve the identification
of African languages.  & Fine-tuning \newline PLP coeff.& 2.1\% EER$\searrow$ & No\\ [0.6cm]

\cite{wilkinson2020semi}&CNN-GMM-HMM& Fully-automatic segmentation, semi-supervised training of ASR systems for five-lingual code-switched speech &Semi-supervised& 1.1\% WER$\searrow$ & No\\[0.6cm]
 \cite{zelasko2022discovering} & {TDNNF} & {Unsupervised cross-lingual modeling for ASR to aggregate 13 languages.} &{Zero-shot} & {Depends on phone} & {Yes\tablefootnote{\url{https://github.com/pzelasko/kaldi/tree/discophone/egs/discophone}}}\\[0.6cm]
 \cite{hassan2022improvement} & {DeepSpeech2} & {Train new model for south Asian accents based native English speakers taken from LibriSpeech dataset.}  & {Fine-tuning} & {24.92\% WER$\searrow$} & No\\[0.6cm]
\cite{deng2022improving} &{BERT, GPT2} & {Improve
CTC model by knowledge transfer from BERT and GPT2. Expriments was on AISHELL-1 corpus.} & {Cross-modal} & {16.1\% CER$\searrow$} & No\\[0.6cm]
 \cite{khurana2022magic} &{Wav2vec-2.0} &{The monolingual version of Wav2vec-2.0 has been adapted to perform similarly to the multilingual model.}& {Self-supervised} & {3.6\% WER$\searrow$} &No\\[0.6cm]
 \cite{tachbelie2022multilingual} & {HMM–DNN} &{ASR systems utilize the GlobalPhone database for the purpose of implementing speech recognition across multiple languages.} & {Multi-task learning} &  {33.21\% WER$\searrow$} & No\\[0.6cm]
 \cite{rolland2022multilingual} & {HMM-DNN} &{Novel approach that combines resources from various languages to improve the effectiveness of ASR systems for children's less-resourced languages.} & {Multi-task learning}& {7.73\% WER$\searrow$}& No\\
\hline
\end{tabular}
\end{table}

\subsubsection{Cross-corpus SER (CC-SER)}
In SER, it is usually assumed that speech utterances in training and testing domains are recorded at the same conditions. However, in real-world, this is not the case since speech data is frequently gathered from different environments or devices. Thus, a discrepancy exists between the two domains negatively impacting recognition performance. To that end, the problem of cross-corpus SER (CC-SER) has recently been investigated, where different DTL models have been deployed. For instance, 
a transfer linear subspace learning (TLSL) scheme is proposed in \cite{song2019transfer} to develop a CC-SER framework, enabling learning of shared feature space for source and TDs. Accordingly, to estimate the similarity between different corpora, a nearest neighbor graph algorithm is utilized. Moreover, with the aim of dividing emotional features into two high transferable part (HTP) and low transferable part (LTP), a feature grouping method is developed. 
Moving on, in \cite{liu2018unsupervised}, unsupervised CC-SER is explored, where only the training data is annotated. Specifically, a domain-adaptive subspace learning (DoSL) technique is introduced to learn a projection matrix that allows transforming the source and target speech data from the initial domain to the labeled domain. In doing so, the classifier learned on the labeled SD data can efficiently forecast the emotional states of the unlabeled TD data. 
Similarly, in \cite{liu2021transfer}, the DoSL-based CC-SER method has further been improved by using a transfer subspace learning (TRaSL). 
In \cite{luo2019cross}, a semi-supervised CC-SER approach is designed using non-negative matrix factorization (NMF). It is built upon the idea of incorporating
the information of training corpus labels into NMF, and seeking a
latent low-rank feature space, where the conditional and marginal distribution differences between the two corpora can be minimized simultaneously.

Moving forward, a transfer sparse discriminant subspace learning (TSDSL) is introduced in \cite{zhang2019transfer} to learn a shared feature subspace of various corpora by initiating the $\ell_{2,1}$-norm penalty and discriminative learning. This has helped in learning the most discriminative characteristics across multiple corpora.
Besides, in \cite{luo2020nonnegative}, a non-negative matrix factorization-based transfer subspace learning (NMFTSL) scheme is proposed, which aims at minimizing the distances in the common subspace between the marginal distributions and conditional distributions. Typically, these distances have been estimated using the maximum mean discrepancy (MMD) criterion. 
In \cite{zhang2021cross}, a joint transfer subspace learning and regression (JTSLR) technique is adopted, which learns a latent subspace using the discriminative MMD as the discrepancy metric. Next, a regression function is put in the latent subspace to model the relationships between features and related annotations. Also, a label graph is considered for better transferring the knowledge from relevant SD data to TD data.

Similarly, a target-adapted subspace learning (TaSL) approach for CC-SER is proposed in \cite{chen2019target}, which aims at finding a projection subspace to enable the features regressing the labels more accurately. Moreover, this helps  efficiently bridge the gap of feature distributions in the TD and SD. 
Moving forward, the $\ell_{1}$-norm and $\ell_{2,1}$-norm penalty terms have been mixed with other regularization terms to get a more optimal projection matrix.
In the same way, Zhang et al. \cite{zhao2021cross} use sparse subspace transfer learning (SSTL) to develop a CC-SER technique by learning a robust common subspace projection using discriminative subspace learning. Then, knowledge from the source corpus is transferred to the target corpus using a sparse reconstruction based on $\ell_{2,1}$-norm. In this case, the target samples are appropriately represented as a linear combination of the SD data.

On the other hand, the impact of of cross-corpus data complementation and data augmentation is investigated in \cite{braunschweiler2021study} on the performance of SER models in from the same corpus and different corpus. The investigations have been conducted on six emotional speech corpora, including (i) single and multiple speakers, and (ii)  variations in emotion style (natural, elicited, and acted).

\subsection{Adversarial TL-based ASR} 
\begin{figure}[ht!]
\centering
\includegraphics[scale=0.6]{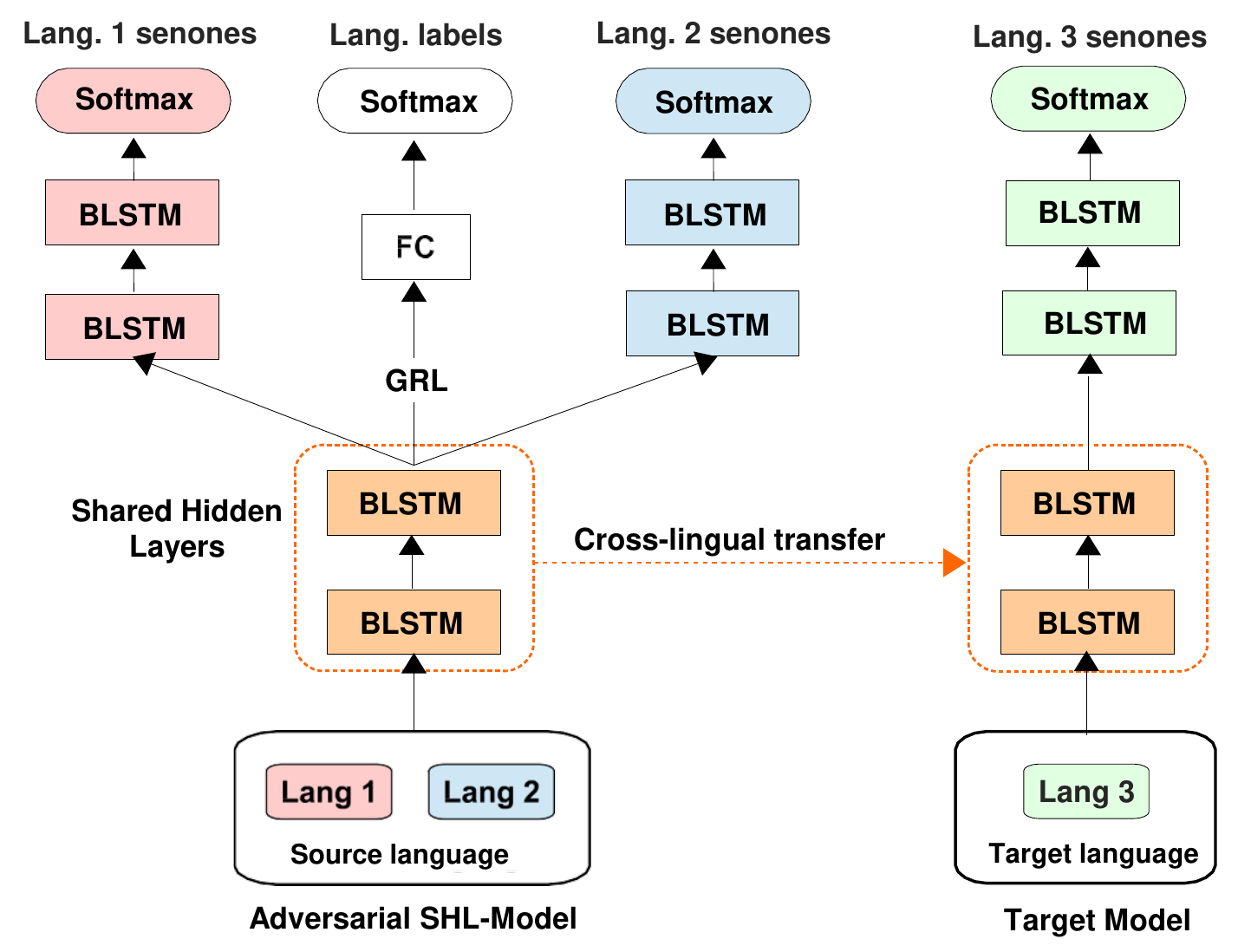}
\caption{ An example of proposed model architecture language-adversarial TL for limited ASR resource \cite{yi2018language}. Senones refers to feature cluster's name, representing similar acoustic states/events.}
\label{fig:adv}
\end{figure}

In most cases, the source model is trained in multiple languages (multilingual training) for which there is a large amount of speech data \cite{qin2018towards,cho2018multilingual}. Multilingual training can be thought of as a series of shared hidden layers (SHL) and language-specific layers or classifier layers for various languages. The source model's SHL serve as a feature converter, converting various language features to a common feature space \cite{liu2019investigation}. However, some language-dependent features may exist in the common feature space, which is not a positive factor for cross-lingual knowledge transfer. Language-adversarial training can effectively address this problem. Adversarial training aids in the creation of a language invariant feature space. After preparing the source model, the first $n$ SHL can be transferred to the target model of an unknown language. Authors in \cite{yi2018language}, proposed language-adversarial TL as a solution to the problem of target model performance degradation caused by shared features that may contain unnecessary language-dependent information. Fig. \ref{fig:adv} illustrate the suggested language-adversarial TL method's architecture \cite{yi2018language}. The adversarial SHL-model, also known as the source model, is on the left. The target model is the correct one. The adversarial SHL Model implies the presence of an exta language discriminator in the SHL-model. The completely connected layer is denoted by the letter FC. The gradient reversal layer (GRL) ensures that the feature distributions across all languages are as similar as feasible for the language discriminator. The language discriminator's output labels are language labels.

To achieve better ASR effectiveness in low-resource situations, the authors in \cite{kumar2021exploration} combined both semi-supervised training and language adversarial TL. The work in \cite{yi2020adversarial} proposes using adversarial TL to improve punctuation prediction performance. A pre-trained BERT model is used to transfer bidirectional representations to punctuation prediction models. The proposed approach has been applied on ASR application as the target task. Table \ref{tab:7} summarises the performance of the most recent work in ASR-based adversarial TL. {Authors in \cite{li2022sequence} proposes a sequence distribution matching approach for unsupervised domain adaptation in ASR. The approach uses sequence pooling to capture and match sequential statistics, which improves the performance of distribution matching. The authors conducted experiments on different accent mismatches and found that the sequence pooling method outperformed the MMD-based approach and slightly improved the domain adversarial training (DAT)-based approach. By combining sequence pooling features and original features, the MMD-based and DAT-based approaches reduced WER up to 14.72\% over the source domain model. DTL was used by fine-tuning the wav2vec  base model pre-trained on the source domain.}

\begin{table}[H]
\caption{A summary of the recent \textit{ASR-based adversarial-language TL} technique. Whereas the marks  ($\nearrow$) and ($\searrow$) indicate improvement and reduction, respectively. If many scenarios has been conducted in one metric, only the best result is mentioned.}
\label{tab:7}
\small
\begin{tabu}{p{0.5cm}p{2cm}p{6cm}p{2cm}p{2.2cm}p{1.6cm}}
\hline
Ref. &Model-based   & ASR Tasks ($\mathbb{T}_{T}$) & Characteristic & Performance &  {Code available?}\\
 \hline
 \cite{kumar2021exploration} & Adversarial SHL-Mode  & Use three Indian languages ( Hindi, Marathi, and Bengali ) cross-lingual to improve Hindi ASR &  Semi-supervised & WER=5.5\% \newline (25.65\% WER $\searrow$) & No\\
\cite{yi2018language}& SHL Model & Improve the performance of low-resource ASR & Cross-lingual & 10.1\% WER $\searrow$ & No\\
\cite{yi2020adversarial}& BERT  &Improve
the performance of punctuation predicting & Multi-task & 9.4\%   F1-score $\nearrow$ & No\\
\cite{zhang2022joint} & LAS and TTS models&Expressive speech synthesis using joint ASR and TTS training &Cross-domain& 0.7\% WER$\searrow$, 1.8 WIL$\searrow$ & No\\
\hline
\end{tabu}
\end{table}


\subsection{DTL-based ASR for medical diagnosis } 
The field of ASR, especially the DTL-based ASR, has provided a qualitative leap in the field of medicine for the early detection of diseases. This progress is in several medical areas summarised in Table \ref{tab:9}, including:

\subsubsection{Heart sound classification} 
The heart sound is made up of several components. the first (S1) and second (S2) heart sounds are considered normal, while the third (S3) and fourth (S4) heart sounds are often associated with murmurs, and ejection clicks are usually associated with some illnesses or abnormalities. Koike et al. \cite{koike2020audio} proposed a novel DTL PANNs-based model pre-trained on large scale audio data for a heart sound classification task. Another scheme for heart sound classification is proposed by Boulares et al. \cite{boulares2020transfer}.  Without any denoising or cleaning steps, DTL is applied to the Pascal public dataset to provide an experimental benchmark. The main goal is to produce a set of experimental results that can be used as a starting point for future cardiovascular disease (CVD) recognition research that uses cardiac cycle vibration sound and is based on phonological cardiogram (PCG). The proposed scheme addresses the lack of a CVD recognition benchmark, as well as the fact that classification results are highly variable and cannot be compared objectively.

\subsubsection{Parkinson disease detection} 
PD is a progressive neurodegenerative disease that affects  millions of people globally. The diagnosis of PD is critical for enhancing the quality of workday activities and  prolonging the patients' lives. Because each person's symptoms and disease progression differ so much, it is difficult to predict how PD symptoms will change over time and how they will affect the patient's life. Traditional PD detection techniques, on the other hand, are frequently handcrafted and require specialized knowledge. Over 90\% of patients with PD have distinctive patterns of language disability and atrophy, which is one of the early symptoms of the disease. The voice becomes hushed, and the speech becomes monotonous and rapid. With time, the patient's voice becomes less audible, and in later stages of the disease, the patient can only whisper. Mumbling can be a warning sign of illness. Karaman et al. \cite{karaman2021robust} developed a robust automated PD detection relies on DTL-based ASR. The pre-trained models SqueezeNet1\_1, ResNet101, and DenseNet161 were fine-tuned and retrained. The proposed scheme presents an acceptable PD detection. In order to solve the scarcity of speech-based PD, and the existence of inconsistency in the distribution between subjects, a novel two-step unsupervised DTL algorithm called two-step sparse transfer learning (TSTL) \cite{li2021insight} is proposed to deal with the above two mentioned problems. The method can assist in extracting useful information from large amounts of unlabeled speech data, aligning the distribution of the training and test sets, and preserving the original structure between samples all at the same time. Another strategy proposed by Qing et al.  \cite{yu2021enhancing} which is based on pre-trained long short-term memory (LSTM) neural network model. The aim of the proposed scheme is to enhance ASR for Parkinson patient. To alleviate the over-fitting problem and reduce the WER, the  frequency spectrogram masking data augmentation method was used in the latter-mentioned scheme.

\subsubsection{Other medical diagnosis} 
The proposed technique in \cite{harati2021speech} investigated a speech-based DTL method that employs a lightweight encoder and only transfers the encoder weights, allowing for a simplified run-time model for speech-based depression prediction. For speakers with dysarthria, an improved DTL framework was applied to robust personalised speech recognition models is proposed by Xiong et al. \cite{xiong2020source}. With the limited data available from target dysarthric speakers, the CNN-TDNN-F ASR AM is adapted onto the TD via neural network weight adaptation. Another framework for dysarthria speaking identification, Takashima et al. \cite{takashima2019knowledge} propose a method for Japanese people with articulation disorders to transfer two types of knowledge corresponding to different datasets: the language-dependent (phonetic and linguistic) characteristic of unimpaired speech, which corresponds to non-dysarthric speech data, and the language-independent characteristic of dysarthric speech, which corresponds to non-Japanese dysarthric speech data. The work in \cite{sertolli2021representation} presents a novel feature representation for end-to-end DTL-based ASR framework for health states identification. The use of ASR DNNs as feature extractors, the fusion of several extracted feature representations using compact bilinear pooling (CBP), and finally inference using a specially optimized RNN classifier are all part of latter proposed approach. {The  authors in \cite{Gruzitis2022267} discuss the development of ASR systems for the radiology domain in the context of low-resourced languages. TL and fine-tuning of transformer models are employed to achieve state-of-the-art results. A 30-hour in-domain speech corpus and a text corpus of over 1GB are used for training. The best results obtained consist of WER ranging from 20\% to 60\% for individual reports. Efforts are made to create a cleaner test dataset for more objective evaluation.}

\begin{table}
\caption{A summary of a DTL-based ASR technique in medical diagnosis, whereas the marks  ($\nearrow$) and ($\searrow$) indicate the improvement and reduction, respectively. If many scenarios have been conducted in one metric, only the best result is mentioned.}
\label{tab:9}
\begin{tabular}{lp{3cm}p{4.5cm}p{2.cm}p{2.8cm}p{1.6cm}}
\hline
Scheme &Model-based   & ASR Tasks ($\mathbb{T}_{T}$) $\nearrow$& DTL Type & Performance& Code available? \\
 \hline
 \cite{karaman2021robust} & DenseNet-161 & PD detection & Fine-tuning  & Accuracy= 91.17\%& No\\
\cite{li2021insight}& TSTL-based CSC\&SF & PD speech diagnosis & Unsupervised   & Accuracy= 97.50\%&Yes\tablefootnote{\url{ https://share.weiyun.com/14a0OH0B}}\\
\cite{yu2021enhancing}& Proposed four layers  &PD
speech& Fine-tuning&13.5\% WER$\searrow$& No\\
\cite{koike2020audio} & PANN CNN14 & Heart sound classification & Fine-tuning & UAR= 89.7\%& No \\

\cite{harati2021speech}&EH-AC&Depression prediction &LHN (encoder weights) &27\% AUC $\nearrow$ & \\
\cite{xiong2020source}&CNN-TDNN-F &Dysarthric speech&Neural weight adapter & 11.6\% WER$\searrow$ & No\\

\cite{boulares2020transfer} & InceptionResNet-v2 & PCG-based CVD classification&  Fine-tuning &  Accuracy= 0.89\% & No\\

\cite{takashima2019knowledge}& LAS  &Dysarthric Speech& Multilingual& 45.9\% PER$\searrow$ & No\\
\cite{sertolli2021representation}& Wav2Letter and \newline DeepSpeech &Health states classification &Transductive&UAR= 73.0\% \newline (8.6\% $\nearrow$)& No \\
\cite{hirevs2022convolutional} &{ResNet, Xception} &{PD detection} & {Fine-tuning} & {Accuracy=99\%} & No\\
\cite{pahar2022covid} &{Resnet50} &{COVID-19 detection in cough} & {Fine-tuning} & {AUC=98\%} & No\\
\cite{harati2022generalization} & {VGG} & {Large-scale depression screening} & {Feature trans.} & {AUC=79\%}  & No\\
\cite{rejaibi2022mfcc} & {RNN} & {Clinical
depression recognition and assessment from speech} & {Feature trans.} &  {Accuracy=76.27\%} & No\\
\cite{yue2022raw} & {CNN} &{Dysarthric speech recognition} & {Feature fusion} & {1.7\% WER$\searrow$} &No\\
\cite{almadhor2023e2e} &{CNN and Attention} &{Dysarthric speech recognition} & {Fine-tuning} & {25.75\%$\nearrow$} & No\\
\cite{hu2021generating} &{DNN} &{Overcome dysarthria
acoustic data scarcity} & {Adaptation} & {5.67\% Accuracy$\nearrow$} & Yes\tablefootnote{\url{ https://github.com/rshahamiri/SpeechVision}}\\
\cite{han2023spatial} &{Wav2vec} & {Depression recognition} & {Map feature} & {F1-score=76\%} & No\\
\hline
\end{tabular}
\end{table}

\subsection{DTL-based ASR attacks and security} 
Adversarial examples are produced by slight perturbation, a valid audio file or speech characteristics, with a few quantities of noise in order to either boost or fool ASR systems. Even though the added noise is imperceptible to humans auditory system or are only perceived as faint background noises by the ASR model, they can cause the inputs to be well-classified or misclassified. For example adding a small disruption to the following speech: \textit{"At the still point, there the dance is"} makes an ASR generate \textit{"At the tail point, there the tense is"}, which is a wrong result \cite{hu2019adversarial}. The key idea behind attacking ASR scheme is that the model can be easily fooled or boosted by adversarial examples. As a result, it naturally motivates speech researchers to create speech adversarial examples. Based of this, researchers can generate adversarial examples for time and/or frequency speech representations, which represent different speech features and can all be used as inputs to neural networks, allowing the developers to enhance or decrease ASR performances. DTL aims to realise the concept of transferability, this latter allows adversarial examples targeting a source model to gain the potential to attack the  target models classifying the same
kind of data. For that, the adversarial attacks for DTL-based ASR models can be divided into two categories as shown in Fig. \ref{fig:7}, which are: 

\begin{figure}[ht!]
\centering
\includegraphics[width=0.9\textwidth]{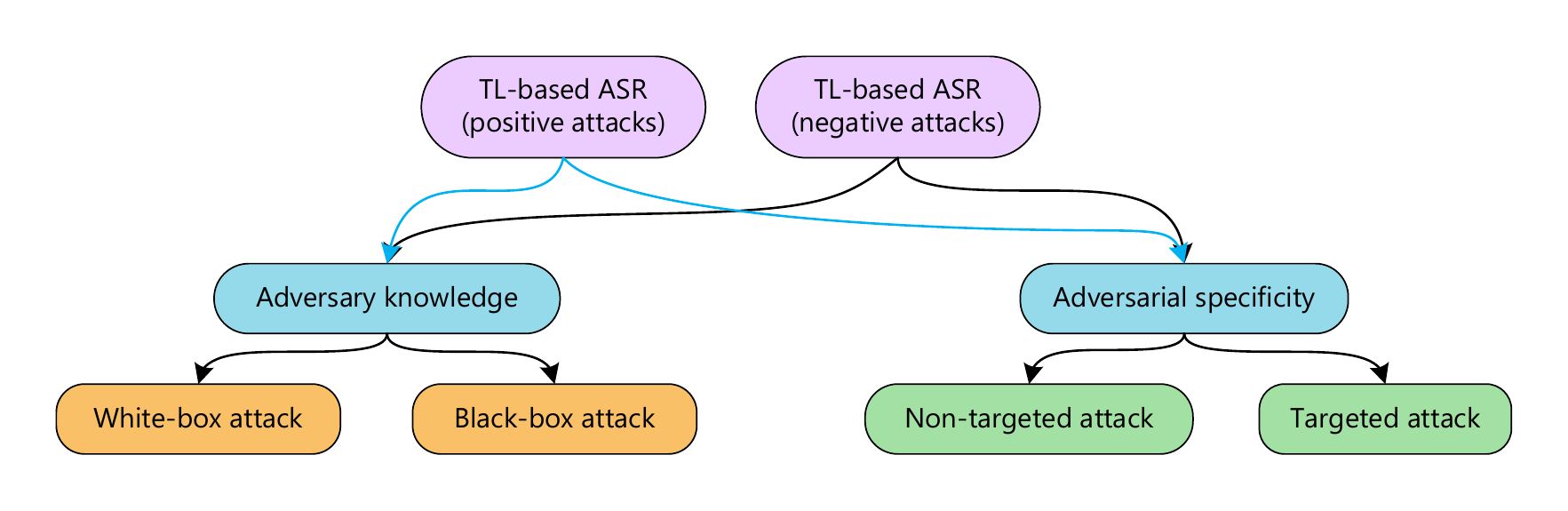}
\caption{ Possible adversarial attacks in DTL-based ASR schemes.}
\label{fig:7}
\end{figure}

\subsubsection{Positive adversarial attacks}
Includes all methods that \textit{increase} or \textit{secure} the effectiveness of existing ASR schemes. In \cite{sun2018training}, the authors proposed, using natural data combined with adversarial data, to train robust AM. They focused on MFCC features and used the gradient sign based method to generate adversarial MFCC features based on the network model and inputs parameters for each mini-batch. The neural network was then trained using teacher/student training concept on natural data that had been supplemented with adversarial data. Experiments on CHiME-4 and  Aurora-4 tasks using a customized CNN validated their scheme.

\subsubsection{Negative adversarial attacks}
Includes all methods that \textit{decrease} or \textit{threatens} the effectiveness of existing ASR schemes. In \cite{abdullah2021hear}, the authors suggest a framework that employs google (Phone) as the source model and examined the impact of adversarial attacks on target model (deep-speech 1). The adversarial attacks have been applied to audio waveform just after signal decomposition and thresholding processes, and the resulting output fed the ASR source model.  
 
According to \cite{hu2019adversarial}, the adversarial attack model can be divided into two sorts based on the adversary’s objective, knowledge, and background, which are (Fig. \ref{fig:7}):

\begin{itemize}
    \item \textbf{Adversary knowledge:} It is divided into:
    \begin{itemize}
        \item  White-box attack: Assumes the adversary has complete knowledge of $M_T$ , including its architecture, type, training weights, and the values of all parameters, among other things.
        
        \item  Black-box attack: Assumes the adversary has no access to $M_T$  and acts as if he or she is a regular user who only knows the model's output.
    \end{itemize}
    \item \textbf{Adversarial specificity:} It is divided into:
    \begin{itemize}
        \item Non-targeted attack:  Aims to make the adversarial example's $M_T$ predict any incorrect class. Its sole goal is to take out the ASR algorithm.
        \item Targeted attack: Its goal is to deceive $M_T$ into assigning the adversarial example to a specific class (selected by the attacker). An attack like this imposes an ASR scheme to carry out specific instructions.
    \end{itemize}
\end{itemize}


Moving on, Table \ref{tab:10} shows a summary of DTL-based adversarial models for existing works. 

\begin{table}[ht!]
\caption{A summary of the recent DTL-based ASR for adversarial attacks. Whereas the marks  ($\nearrow$)and  ($\searrow$), indicate improvement and reduction respectively.}
\label{tab:10}
\small
\begin{tabular}{p{0.5cm}p{2.5cm}p{2.5cm}p{2cm}p{1.5cm}p{2cm}p{2.5cm}p{1.6cm}}
\hline
Ref. &Model-based   & ASR Tasks ($\mathbb{T}_{T}$) & Target object & Adversary \newline knowledge & Adversarial \newline specificity & Performance& Code available? \\
 \hline

\cite{sun2018training} & Aurora-4 & Boost & MFCC  & White-box & Targeted &23\% WER $\searrow$& No\\
\cite{schonherr2018adversarial} & DNN-HMM (Kaldi) & Boost &  Waveform  & White-box & Targeted &Accuracy= 98\% & No\\
\cite{zelasko2021adversarial}&DeepSpeech&Fool&Waveform&White-box&Targeted&4-5\% WER$\nearrow$ & Yes\tablefootnote{\url{https://github.com/pzelasko/espresso/tree/feature/librispeech-wav-model}} \\
\cite{subramanian2020study}&VGG13&Fool (Dense\_mel)&Mel-spectro.&White-box&Non-targeted&SNR=29.06 dB& No\\
\cite{carlini2018audio}&DeepSpeech&Fool (Speech-to-Text) &Waveform&White-box&Targeted&Attack success rate= 100\%& Yes\tablefootnote{\url{Audio adversarial examples: Targeted attacks on speech-to-text}}\\
\cite{abdullah2021hear}&Google \newline
(Phone)&Fool \newline (Deep-Speech 1)&Waveform&Black-box&Non-targeted&Attack success rate=87\% & No\\
\cite{kwon2019selective}&DeepSpeech&Fool (victim DeepSpeech) &Mel-frequency
cepstrum &White-box &Targeted&Attack success rate=91.67\%)& No\\
\hline
\end{tabular}
\end{table}

\subsection{DTL-based ASR for other applications} 
DTL-based ASR is applied in different fields other than those mentioned above. For example, the work in \cite{boes2021audiovisual} examines the value of DTL for two types of sound recognition tasks: audio tagging and sound event detection. 
The authors adapt a baseline system that only uses spectral acoustic inputs to include pre-trained auditory and visual features extracted from networks built for different tasks and trained with external data using feature fusion. Another work in \cite{arora2017study} employ the concept of DTL to address the lack of large annotated databases for real-life audio event detection. For deep speech enhancement application, the scheme in \cite{wang2020cross} proposes an environment adaptation technique to enhance deep speech enhancement models by minimizing the KLD between posterior probabilities obtained by a multi-condition senone classifier (teacher) fed with noisy speech features and a clean-condition senone classifier (student) injected with clean speech characteristics. In \cite{chen2018transfer}, DTL has been employed for a wearable device for long-term social speech evaluations. The authors use social sensing to measure a person's mental health by extracting and analyzing speech characteristics in completely natural daily situation. DTL is used to transfer the model to the audio segmentation process using the following characteristics: formant, energy, brightness, and entropy. The output results showed promise in classifying several acoustic scenarios in normal conditions. 

In \cite{wu2020self}, the authors investigate whether self-supervised pre-trained speech can help with the ST in both high and low resource features, i.e., whether they can be transferred to other languages, and whether they can be effectively merged with other schemes for improving low-resource end-to-end ST, e.g., using a pre-trained high-resource ASR framework. 
Results show that self-supervised pre-trained features can always enhance ST performance, and cross-lingual DTL allows the ease of expansion to many languages with little or no tuning. Most end-to-end ST models performed poorly in the absence of source speech information. As a result, the authors in \cite{zhu2021conwst} propose a self-supervised bidirectional distillation processing system for low-resource non-native ST. It improves speech ST performance by combining a large amount of untagged speech and text with source information in a complementary manner. The framework is based on an seq2seq model that guides the encoder in reconstructing the acoustic representation using wav2vec2.0 pre-training model. For SV field, the authors in \cite{hong2017transfer} make the assumption that SV of short utterances, in particular, can be thought of as a task in a domain with a limited number of long utterances. DTL for probabilistic linear
discriminant analysis (PLDA) can thus be used to learn discriminative information from domains with a large number of long utterances. AlexNet pre-trained model was used in \cite{lu2021detection} to efficiently recognize and classify killer whale noises, pilot whales, harp seals, and long-finned, in very overlapping living areas using DTL. Because the training samples were insufficient, DTL was employed to prevent the over-fitting problem of deep networks. The proposed method was tested using a challenging dataset containing both target and non-target sounds. Even though the sounds used in the test dataset were completely independent of the sounds used in the training dataset, the proposed method identifies well the real distinctions between the sounds of several marine mammals. The paper in \cite{azizah2020hierarchical} uses hierarchical DTL to implement multilingual TTS using DNNs for low-resource languages. Using the same model architecture, a pre-trained monolingual TTS on the high-resource language is fine-tuned on the low-resource language. Then, A partial network-based DTL is used on a pre-trained monolingual TTS model to generate a multilingual TTS model.

The authors of paper \cite{luo2022physics} introduces the so-called PhyAug scheme\footnote{\url{https://github.com/jiegev5/PhyAug}}, a physics-directed data augmentation approach for deep model transfer to specific sensors. PhyAug leverages the first principle governing domain shift to transform source-domain training data into augmented target-domain data. It effectively calibrates deep neural networks and reduces the need for target-domain data. The results show that PhyAug achieves the lowest WERs for all tested microphones, reducing WER up to 70\%. It outperforms alternative approaches such as data calibration,  and Cycle-GAN methods. PhyAug enhances recognition accuracy by mitigating the impact of smartphone microphone variations, resulting in a significant improvement ranging from 33\% to 80\%. Tropea et al.  \cite{s22166292} propose an automatic stone recognition system based on a Two-Stage Hybrid Model that combines CNNs and ML algorithms. Four pre-trained CNN models were utilized as feature extractors, and the output of the penultimate layer, representing the feature vector, was used as input to the ML classifier to perform stone recognition. The proposed system was tested on a dataset of stones from Calabrian quarries, and the authors compared different DL and ML combinations to evaluate the performance of their proposed model. The results showed that the two-stage hybrid model achieved high accuracy, up to 99.9\% for ResNet50 CNN in the first stage and a kNN ML in the second stage, in stone classification, providing a useful tool for non-geologists to identify and classify different types of stones.

For knowledge distillation (KD) application, \cite{yoon2023inter} discusses the use of intermediate knowledge distillation (Inter-KD) in the CTC-based ASR framework. Inter-KD transfers the source model's knowledge to the intermediate CTC layers of the target network, improving performance without the need for LMs or data augmentation. Experimental results on LibriSpeech show that Inter-KD significantly reduces the WER of the target model compared to other KD methods. The distilled student model achieved a WER of 6.24\% on dev-clean and 6.3\% on test-clean, outperforming conventional KD approaches.  Similarly, \cite{lee2022knowledge} focuses on proposing a new AM training method that combines multi-task learning and KD. It experimentally demonstrates the effectiveness of the proposed method in compensating for the weaknesses of the interpolation-based KD method. Additionally, a hierarchical distillation method is proposed, which reduces the relative WER of the speech recognition system by 9\%.

Spoken hesitations like "um," "uh," and "er" commonly occur in speech, impacting ASR system performance in human-machine interactions. Chatziagapi et al. \cite{chatziagapi2022audio} proposed a DTL approach using CNNs for audio and an RNN method for text, leveraging ASR-derived textual information. Pre-training the CNN on speaking rate and fine-tuning for filled pause detection improved results. The framework achieved a 73.9\% F-score (4\% improvement) on the internal dataset, and 82.3\% (4.1\% improvement) on the Switchboard corpus with fusion (Fine-tuning) on the internal dataset. Moving on, \cite{sahoo2022mic_fuzzynet} discusses a proposed framework, called MIC\_FuzzyNET, for the automatic classification of musical instruments using DTL and fuzzy integral-based ensemble techniques. The fuzzy rank-based ensemble approach compensates for individual classifier faults and reduces errors and biases. The use of pre-trained models, such as EfficientNetV2 and ResNet18, allowed for better classification accuracy with smaller datasets. The validation and testing accuracies are 98.74\% and 96\%, respectively, for the IRMAS dataset. For the PCMIR dataset, the validation and testing accuracies are 98.74\% and 98\%, respectively.



\section{Discussion of key challenges} \label{sec5}
The studies reviewed in this document underscore the effectiveness of DTL-based ASR methods across various application scenarios, attributed to their computational efficiency and the capability to outperform existing ML algorithms. This superiority becomes especially evident when the TD samples diverge significantly from those deployed in the training of ML models.

However, there are other pressing challenges that warrant attention to enhance the performance and generalization of DTL-based ASR systems. A number of DTL studies, for instance, have overlooked describing the selection process of SD samples that could potentially bolster learning at the TD. The effectiveness of DTL algorithms commonly hinges upon accurately defining the similarity between the SD and TD. Consequently, an insufficient level of similarity between the SD and TD may drastically limit the advantages of DTL, or worse, lead to negative transfer \cite{weiss2016survey,xu2020hybrid}.

In addition, most DL frameworks necessitate pre-processing, such as time-scale domain feature calculation, time-frequency domain transformation, or frequency-domain analysis. These frameworks demand a high level of similarity between the SD and TD and consistency in their dimensions. Regrettably, limited research has focused on addressing the inconsistency between the dimensions of SD and TD data. In this regard, Hu et al. \cite{hu2019heterogeneous} proposed a DTL approach to examine knowledge transfer across heterogeneous domains.
This section aims to illuminate the most urgent issues currently gaining substantial attention in the field.

\subsection{The problem of NT}
However, NT, which signifies a decline in learning performance when data/knowledge is transferred from the SDs to TD, can affect the effectiveness of DTL.
Experiments in \cite{rosenstein2005to} first uncovered NT. The author showcased that a significant discrepancy between the SD and TD can lead to a drop in DTL performance. Additionally, the study proved that the performance of the target task could be negatively influenced by the "inductive bias" learned from the additional tasks.
Subsequently, a mathematical definition of NT was provided in \cite{wang2019characterizing}, wherein the concept of the negative transfer gap (NTG) was introduced to determine the occurrence of NT.

Moreover, \cite{meftah2021hidden} presented the authors' detailed quantitative and qualitative analyses investigating the covert NT resulting from the knowledge transfer from the "News" domain to the "Tweets" domain for Natural Language Processing (NLP) applications.
The exploration of positive and NT in a multi-domain ASR scenario was undertaken in \cite{doulaty2015data}. The study employed submodular functions based on the acoustic similarities between the source and target sets, consequently utilizing the positive transfer to enhance performance across domains while concurrently mitigating the effects of NT.
Lastly, the link between the quality and performance of TL and the estimated Kullback-Leibler divergence between the SDs and TD was demonstrated in \cite{sousa2014transfer}.



\subsection{The problem of overfitting}
Overcoming overfitting when developing DTL-based ASR schemes is among the key goals and challenges. Although DTL can manage overfitting better than other ML and DL models, this issue can be significant when developed models learn noises from the SDs, which negatively impacts their outputs \cite{delfosse2020deep}.
Typically, in DTL, we can not remove the network layers for identifying with confidence the appropriate classification/prediction parameters of DL models. Accordingly, if the first layers are removed, the dense layers can negatively be impacted as the number of trainable parameters will change. 
On another hand, we can reduce the number of dense layers, although analyzing the number of layers to be removed while alleviating model overfitting is computationally costly and laborious \cite{sayed2022deep}.

Besides, it is doable to partially overcome overfitting in DTL by introducing regularization methods, including dropout techniques for DL models \cite{fan2020statistical,alghamdi2020detection} and least absolute shrinkage and selection operator (LASSO) regularization for multiple linear regressions \cite{das2018evaluation}. Additionally, overfitting can be alleviated by elaborately designing a model development scheme. This means using cross-validation to optimize model parameters, where a whole dataset is divided into three groups for training, validation, and testing \cite{himeur2021artificial}. 
Additionally, it is possible to address overfitting using data augmentation techniques to generate synthetic data for better training DL models \cite{jha2019enhancing,zhao2017research}.

\subsection{Reproducibility of DTL-based ASR results}
With the increasing attention put into developing DTL-based solutions, there are new challenges that still hinder the extensive use of DTL models and largely affect their reproducibility \cite{copiaco2023innovative}. 
Consequently, empirical comparisons of the performance of DTL-based ASR techniques is still a challenge because of (i) the difficulties encountered when evaluating the generalisability of DTL networks since some studies have been tested on private customized datasets which are not available online; (ii) the absence of an online platforms that integrate previous DTL-based ASR algorithms and corresponding datasets. This is because of the limited number of existing open-source, benchmarked datasets; and (iii) the diversity of the evaluation metrics and parameters used to measure the performance of DTL models or quantify the distances between the SDs and TD \cite{himeur2022recent}. 
Therefore, the aforementioned issues make a fair comparison between DTL-based ASR solutions a complex, even impossible task.


\subsection{Measuring knowledge gains}
Measuring the knowledge gained when adopting a DTL model for specific ASR tasks holds paramount importance, yet the problem remains underexplored. To date, only a handful of research efforts have been directed toward this issue.

An analysis of how we quantify DTL gain was conducted in \cite{glorot2011domain}, where the authors proposed four metrics for assessing knowledge gain: the transfer ratio, transfer loss, transfer error, and in-domain ratio. Despite these metrics' ability to circumvent interpretation issues associated with performance results derived from varying SDs, predicting their behavior in other DTL-based techniques can be challenging. This is particularly evident in ASR tasks where class sets differ between problems. Furthermore, deriving a perfect baseline network can lead to indefinite performance results.

In light of these complexities, simpler evaluation metrics, such as accuracy, MSE, RMSE, MAE, PIR, F1 score, and other statistical measures (including class agreement), are often used to gauge the performance of DTL-based ASR solutions. Furthermore, \cite{cook2013transfer} investigates a practical approach to quantifying the similarity between two domains using a universal and domain-independent distance. While this provides a robust means of intelligently selecting suitable models and evaluating DTL techniques' performance, its application remains limited.

For instance, \cite{sullivan2022improving} uses WER to quantify the knowledge gain when a pre-trained wav2vec 2.0 \cite{baevski2020wav2vec,xu2021self} model is fine-tuned under a range of L1 and L2 training conditions. Similarly, WER is employed in \cite{xiong2020source} to measure the knowledge gain of a CNN-TDNN-F ASR acoustic model, which is initially trained on SD data before transferring its knowledge to a target dysarthric speaker dataset using neural network weight adaptation.

\subsection{Unification of DTL}
The development of effective and reliable DTL-based ASR strategies is often hindered by the broad range of mathematical formulations used to explain the underpinnings of DTL-based SRT techniques. For instance, heterogeneous DTL is advocated by Hu et al. in \cite{hu2019heterogeneous}, whereas Fan et al. \cite{fan2020statistical} pursue statistical examinations of DTL-based methodologies. Meanwhile, works like \cite{lin2021deep, zhang2019domain,zhang2020semi} focus on deep DA.

Although the studies covered in this review share the central concept of DTL, their definitions and implementations vary based on the scenarios considered. In other words, the employment of diverse terminologies and their variants can lead to reader confusion.
Therefore, it is crucial to standardize DTL definitions and background formulations to avoid such confusion. An attempt to unify DTL formulation and definitions has been made by Patricia et al. \cite{patricia2014learning}, though further efforts are still needed in this direction.

\subsection{Other challenges}
Processing DTL in a speech-based context appears more challenging than its image-based counterpart, largely due to the potential for extensive discrepancies between source and target databases, which could be attributed to language, speaker variation, age groups, ethnicity, and most notably, acoustic surroundings \cite{lu2021detection,karaman2021robust}. Thus, even with DTL implementation, some instances of ASR continue to struggle under limited resource conditions and the lack of clean data.

While the CTC has demonstrated significant potential in end-to-end speech recognition, it is bounded by frame independence assumptions where the output of one frame does not influence subsequent frame outputs, mirroring the unary potential of conditional random fields \cite{chan2016listen}.

The challenges of cross-lingual DTL can be bifurcated into two categories:
(i) For DTL, multilingual shared knowledge should encapsulate differing degrees of linguistic characteristics from multiple sources. The definition and acquisition of such information in a consistent manner is pivotal. (ii) A pursuit for integration at all levels of the knowledge hierarchy is necessary while accounting for linguistic variations \cite{he2020multi}.
Moreover, the computational load presents a substantial hurdle in the DTL deep DA processes. Transfers between SD and TD can lead to an increase in computational expenses. Adding to this, the deep architectures upon which DTL techniques rest inherently contribute to further computational burden.

\section{Future directions in DTL-based ASR} \label{sec6}

\subsection{Overcoming NT and measuring the cross-domain transferability}
While different research perspectives can be derived to improve DTL-based ASR solutions and facilitate their implementation in smart cities, two important directions are (i) to overcome the NT and (ii) to measure the cross-domain transferability, especially when there is a discrepancy between SDs and TD. It is expected that addressing these challenges will attract significant research and development in the near and far future.
Besides, there is increasing interest from the AI research community to investigate NT from different sides, such as NT mitigation, distant transfer, domain similarity estimation, and secure transfer \cite{paul2018comparison,wang2019characterizing,minoofam2021trcla}.
Additionally, some approaches have been proposed to alleviate the impact of NT; for example, when the domain similarity is measured, different methods can be selected according to the similarity level.
First, if the similarity level is high, the SD and TD data can be directly concatenated to train DTL algorithms. Second, if the similarity level is medium, an NT mitigation scheme can be adopted. Third, if the similarity level is low, a distant transfer or no transfer should be considered \cite{zhang2020survey}.

\subsection{Privacy preservation}
With ML, DL, and DTL advances, ASR systems have become more practical and more scalable. However, serious privacy threats can be posed since speech is a rich source of sensitive acoustic and textual information. 
Although open-source and offline ASR systems can eliminate privacy risks, online DTL-based systems can increase these threats. Moreover, offline and open-source ASR systems' transcription performance is inferior to that of cloud-based ASR systems, particularly in real-world scenarios \cite{ahmed2020preech}. 
Besides, in this case, the SD data can encompass vulnerable data that might be safeguarded. Accordingly, the knowledge transfer from the SD to the TD that can preserve users' privacy is a primary issue. 
Future research efforts must be considered by suggesting to integrate effective security and privacy protection strategies, e.g., decentralized DTL using blockchain \cite{ul2020decentralized,wang2021enabling} and federated DTL \cite{zhang2021federated,maurya2021federated}.

\subsection{Interpretation of DTL models}
Although DTL-based ASR models have achieved great success, they are still perceived as \enquote{black box} schemes that lack interpretation. This does not provide convincing insights into \enquote{how} and \enquote{why} they can reach final decisions \cite{arefeen2021transjury}. This can doubt the credibility of reached decisions and lack compelling evidence for convincing users or companies that these algorithms can work repeatedly. 
Moreover, applications of speech processing in general, and ASR in particular, have strict requirements for privacy preservation and accuracy. Thus, explaining the reasonableness of the prediction decisions is essential. 
Recently, the theory of explainable and interpretable ML/DL is attracting the growing interest of academic scientists not only for speech processing but also for other applications \cite{kirchhof2021chances,incahuanaco2022coffeese}. 
For instance, the study in \cite{ramakrishnan2016towards} presents the first attempt to introduce interpretable explanations for DTL in sequential tasks. Accordingly, an agent was set to explain how it learns a new task given prior common knowledge. 
This can then help in enhancing users' trust and acceptance of the system results and enabling iterative feedback to improve the system. 
Moving on, interpretable features are defined in \cite{kim2019the} and used to train a DTL algorithm for a new task. Typically, the relationship between the SD and TD in a DTL task has been explained, and the interpretability of the pretrained DTL has been examined.

Besides, lee et al. \cite{lee2021interpretable} introduce a knowledge distillation approach that (i) generates interpretable embedding procedure (IEP) knowledge based on PCA, and (ii) distills it in a form that can be transferred to the student network using a message passing neural network. Following, the teacher's knowledge and student network's target task have been trained via multi-task learning. 
Moreover, a DTL scheme that provides accurate, explainable classification results of optical coherence tomography (OCT) scans using a small sample size of images is proposed in \cite{carr2021interpretable}. It has been named interpretable staged TL (iSTL). The latter has significantly outperformed DTL techniques for unseen data, for which attention maps have shown that iSTL utilized clinical features for making predictions (and not uninterpretable abstractions).


\subsection{Online DTL}
An important issue that can be raised in ASR is related to how to transfer useful knowledge from the universal classifier trained on the SD data to personalize the speech recognition of each user in an online learning manner (TD data). To that end, online deep transfer learning (ODTL), which aims at transferring knowledge from an offline SD to an online TD learning task (represented by a similar or different feature space), can be explored. ODTL is challenging, especially if the data in the SD and TD can be different in terms of their class distributions as well as their feature representations. In this case, ODTL will assume that the SD feature space is a subset of the TD \cite{zhao2014online}.

To overcome these problems, some studies have investigated two different solutions based on ODTL. The first is based on homogeneous ODTL that relies upon a common feature space for SD and TD. In contrast, the second refers to the heterogeneous ODTL, which considers domains of different feature spaces \cite{wu2017online}. For instance, to overcome the problem of heterogeneous ODTL in \cite{wu2019online}, unlabeled co-occurrence instances are considered as intermediate supplementary data to link the SD and TD before performing knowledge transition.
In the same context, an ODTL scheme with an extreme learning machine is introduced in \cite{alasbahi2022an}. Typically, to address the data scarcity problem in the TD, a transfer learning with lag (TLL) technique that relies on embedded shallow neural networks is adopted. The latter enables knowledge transfer when the number of active features changes.

All in all, when applied in an online or incremental learning setting, ODTL is able to handle situations where the data distribution changes over time (concept drift) or where new tasks appear over time. The idea is to update the model continually as new data comes in, often by using a fraction of the incoming data to fine-tune the existing model while retaining the knowledge from previous tasks.
Typically, the challenge in ODTL, similar to other online learning scenarios, is to balance between the adaptation to new data (plasticity) and the preservation of previously learned knowledge (stability), often referred to as the stability-plasticity dilemma.


\subsection{DTL-based large language models (LLMs)}

\subsubsection{Using chatGPT as an ASR source model}

DTL-based LLM, such as chatGPT models, have shown promising potential for ASR tasks, specifically for both AM and LM components. The utilization of TL techniques in ChatGPT models allows for the transfer of knowledge learned from large-scale pre-training tasks to improve ASR performance. In the case of the AM, DTL-based ChatGPT models can benefit from the knowledge learned from pre-trained models that have been exposed to vast amounts of acoustic data \cite{qiu2021easytransfer,gruetzemacher2022deep}. The AM component of the ChatGPT model can learn to extract acoustic features, such as MFCCs or spectrograms, and leverage the pre-trained knowledge to improve its capability to recognize and transcribe speech accurately. By fine-tuning the AM component using TL, the model can adapt and specialize in specific acoustic domains or datasets, leading to improved performance in ASR tasks \cite{kasneci2023chatgpt}.

Similarly, in the LM component, DTL-based ChatGPT models can leverage TL to enhance the language modeling capability of the ASR system. By pre-trained on large text corpora, ChatGPT models have already learned rich language representations, which can be useful for ASR \cite{hu2023advancing}. The LM component can incorporate these pre-learned language features to better handle language-related challenges, such as handling different accents, dialects, or out-of-vocabulary words. Fine-tuning the LM using TL allows the model to adapt to the specific language characteristics of the ASR task, leading to more accurate and contextually appropriate transcriptions. Furthermore, DTL-based ChatGPT models for AM and LM can also benefit from domain adaptation \cite{sohail2023future}. By using auxiliary data from the target domain, such as additional labeled or unlabeled speech data, the models can be further fine-tuned to the specific characteristics and vocabulary of the ASR task. This helps to reduce domain mismatch and improve the overall performance and generalization of the ASR system.

All in all, DTL-based ChatGPT models for both AM and LM components offer the advantage of leveraging TL and domain adaptation to enhance ASR performance. By incorporating pre-learned knowledge from large-scale pre-training and fine-tuning on domain-specific data, these models can improve acoustic and language modeling capabilities, leading to more accurate and robust speech recognition.

\subsubsection{Using chatGPT for ASR assessment tools}

The utilization of the ChatGPT model for objective testing in ASR and MOS scale evaluation involves several steps. Firstly, a dataset comprising objective test cases for ASR and MOS evaluation, including audio recordings, transcripts, and MOS scores, should be assembled. The dataset should be diverse in terms of speech patterns, accents, and background noise \cite{wu2023brief}. Next, the ChatGPT model is fine-tuned using this objective test dataset to generate accurate transcriptions and MOS scores for the given audio inputs. The fine-tuned ChatGPT model is then integrated into the ASR system, either as a component in the ASR pipeline for refining transcriptions or for directly generating MOS scores for MOS evaluation based on the audio inputs \cite{strzelecki2023use}.

The performance of the integrated system can be evaluated by comparing the refined transcriptions and generated MOS scores from the ChatGPT model with the ground truth transcriptions and human-rated MOS scores. Evaluation metrics such as WER for ASR and correlation coefficients, like Pearson's correlation, for MOS evaluation can be utilized \cite{laskar2023systematic}. After analyzing the results, areas for improvement in the ASR system and MOS scale generation can be identified. The ChatGPT model can be further fine-tuned if necessary, and the evaluation process can be repeated until satisfactory performance is achieved \cite{cheng2023potential,sohail2023using}.

\section{Conclusion} \label{sec7}
This comprehensive review has explored the vast landscape of DTL within the sphere of ASR. Beginning with foundational aspects such as the literature search strategy, selection study, and quantitative analysis, the survey has unraveled the intricate threads of this emerging technology, illuminating its immense potential.
Delving into the conceptual background of DTL and ASR, we dissected the structure of ASR systems, evaluation criteria, and the pivotal role of datasets. We established a taxonomy for existing DTL techniques, from inductive and transductive DTL to adversarial DTL, all significant in their contributions to speech recognition.
The examination of DTL's application to AMs and LMs demonstrated its versatility. From feature normalization based-DTL, conservative training, and subspace-based DTL to BERT, LDA, NNLM, and LSTM-based DTL, DTL's multifaceted approaches to AMs and LMs showcased the method's breadth.
The review's exploration of cross-domain ASR and medical diagnosis illuminated how DTL has been instrumental in emotion recognition, cross-language DTL, cross-corpus SER (CC-SER), and adversarial TL-based ASR. The role of DTL in heart sound classification, Parkinson's disease detection, and other medical diagnosis applications further underlined its real-world implications.

While acknowledging the advancements in DTL-based ASR, we addressed key challenges like NT, overfitting, reproducibility, measuring knowledge gains, and the need for unification in DTL, shedding light on the roadblocks impeding the full realization of DTL's potential.
Turning our gaze to the future, we discussed promising directions in DTL-based ASR, such as overcoming NT, privacy preservation, interpretation of DTL models, online DTL, and more.
The final leg of our exploration ventured into DTL-based LLMs, focusing on ChatGPT's potential as both an ASR source model and an assessment tool.
In its entirety, this survey has not only provided a comprehensive overview of the current state of DTL and ASR but also charted the course for future exploration and innovation. As we navigate through these uncharted waters, it is crucial to remember that each challenge represents an opportunity for further growth and discovery. As we continue to refine and expand our understanding, we edge closer to fully harnessing the transformative potential of DTL in Automatic Speech Recognition.



Overall, it has been seen that the intersection between DL/DTL and ASR in this era will significantly speed up the research advancement of speech technology in general and particularly ASR and NLP. Typically, the speech technology and AI scientific communities look forward to the substantial advances and progressive significance of DTL technology not only for the ASR problematic but also for other research fields, including medical diagnosis, energy, smart cities, fault and anomaly detection, etc.

Lastly, it is worth noting that DTL, as a promising area in DL, has shown an ensemble of benefits over conventional ML and DL, including less computational cost in some scenarios, less data dependence, less label dependence, and better performance in some case studies. However, further research and development efforts still need to be devoted to improve the generalizability and overall performance of DTL models. We hope this study will help the AI and speech-processing communities better understand the research status and the research ideas for using DTL for ASR applications.


\bibliographystyle{elsarticle-num}
\bibliography{references}

\begin{thebibliography}{100}
\expandafter\ifx\csname url\endcsname\relax
  \def\url#1{\texttt{#1}}\fi
\expandafter\ifx\csname urlprefix\endcsname\relax\def\urlprefix{URL }\fi
\expandafter\ifx\csname href\endcsname\relax
  \def\href#1#2{#2} \def\path#1{#1}\fi

\bibitem{nedjah2023automatic}
N.~Nedjah, A.~D. Bonilla, L.~de~Macedo~Mourelle, Automatic speech recognition
  of portuguese phonemes using neural networks ensemble, Expert Systems with
  Applications 229 (2023) 120378.

\bibitem{anoop2023suitability}
C.~S. Anoop, A.~G. Ramakrishnan, Suitability of syllable-based modeling units
  for end-to-end speech recognition in sanskrit and other indian languages,
  Expert Systems with Applications 220 (2023) 119722.

\bibitem{haneche2021compressed}
H.~Haneche, A.~Ouahabi, B.~Boudraa, Compressed sensing-speech coding scheme for
  mobile communications, Circuits, Systems, and Signal Processing (2021) 1--21.

\bibitem{michelsanti2021overview}
D.~Michelsanti, Z.-H. Tan, S.-X. Zhang, Y.~Xu, M.~Yu, D.~Yu, J.~Jensen, An
  overview of deep-learning-based audio-visual speech enhancement and
  separation, IEEE/ACM Transactions on Audio, Speech, and Language Processing
  (2021).

\bibitem{luo2021group}
Y.~Luo, C.~Han, N.~Mesgarani, Group communication with context codec for
  lightweight source separation, IEEE/ACM Transactions on Audio, Speech, and
  Language Processing 29 (2021) 1752--1761.

\bibitem{kheddar2019pitch}
H.~Kheddar, M.~Bouzid, D.~Meg{\'\i}as, Pitch and fourier magnitude based
  steganography for hiding 2.4 kbps melp bitstream, IET Signal Processing
  13~(3) (2019) 396--407.

\bibitem{kheddar2022speech}
H.~Kheddar, A.~C. Mazari, G.~H. Ilk, Speech steganography based on double
  approximation of lsfs parameters in amr coding, in: 2022 7th International
  Conference on Image and Signal Processing and their Applications (ISPA),
  IEEE, 2022, pp. 1--8.

\bibitem{kheddar2018fourier}
H.~Kheddar, D.~Megias, M.~Bouzid, Fourier magnitude-based steganography for
  hiding 2.4 kbpsmelp secret speech, in: 2018 International Conference on
  Applied Smart Systems (ICASS), IEEE, 2018, pp. 1--5.

\bibitem{yassine2012secure}
H.~Yassine, B.~Bachir, K.~Aziz, A secure and high robust audio watermarking
  system for copyright protection, International Journal of Computer
  Applications 53~(17) (2012) 33--39.

\bibitem{yamni2022efficient}
M.~Yamni, H.~Karmouni, M.~Sayyouri, H.~Qjidaa, Efficient watermarking algorithm
  for digital audio/speech signal, Digital Signal Processing 120 (2022) 103251.

\bibitem{chen2020specmark}
H.~Chen, B.~D. Rouhani, F.~Koushanfar, Specmark: A spectral watermarking
  framework for ip protection of speech recognition systems., in: INTERSPEECH,
  2020, pp. 2312--2316.

\bibitem{olivieri2021audio}
M.~Olivieri, R.~Malvermi, M.~Pezzoli, M.~Zanoni, S.~Gonzalez, F.~Antonacci,
  A.~Sarti, Audio information retrieval and musical acoustics, IEEE
  Instrumentation \& Measurement Magazine 24~(7) (2021) 10--20.

\bibitem{wold1996content}
E.~Wold, T.~Blum, D.~Keislar, J.~Wheaten, Content-based classification, search,
  and retrieval of audio, IEEE multimedia 3~(3) (1996) 27--36.

\bibitem{boes2021audiovisual}
W.~Boes, et~al., Audiovisual transfer learning for audio tagging and sound
  event detection, Proceedings Interspeech 2021 (2021).

\bibitem{tang2021general}
Y.~Tang, J.~Pino, C.~Wang, X.~Ma, D.~Genzel, A general multi-task learning
  framework to leverage text data for speech to text tasks, in: ICASSP
  2021-2021 IEEE International Conference on Acoustics, Speech and Signal
  Processing (ICASSP), IEEE, 2021, pp. 6209--6213.

\bibitem{plaza2021comparing}
F.~M. Plaza-del Arco, M.~D. Molina-Gonz{\'a}lez, L.~A. Ure{\~n}a-L{\'o}pez,
  M.~T. Mart{\'\i}n-Valdivia, Comparing pre-trained language models for spanish
  hate speech detection, Expert Systems with Applications 166 (2021) 114120.

\bibitem{meghraoui2021novel}
D.~Meghraoui, B.~Boudraa, T.~Merazi, P.~G. Vilda, A novel pre-processing
  technique in pathologic voice detection: Application to parkinson’s disease
  phonation, Biomedical Signal Processing and Control 68 (2021) 102604.

\bibitem{lin2021speech}
Y.-Y. Lin, W.-Z. Zheng, W.~C. Chu, J.-Y. Han, Y.-H. Hung, G.-M. Ho, C.-Y.
  Chang, Y.-H. Lai, A speech command control-based recognition system for
  dysarthric patients based on deep learning technology, Applied Sciences
  11~(6) (2021) 2477.

\bibitem{kumar2022novel}
Y.~Kumar, S.~Gupta, W.~Singh, A novel deep transfer learning models for
  recognition of birds sounds in different environment, Soft Computing 26~(3)
  (2022) 1003--1023.

\bibitem{padi2021improved}
S.~Padi, S.~O. Sadjadi, R.~D. Sriram, D.~Manocha, Improved speech emotion
  recognition using transfer learning and spectrogram augmentation, in:
  Proceedings of the 2021 International Conference on Multimodal Interaction,
  2021, pp. 645--652.

\bibitem{himeur2022next}
Y.~Himeur, M.~Elnour, F.~Fadli, N.~Meskin, I.~Petri, Y.~Rezgui, F.~Bensaali,
  A.~Amra, Next-generation energy systems for sustainable smart cities: Roles
  of transfer learning, Sustainable Cities and Society (2022) 1--35.

\bibitem{niu2020decade}
S.~Niu, Y.~Liu, J.~Wang, H.~Song, A decade survey of transfer learning
  (2010--2020), IEEE Transactions on Artificial Intelligence 1~(2) (2020)
  151--166.

\bibitem{sayed2023time}
A.~N. Sayed, Y.~Himeur, F.~Bensaali, From time-series to 2d images for building
  occupancy prediction using deep transfer learning, Engineering Applications
  of Artificial Intelligence 119 (2023) 105786.

\bibitem{himeur2023face}
Y.~Himeur, S.~Al-Maadeed, I.~Varlamis, N.~Al-Maadeed, K.~Abualsaud, A.~Mohamed,
  Face mask detection in smart cities using deep and transfer learning: lessons
  learned from the covid-19 pandemic, Systems 11~(2) (2023) 107.

\bibitem{kheddar2022high}
H.~Kheddar, D.~Meg{\'\i}as, High capacity speech steganography for the g723. 1
  coder based on quantised line spectral pairs interpolation and cnn
  auto-encoding, Applied Intelligence (2022) 1--19.

\bibitem{jia2018transfer}
Y.~Jia, Y.~Zhang, R.~J. Weiss, Q.~Wang, J.~Shen, F.~Ren, Z.~Chen, P.~Nguyen,
  R.~Pang, I.~L. Moreno, et~al., Transfer learning from speaker verification to
  multispeaker text-to-speech synthesis, arXiv preprint arXiv:1806.04558
  (2018).

\bibitem{malik2021automatic}
M.~Malik, M.~K. Malik, K.~Mehmood, I.~Makhdoom, Automatic speech recognition: a
  survey, Multimedia Tools and Applications 80~(6) (2021) 9411--9457.

\bibitem{hazarika2021conversational}
D.~Hazarika, S.~Poria, R.~Zimmermann, R.~Mihalcea, Conversational transfer
  learning for emotion recognition, Information Fusion 65 (2021) 1--12.

\bibitem{vryzas2021web}
N.~Vryzas, L.~Vrysis, R.~Kotsakis, C.~Dimoulas, A web crowdsourcing framework
  for transfer learning and personalized speech emotion recognition, Machine
  Learning with Applications 6 (2021) 100132.

\bibitem{malhotra2021bidirectional}
S.~Malhotra, V.~Kumar, A.~Agarwal, Bidirectional transfer learning model for
  sentiment analysis of natural language, Journal of Ambient Intelligence and
  Humanized Computing (2021) 1--21.

\bibitem{hettiarachchi2021novel}
R.~Hettiarachchi, U.~Haputhanthri, K.~Herath, H.~Kariyawasam, S.~Munasinghe,
  K.~Wickramasinghe, D.~Samarasinghe, A.~De~Silva, C.~U. Edussooriya, A novel
  transfer learning-based approach for screening pre-existing heart diseases
  using synchronized ecg signals and heart sounds, in: 2021 IEEE International
  Symposium on Circuits and Systems (ISCAS), IEEE, 2021, pp. 1--5.

\bibitem{karaman2021robust}
O.~Karaman, H.~{\c{C}}ak{\i}n, A.~Alhudhaif, K.~Polat, Robust automated
  parkinson disease detection based on voice signals with transfer learning,
  Expert Systems with Applications 178 (2021) 115013.

\bibitem{harati2021speech}
A.~Harati, E.~Shriberg, T.~Rutowski, P.~Chlebek, Y.~Lu, R.~Oliveira,
  Speech-based depression prediction using encoder-weight-only transfer
  learning and a large corpus, in: ICASSP 2021-2021 IEEE International
  Conference on Acoustics, Speech and Signal Processing (ICASSP), IEEE, 2021,
  pp. 7273--7277.

\bibitem{lu2015transfer}
J.~Lu, V.~Behbood, P.~Hao, H.~Zuo, S.~Xue, G.~Zhang, Transfer learning using
  computational intelligence: A survey, Knowledge-Based Systems 80 (2015)
  14--23.

\bibitem{weiss2016survey}
K.~Weiss, T.~M. Khoshgoftaar, D.~Wang, A survey of transfer learning, Journal
  of Big data 3~(1) (2016) 1--40.

\bibitem{himeur2023video}
Y.~Himeur, S.~Al-Maadeed, H.~Kheddar, N.~Al-Maadeed, K.~Abualsaud, A.~Mohamed,
  T.~Khattab, Video surveillance using deep transfer learning and deep domain
  adaptation: Towards better generalization, Engineering Applications of
  Artificial Intelligence 119 (2023) 105698.

\bibitem{zhuang2020comprehensive}
F.~Zhuang, Z.~Qi, K.~Duan, D.~Xi, Y.~Zhu, H.~Zhu, H.~Xiong, Q.~He, A
  comprehensive survey on transfer learning, Proceedings of the IEEE 109~(1)
  (2020) 43--76.

\bibitem{durrani2021transfer}
S.~Durrani, U.~Arshad, Transfer learning from high-resource to low-resource
  language improves speech affect recognition classification accuracy, arXiv
  preprint arXiv:2103.11764 (2021).

\bibitem{wan2021review}
Z.~Wan, R.~Yang, M.~Huang, N.~Zeng, X.~Liu, A review on transfer learning in
  eeg signal analysis, Neurocomputing 421 (2021) 1--14.

\bibitem{bashath2022data}
S.~Bashath, N.~Perera, S.~Tripathi, K.~Manjang, M.~Dehmer, F.~E. Streib, A
  data-centric review of deep transfer learning with applications to text data,
  Information Sciences 585 (2022) 498--528.

\bibitem{kheddar2023deep}
H.~Kheddar, Y.~Himeur, A.~I. Awad, Deep transfer learning applications in
  intrusion detection systems: A comprehensive review, arXiv preprint
  arXiv:2304.10550 (2023).

\bibitem{lu2021general}
Y.~Lu, Z.~Tian, R.~Zhou, W.~Liu, A general transfer learning-based framework
  for thermal load prediction in regional energy system, Energy 217 (2021)
  119322.

\bibitem{ramirez2019learning}
P.~Z. Ramirez, A.~Tonioni, S.~Salti, L.~D. Stefano, Learning across tasks and
  domains, in: Proceedings of the IEEE/CVF International Conference on Computer
  Vision, 2019, pp. 8110--8119.

\bibitem{li2020transfer}
S.~Li, T.~T. Cai, H.~Li, Transfer learning in large-scale gaussian graphical
  models with false discovery rate control, arXiv preprint arXiv:2010.11037
  (2020).

\bibitem{tuia2016domain}
D.~Tuia, C.~Persello, L.~Bruzzone, Domain adaptation for the classification of
  remote sensing data: An overview of recent advances, IEEE geoscience and
  remote sensing magazine 4~(2) (2016) 41--57.

\bibitem{alyafeai2020survey}
Z.~Alyafeai, M.~S. AlShaibani, I.~Ahmad, A survey on transfer learning in
  natural language processing, arXiv preprint arXiv:2007.04239 (2020).

\bibitem{wang2020transfer}
J.~Wang, Y.~Chen, W.~Feng, H.~Yu, M.~Huang, Q.~Yang, Transfer learning with
  dynamic distribution adaptation, ACM Transactions on Intelligent Systems and
  Technology (TIST) 11~(1) (2020) 1--25.

\bibitem{ganin2016domain}
Y.~Ganin, E.~Ustinova, H.~Ajakan, P.~Germain, H.~Larochelle, F.~Laviolette,
  M.~Marchand, V.~Lempitsky, Domain-adversarial training of neural networks,
  The journal of machine learning research 17~(1) (2016) 2096--2030.

\bibitem{bousmalis2016domain}
K.~Bousmalis, G.~Trigeorgis, N.~Silberman, D.~Krishnan, D.~Erhan, Domain
  separation networks, Advances in neural information processing systems 29
  (2016).

\bibitem{chen2019joint}
C.~Chen, Z.~Chen, B.~Jiang, X.~Jin, Joint domain alignment and discriminative
  feature learning for unsupervised deep domain adaptation, in: Proceedings of
  the AAAI conference on artificial intelligence, Vol.~33, 2019, pp.
  3296--3303.

\bibitem{long2017deep}
M.~Long, H.~Zhu, J.~Wang, M.~I. Jordan, Deep transfer learning with joint
  adaptation networks, in: International conference on machine learning, PMLR,
  2017, pp. 2208--2217.

\bibitem{zhang2018collaborative}
W.~Zhang, W.~Ouyang, W.~Li, D.~Xu, Collaborative and adversarial network for
  unsupervised domain adaptation, in: Proceedings of the IEEE conference on
  computer vision and pattern recognition, 2018, pp. 3801--3809.

\bibitem{filippidou2020alpha}
F.~Filippidou, L.~Moussiades, A benchmarking of ibm, google and wit automatic
  speech recognition systems, in: IFIP International Conference on Artificial
  Intelligence Applications and Innovations, Springer, 2020, pp. 73--82.

\bibitem{jiang2021gdpr}
D.~Jiang, C.~Tan, J.~Peng, C.~Chen, X.~Wu, W.~Zhao, Y.~Song, Y.~Tong, C.~Liu,
  Q.~Xu, et~al., A gdpr-compliant ecosystem for speech recognition with
  transfer, federated, and evolutionary learning, ACM Transactions on
  Intelligent Systems and Technology (TIST) 12~(3) (2021) 1--19.

\bibitem{kumar2021exploration}
A.~Kumar, R.~K. Aggarwal, An exploration of semi-supervised and
  language-adversarial transfer learning using hybrid acoustic model for hindi
  speech recognition, Journal of Reliable Intelligent Environments (2021)
  1--16.

\bibitem{chan2016listen}
W.~Chan, N.~Jaitly, Q.~Le, O.~Vinyals, Listen, attend and spell: A neural
  network for large vocabulary conversational speech recognition, in: 2016 IEEE
  International Conference on Acoustics, Speech and Signal Processing (ICASSP),
  IEEE, 2016, pp. 4960--4964.

\bibitem{yu2021enhancing}
Q.~Yu, Y.~Ma, Y.~Li, Enhancing speech recognition for parkinson’s disease
  patient using transfer learning technique, Journal of Shanghai Jiaotong
  University (Science) (2021) 1--9.

\bibitem{bai2021fast}
Y.~Bai, J.~Yi, J.~Tao, Z.~Tian, Z.~Wen, S.~Zhang, Fast end-to-end speech
  recognition via non-autoregressive models and cross-modal knowledge
  transferring from bert, IEEE/ACM Transactions on Audio, Speech, and Language
  Processing 29 (2021) 1897--1911.

\bibitem{zhang2017towards}
Y.~Zhang, M.~Pezeshki, P.~Brakel, S.~Zhang, C.~L.~Y. Bengio, A.~Courville,
  Towards end-to-end speech recognition with deep convolutional neural
  networks, arXiv preprint arXiv:1701.02720 (2017).

\bibitem{qin2018towards}
C.-X. Qin, D.~Qu, L.-H. Zhang, Towards end-to-end speech recognition with
  transfer learning, EURASIP Journal on Audio, Speech, and Music Processing
  2018~(1) (2018) 1--9.

\bibitem{zhu2020end}
X.~Zhu, H.~Huang, End-to-end amdo-tibetan speech recognition based on knowledge
  transfer, IEEE Access 8 (2020) 170991--171000.

\bibitem{errattahi2018automatic}
R.~Errattahi, A.~El~Hannani, H.~Ouahmane, Automatic speech recognition errors
  detection and correction: A review, Procedia Computer Science 128 (2018)
  32--37.

\bibitem{li2021insight}
Y.~Li, X.~Zhang, P.~Wang, X.~Zhang, Y.~Liu, Insight into an unsupervised
  two-step sparse transfer learning algorithm for speech diagnosis of
  parkinson’s disease, Neural Computing and Applications (2021) 1--18.

\bibitem{ramadan2021detecting}
R.~A. Ramadan, Detecting adversarial attacks on audio-visual speech recognition
  using deep learning method, International Journal of Speech Technology (2021)
  1--7.

\bibitem{zhang2021multi}
L.~Zhang, Q.~Wang, K.~A. Lee, L.~Xie, H.~Li, Multi-level transfer learning from
  near-field to far-field speaker verification, arXiv preprint arXiv:2106.09320
  (2021).

\bibitem{hong2017transfer}
Q.~Hong, L.~Li, J.~Zhang, L.~Wan, H.~Guo, Transfer learning for plda-based
  speaker verification, Speech Communication 92 (2017) 90--99.

\bibitem{yusuf2019low}
B.~Yusuf, B.~Gundogdu, M.~Saraclar, Low resource keyword search with
  synthesized crosslingual exemplars, IEEE/ACM Transactions on Audio, Speech,
  and Language Processing 27~(7) (2019) 1126--1135.

\bibitem{lu2021detection}
T.~Lu, B.~Han, F.~Yu, Detection and classification of marine mammal sounds
  using alexnet with transfer learning, Ecological Informatics 62 (2021)
  101277.

\bibitem{arora2017study}
P.~Arora, R.~Haeb-Umbach, A study on transfer learning for acoustic event
  detection in a real life scenario, in: 2017 IEEE 19th International Workshop
  on Multimedia Signal Processing (MMSP), IEEE, 2017, pp. 1--6.

\bibitem{kumar2021development}
G.~A. Kumar, J.~H. William, Development of visual-only speech recognition
  system for mute people, Circuits, Systems, and Signal Processing (2021)
  1--21.

\bibitem{koike2020audio}
T.~Koike, K.~Qian, Q.~Kong, M.~D. Plumbley, B.~W. Schuller, Y.~Yamamoto, Audio
  for audio is better? an investigation on transfer learning models for heart
  sound classification, in: 2020 42nd Annual International Conference of the
  IEEE Engineering in Medicine \& Biology Society (EMBC), IEEE, 2020, pp.
  74--77.

\bibitem{markitantov2020transfer}
M.~Markitantov, Transfer learning in speaker’s age and gender recognition,
  in: International Conference on Speech and Computer, Springer, 2020, pp.
  326--335.

\bibitem{fahmy2020transfer}
F.~K. Fahmy, M.~I. Khalil, H.~M. Abbas, A transfer learning end-to-end arabic
  text-to-speech (tts) deep architecture, in: IAPR Workshop on Artificial
  Neural Networks in Pattern Recognition, Springer, 2020, pp. 266--277.

\bibitem{oord2016wavenet}
A.~v.~d. Oord, S.~Dieleman, H.~Zen, K.~Simonyan, O.~Vinyals, A.~Graves,
  N.~Kalchbrenner, A.~Senior, K.~Kavukcuoglu, Wavenet: A generative model for
  raw audio, arXiv preprint arXiv:1609.03499 (2016).

\bibitem{recommendation2001perceptual}
I.-T. Recommendation, Perceptual evaluation of speech quality (pesq): An
  objective method for end-to-end speech quality assessment of narrow-band
  telephone networks and speech codecs, Rec. ITU-T P. 862 (2001).

\bibitem{siddiqui2020using}
S.~Siddiqui, G.~Rasool, R.~P. Ramachandran, N.~C. Bouaynaya, Using deep speech
  recognition to evaluate speech enhancement methods, in: 2020 International
  Joint Conference on Neural Networks (IJCNN), IEEE, 2020, pp. 1--7.

\bibitem{peng2019security}
J.~Peng, Y.~Jiang, S.~Tang, F.~Meziane, Security of streaming media
  communications with logistic map and self-adaptive detection-based
  steganography, IEEE Transactions on Dependable and Secure Computing (2019).

\bibitem{vu2020deep}
L.~Vu, Q.~U. Nguyen, D.~N. Nguyen, D.~T. Hoang, E.~Dutkiewicz, Deep transfer
  learning for {IoT} attack detection, IEEE Access 8 (2020) 107335--107344.

\bibitem{tendle2021study}
A.~Tendle, M.~R. Hasan, A study of the generalizability of self-supervised
  representations, Machine Learning with Applications 6 (2021) 100124.

\bibitem{liu2022improved}
Z.~Liu, Y.~Xu, Y.~Xu, Q.~Qian, H.~Li, X.~Ji, A.~Chan, R.~Jin, Improved
  fine-tuning by better leveraging pre-training data, Advances in Neural
  Information Processing Systems 35 (2022) 32568--32581.

\bibitem{liu2023dropout}
Z.~Liu, Z.~Xu, J.~Jin, Z.~Shen, T.~Darrell, Dropout reduces underfitting, arXiv
  preprint arXiv:2303.01500 (2023).

\bibitem{bu2023achieving}
J.~Bu, Achieving more with less: Learning generalizable neural networks with
  less labeled data and computational overheads, Ph.D. thesis, Virginia Tech
  (2023).

\bibitem{garofolo1993darpa}
J.~S. Garofolo, L.~F. Lamel, W.~M. Fisher, J.~G. Fiscus, D.~S. Pallett, Darpa
  timit acoustic-phonetic continous speech corpus cd-rom. nist speech disc
  1-1.1, NASA STI/Recon technical report n 93 (1993) 27403.

\bibitem{wang2020cross}
S.~Wang, W.~Li, S.~M. Siniscalchi, C.-H. Lee, A cross-task transfer learning
  approach to adapting deep speech enhancement models to unseen background
  noise using paired senone classifiers, in: ICASSP 2020-2020 IEEE
  International Conference on Acoustics, Speech and Signal Processing (ICASSP),
  IEEE, 2020, pp. 6219--6223.

\bibitem{panayotov2015librispeech}
V.~Panayotov, G.~Chen, D.~Povey, S.~Khudanpur, Librispeech: an asr corpus based
  on public domain audio books, in: 2015 IEEE international conference on
  acoustics, speech and signal processing (ICASSP), IEEE, 2015, pp. 5206--5210.

\bibitem{yu2022enhancing}
Q.~Yu, Y.~Ma, Y.~Li, Enhancing speech recognition for parkinson’s disease
  patient using transfer learning technique, Journal of Shanghai Jiaotong
  University (Science) 27~(1) (2022) 90--98.

\bibitem{di2019must}
M.~A. Di~Gangi, R.~Cattoni, L.~Bentivogli, M.~Negri, M.~Turchi, Must-c: a
  multilingual speech translation corpus, in: 2019 Conference of the North
  American Chapter of the Association for Computational Linguistics: Human
  Language Technologies, Association for Computational Linguistics, 2019, pp.
  2012--2017.

\bibitem{veaux2017superseded}
C.~Veaux, J.~Yamagishi, K.~MacDonald, et~al., Superseded-cstr vctk corpus:
  English multi-speaker corpus for cstr voice cloning toolkit (2017).

\bibitem{schultz2013globalphone}
T.~Schultz, N.~T. Vu, T.~Schlippe, Globalphone: A multilingual text \& speech
  database in 20 languages, in: 2013 IEEE International Conference on
  Acoustics, Speech and Signal Processing, IEEE, 2013, pp. 8126--8130.

\bibitem{sahraeian2018cross}
R.~Sahraeian, D.~Van~Compernolle, Cross-entropy training of dnn ensemble
  acoustic models for low-resource asr, IEEE/ACM Transactions on Audio, Speech,
  and Language Processing 26~(11) (2018) 1991--2001.

\bibitem{hartmann2017analysis}
W.~Hartmann, D.~Karakos, R.~Hsiao, L.~Zhang, T.~Alum{\"a}e, S.~Tsakalidis,
  R.~Schwartz, Analysis of keyword spotting performance across iarpa babel
  languages, in: 2017 IEEE International Conference on Acoustics, Speech and
  Signal Processing (ICASSP), IEEE, 2017, pp. 5765--5769.

\bibitem{yi2018language}
J.~Yi, J.~Tao, Z.~Wen, Y.~Bai, Language-adversarial transfer learning for
  low-resource speech recognition, IEEE/ACM Transactions on Audio, Speech, and
  Language Processing 27~(3) (2018) 621--630.

\bibitem{liu2016open}
C.~Liu, D.~Springer, Q.~Li, B.~Moody, R.~A. Juan, F.~J. Chorro, F.~Castells,
  J.~M. Roig, I.~Silva, A.~E. Johnson, et~al., An open access database for the
  evaluation of heart sound algorithms, Physiological Measurement 37~(12)
  (2016) 2181.

\bibitem{kim2008dysarthric}
H.~Kim, M.~Hasegawa-Johnson, A.~Perlman, J.~Gunderson, T.~S. Huang, K.~Watkin,
  S.~Frame, Dysarthric speech database for universal access research, in: Ninth
  Annual Conference of the International Speech Communication Association,
  2008.

\bibitem{xiong2020source}
F.~Xiong, J.~Barker, Z.~Yue, H.~Christensen, Source domain data selection for
  improved transfer learning targeting dysarthric speech recognition, in:
  ICASSP 2020-2020 IEEE International Conference on Acoustics, Speech and
  Signal Processing (ICASSP), IEEE, 2020, pp. 7424--7428.

\bibitem{shahamiri2021speech}
S.~R. Shahamiri, Speech vision: An end-to-end deep learning-based dysarthric
  automatic speech recognition system, IEEE Transactions on Neural Systems and
  Rehabilitation Engineering 29 (2021) 852--861.

\bibitem{mesaros2016tut}
A.~Mesaros, T.~Heittola, T.~Virtanen, Tut database for acoustic scene
  classification and sound event detection, in: 2016 24th European Signal
  Processing Conference (EUSIPCO), IEEE, 2016, pp. 1128--1132.

\bibitem{chen2018transfer}
Y.~Chen, B.~Gao, L.~Jiang, K.~Yin, J.~Gu, W.~L. Woo, Transfer learning for
  wearable long-term social speech evaluations, IEEE Access 6 (2018)
  61305--61316.

\bibitem{devlin2019bert}
J.~Devlin, M.-W. Chang, K.~Lee, K.~Toutanova, Bert: Pre-training of deep
  bidirectional transformers for language understanding (2019).
\newblock \href {http://arxiv.org/abs/1810.04805} {\path{arXiv:1810.04805}}.

\bibitem{wang2019overview}
D.~Wang, X.~Wang, S.~Lv, An overview of end-to-end automatic speech
  recognition, Symmetry 11~(8) (2019) 1018.

\bibitem{novoa2018uncertainty}
J.~Novoa, J.~Fredes, V.~Poblete, N.~B. Yoma, Uncertainty weighting and
  propagation in dnn--hmm-based speech recognition, Computer Speech \& Language
  47 (2018) 30--46.

\bibitem{fahad2021dnn}
M.~S. Fahad, A.~Deepak, G.~Pradhan, J.~Yadav, Dnn-hmm-based speaker-adaptive
  emotion recognition using mfcc and epoch-based features, Circuits, Systems,
  and Signal Processing 40~(1) (2021) 466--489.

\bibitem{nakatani2019improving}
T.~Nakatani, Improving transformer-based end-to-end speech recognition with
  connectionist temporal classification and language model integration, in:
  Proc. Interspeech 2019, 2019, pp. 1408--1412.

\bibitem{salazar2019self}
J.~Salazar, K.~Kirchhoff, Z.~Huang, Self-attention networks for connectionist
  temporal classification in speech recognition, in: ICASSP 2019-2019 IEEE
  International Conference on Acoustics, Speech and Signal Processing (ICASSP),
  IEEE, 2019, pp. 7115--7119.

\bibitem{chiu2018state}
C.-C. Chiu, T.~N. Sainath, Y.~Wu, R.~Prabhavalkar, P.~Nguyen, Z.~Chen,
  A.~Kannan, R.~J. Weiss, K.~Rao, E.~Gonina, et~al., State-of-the-art speech
  recognition with sequence-to-sequence models, in: 2018 IEEE International
  Conference on Acoustics, Speech and Signal Processing (ICASSP), IEEE, 2018,
  pp. 4774--4778.

\bibitem{sukhadia2023domain}
V.~N. Sukhadia, S.~Umesh, Domain adaptation of low-resource target-domain
  models using well-trained asr conformer models, in: 2022 IEEE Spoken Language
  Technology Workshop (SLT), IEEE, 2023, pp. 295--301.

\bibitem{fan2022towards}
R.~Fan, Y.~Zhu, J.~Wang, A.~Alwan, Towards better domain adaptation for
  self-supervised models: A case study of child asr, IEEE Journal of Selected
  Topics in Signal Processing 16~(6) (2022) 1242--1252.

\bibitem{thomas2022efficient}
B.~Thomas, S.~Kessler, S.~Karout, Efficient adapter transfer of self-supervised
  speech models for automatic speech recognition, in: ICASSP 2022-2022 IEEE
  International Conference on Acoustics, Speech and Signal Processing (ICASSP),
  IEEE, 2022, pp. 7102--7106.

\bibitem{mridha2021study}
M.~Mridha, A.~Q. Ohi, M.~A. Hamid, M.~M. Monowar, A study on the challenges and
  opportunities of speech recognition for bengali language, Artificial
  Intelligence Review (2021) 1--25.

\bibitem{huang2015maximum}
Z.~Huang, S.~M. Siniscalchi, I.-F. Chen, J.~Wu, C.-H. Lee, Maximum a posteriori
  adaptation of network parameters in deep models, arXiv preprint
  arXiv:1503.02108 (2015).

\bibitem{elaraby2016deep}
M.~S. Elaraby, M.~Abdallah, S.~Abdou, M.~Rashwan, A deep neural networks (dnn)
  based models for a computer aided pronunciation learning system, in:
  International Conference on Speech and Computer, Springer, 2016, pp. 51--58.

\bibitem{mimura2016joint}
M.~Mimura, S.~Sakai, T.~Kawahara, Joint optimization of denoising autoencoder
  and dnn acoustic model based on multi-target learning for noisy speech
  recognition., in: Interspeech, 2016, pp. 3803--3807.

\bibitem{ma2017approaches}
M.~Ma, M.~Nirschl, F.~Biadsy, S.~Kumar, Approaches for neural-network language
  model adaptation., in: INTERSPEECH, 2017, pp. 259--263.

\bibitem{kadyan2022transfer}
V.~Kadyan, P.~Bawa, Transfer learning through perturbation-based in-domain
  spectrogram augmentation for adult speech recognition, Neural Computing and
  Applications 34~(23) (2022) 21015--21033.

\bibitem{weninger2019listen}
F.~Weninger, J.~Andr{\'e}s-Ferrer, X.~Li, P.~Zhan, Listen, attend, spell and
  adapt: Speaker adapted sequence-to-sequence asr, arXiv preprint
  arXiv:1907.04916 (2019).

\bibitem{devlin2018bert}
J.~Devlin, M.-W. Chang, K.~Lee, K.~Toutanova, Bert: Pre-training of deep
  bidirectional transformers for language understanding, arXiv preprint
  arXiv:1810.04805 (2018).

\bibitem{deena2018recurrent}
S.~Deena, M.~Hasan, M.~Doulaty, O.~Saz, T.~Hain, Recurrent neural network
  language model adaptation for multi-genre broadcast speech recognition and
  alignment, IEEE/ACM Transactions on Audio, Speech, and Language Processing
  27~(3) (2018) 572--582.

\bibitem{song2019topic}
Y.~Song, D.~Jiang, X.~Wu, Q.~Xu, R.~C.-W. Wong, Q.~Yang, Topic-aware dialogue
  speech recognition with transfer learning., in: INTERSPEECH, 2019, pp.
  829--833.

\bibitem{hentschel2018feature}
M.~Hentschel, M.~Delcroix, A.~Ogawa, T.~Nakatani, Feature-based learning hidden
  unit contributions for domain adaptation of rnn-lms, in: 2018 Asia-Pacific
  Signal and Information Processing Association Annual Summit and Conference
  (APSIPA ASC), IEEE, 2018, pp. 1692--1696.

\bibitem{ng2020cuhk}
S.-I. Ng, W.~Liu, Z.~Peng, S.~Feng, H.-P. Huang, O.~Scharenborg, T.~Lee, The
  cuhk-tudelft system for the slt 2021 children speech recognition challenge,
  arXiv preprint arXiv:2011.06239 (2020).

\bibitem{chen2020darts}
Y.-C. Chen, J.-Y. Hsu, C.-K. Lee, H.-y. Lee, Darts-asr: Differentiable
  architecture search for multilingual speech recognition and adaptation, arXiv
  preprint arXiv:2005.07029 (2020).

\bibitem{sun2017unsupervised}
S.~Sun, B.~Zhang, L.~Xie, Y.~Zhang, An unsupervised deep domain adaptation
  approach for robust speech recognition, Neurocomputing 257 (2017) 79--87.

\bibitem{ghahremani2017investigation}
P.~Ghahremani, V.~Manohar, H.~Hadian, D.~Povey, S.~Khudanpur, Investigation of
  transfer learning for asr using lf-mmi trained neural networks, in: 2017 IEEE
  Automatic Speech Recognition and Understanding Workshop (ASRU), IEEE, 2017,
  pp. 279--286.

\bibitem{huang2016unified}
Z.~Huang, S.~M. Siniscalchi, C.-H. Lee, A unified approach to transfer learning
  of deep neural networks with applications to speaker adaptation in automatic
  speech recognition, Neurocomputing 218 (2016) 448--459.

\bibitem{turan2021improving}
M.~T. Turan, E.~Erzin, Improving phoneme recognition of throat microphone
  speech recordings using transfer learning, Speech Communication 129 (2021)
  25--32.

\bibitem{shivakumar2020transfer}
P.~G. Shivakumar, P.~Georgiou, Transfer learning from adult to children for
  speech recognition: Evaluation, analysis and recommendations, Computer speech
  \& language 63 (2020) 101077.

\bibitem{sayed2021bimodal}
H.~M. Sayed, H.~E. ElDeeb, S.~A. Taie, Bimodal variational autoencoder for
  audiovisual speech recognition, Machine Learning (2021) 1--26.

\bibitem{chen2017progressive}
Z.~Chen, J.~Droppo, J.~Li, W.~Xiong, Progressive joint modeling in unsupervised
  single-channel overlapped speech recognition, IEEE/ACM Transactions on Audio,
  Speech, and Language Processing 26~(1) (2017) 184--196.

\bibitem{cho2018multilingual}
J.~Cho, M.~K. Baskar, R.~Li, M.~Wiesner, S.~H. Mallidi, N.~Yalta, M.~Karafiat,
  S.~Watanabe, T.~Hori, Multilingual sequence-to-sequence speech recognition:
  architecture, transfer learning, and language modeling, in: 2018 IEEE Spoken
  Language Technology Workshop (SLT), IEEE, 2018, pp. 521--527.

\bibitem{he2020multi}
K.~He, W.~Xu, Y.~Yan, Multi-level cross-lingual transfer learning with language
  shared and specific knowledge for spoken language understanding, IEEE Access
  8 (2020) 29407--29416.

\bibitem{lin2021improving}
Y.~Lin, Q.~Li, B.~Yang, Z.~Yan, H.~Tan, Z.~Chen, Improving speech recognition
  models with small samples for air traffic control systems, Neurocomputing 445
  (2021) 287--297.

\bibitem{schneider2019wav2vec}
S.~Schneider, A.~Baevski, R.~Collobert, M.~Auli, wav2vec: Unsupervised
  pre-training for speech recognition, arXiv preprint arXiv:1904.05862 (2019).

\bibitem{manohar2017jhu}
V.~Manohar, D.~Povey, S.~Khudanpur, Jhu kaldi system for arabic mgb-3 asr
  challenge using diarization, audio-transcript alignment and transfer
  learning, in: 2017 IEEE Automatic Speech Recognition and Understanding
  Workshop (ASRU), IEEE, 2017, pp. 346--352.

\bibitem{kim2017cross}
J.-K. Kim, Y.-B. Kim, R.~Sarikaya, E.~Fosler-Lussier, Cross-lingual transfer
  learning for pos tagging without cross-lingual resources, in: Proceedings of
  the 2017 conference on empirical methods in natural language processing,
  2017, pp. 2832--2838.

\bibitem{wang2022arobert}
C.~Wang, S.~Dai, Y.~Wang, F.~Yang, M.~Qiu, K.~Chen, W.~Zhou, J.~Huang, Arobert:
  An asr robust pre-trained language model for spoken language understanding,
  IEEE/ACM Transactions on Audio, Speech, and Language Processing (2022).

\bibitem{song2019speech}
X.~Song, G.~Wang, Z.~Wu, Y.~Huang, D.~Su, D.~Yu, H.~Meng, Speech-xlnet:
  Unsupervised acoustic model pretraining for self-attention networks, arXiv
  preprint arXiv:1910.10387 (2019).

\bibitem{Tian20225438}
X.~Tian, K.~Fu, S.~Gao, Y.~Gu, K.~Wang, W.~Li, Z.~Ma, A multi-task and transfer
  learning based approach for mos prediction, Vol. 2022-September, 2022, p.
  5438 – 5442.

\bibitem{jain2022text}
R.~Jain, M.~Y. Yiwere, D.~Bigioi, P.~Corcoran, H.~Cucu, A text-to-speech
  pipeline, evaluation methodology, and initial fine-tuning results for child
  speech synthesis, IEEE Access 10 (2022) 47628--47642.

\bibitem{sancinetti2022transfer}
M.~Sancinetti, J.~Vidal, C.~Bonomi, L.~Ferrer, A transfer learning approach for
  pronunciation scoring, in: ICASSP 2022-2022 IEEE International Conference on
  Acoustics, Speech and Signal Processing (ICASSP), IEEE, 2022, pp. 6812--6816.

\bibitem{monica2022comparison}
G.~M. Monica, M.~T. Rafael, A comparison of feature-based classifiers and
  transfer learning approaches for cognitive impairment recognition in
  language, in: Artificial Intelligence in Neuroscience: Affective Analysis and
  Health Applications: 9th International Work-Conference on the Interplay
  Between Natural and Artificial Computation, IWINAC 2022, Puerto de la Cruz,
  Tenerife, Spain, May 31--June 3, 2022, Proceedings, Part I, Springer, 2022,
  pp. 426--435.

\bibitem{9900378}
Z.~Yue, E.~Loweimi, H.~Christensen, J.~Barker, Z.~Cvetkovic, Acoustic modelling
  from raw source and filter components for dysarthric speech recognition,
  IEEE/ACM Transactions on Audio, Speech, and Language Processing 30 (2022)
  2968--2980.

\bibitem{9747374}
S.~Kessler, B.~Thomas, S.~Karout, An adapter based pre-training for efficient
  and scalable self-supervised speech representation learning, in: ICASSP 2022
  - 2022 IEEE International Conference on Acoustics, Speech and Signal
  Processing (ICASSP), 2022, pp. 3179--3183.

\bibitem{10023147}
M.~Huzaifah, I.~Kukanov, An analysis of semantically-aligned speech-text
  embeddings, in: 2022 IEEE Spoken Language Technology Workshop (SLT), 2023,
  pp. 747--754.

\bibitem{qin2022improving}
S.~Qin, L.~Wang, S.~Li, J.~Dang, L.~Pan, Improving low-resource tibetan
  end-to-end asr by multilingual and multilevel unit modeling, EURASIP Journal
  on Audio, Speech, and Music Processing 2022~(1) (2022) 1--10.

\bibitem{schlotterbeck2022teacher}
D.~Schlotterbeck, A.~Jim{\'e}nez, R.~Araya, D.~Caballero, P.~Uribe, J.~Van~der
  Molen~Moris, “teacher, can you say it again?" improving automatic speech
  recognition performance over classroom environments with limited data, in:
  Artificial Intelligence in Education: 23rd International Conference, AIED
  2022, Durham, UK, July 27--31, 2022, Proceedings, Part I, Springer, 2022, pp.
  269--280.

\bibitem{medeiros2023domain}
E.~Medeiros, L.~Corado, L.~Rato, P.~Quaresma, P.~Salgueiro, Domain adaptation
  speech-to-text for low-resource european portuguese using deep learning,
  Future Internet 15~(5) (2023) 159.

\bibitem{song2019l2rs}
Y.~Song, D.~Jiang, X.~Zhao, Q.~Xu, R.~C.-W. Wong, L.~Fan, Q.~Yang, L2rs: a
  learning-to-rescore mechanism for automatic speech recognition, arXiv
  preprint arXiv:1910.11496 (2019).

\bibitem{kubo2022knowledge}
Y.~Kubo, S.~Karita, M.~Bacchiani, Knowledge transfer from large-scale
  pretrained language models to end-to-end speech recognizers, in: ICASSP
  2022-2022 IEEE International Conference on Acoustics, Speech and Signal
  Processing (ICASSP), IEEE, 2022, pp. 8512--8516.

\bibitem{parthasarathy2019long}
S.~Parthasarathy, W.~Gale, X.~Chen, G.~Polovets, S.~Chang, Long-span language
  modeling for speech recognition, arXiv preprint arXiv:1911.04571 (2019).

\bibitem{tuske2018investigation}
Z.~T{\"u}ske, R.~Schl{\"u}ter, H.~Ney, Investigation on lstm recurrent n-gram
  language models for speech recognition, in: Interspeech, 2018, pp.
  3358--3362.

\bibitem{winata2019code}
G.~I. Winata, A.~Madotto, C.-S. Wu, P.~Fung, Code-switched language models
  using neural based synthetic data from parallel sentences, arXiv preprint
  arXiv:1909.08582 (2019).

\bibitem{dong2018speech}
L.~Dong, S.~Xu, B.~Xu, Speech-transformer: a no-recurrence sequence-to-sequence
  model for speech recognition, in: 2018 IEEE International Conference on
  Acoustics, Speech and Signal Processing (ICASSP), IEEE, 2018, pp. 5884--5888.

\bibitem{winata2020lightweight}
G.~I. Winata, S.~Cahyawijaya, Z.~Lin, Z.~Liu, P.~Fung, Lightweight and
  efficient end-to-end speech recognition using low-rank transformer, in:
  ICASSP 2020-2020 IEEE International Conference on Acoustics, Speech and
  Signal Processing (ICASSP), IEEE, 2020, pp. 6144--6148.

\bibitem{kim2018towards}
S.~Kim, M.~L. Seltzer, Towards language-universal end-to-end speech
  recognition, in: 2018 IEEE International Conference on Acoustics, Speech and
  Signal Processing (ICASSP), IEEE, 2018, pp. 4914--4918.

\bibitem{milde2017multitask}
B.~Milde, C.~Schmidt, J.~K{\"o}hler, Multitask sequence-to-sequence models for
  grapheme-to-phoneme conversion., in: INTERSPEECH, 2017, pp. 2536--2540.

\bibitem{tits2018asr}
N.~Tits, K.~E. Haddad, T.~Dutoit, Asr-based features for emotion recognition: A
  transfer learning approach, arXiv preprint arXiv:1805.09197 (2018).

\bibitem{ananthram2020multi}
A.~Ananthram, K.~K. Saravanakumar, J.~Huynh, H.~Beigi, Multi-modal emotion
  detection with transfer learning, arXiv preprint arXiv:2011.07065 (2020).

\bibitem{boateng2020speech}
G.~Boateng, L.~Sels, P.~Kuppens, P.~Hilpert, T.~Kowatsch, Speech emotion
  recognition among couples using the peak-end rule and transfer learning, in:
  Companion Publication of the 2020 International Conference on Multimodal
  Interaction, 2020, pp. 17--21.

\bibitem{liu2019investigation}
D.~Liu, J.~Xu, P.~Zhang, Y.~Yan, Investigation of knowledge transfer approaches
  to improve the acoustic modeling of vietnamese asr system, IEEE/CAA Journal
  of Automatica Sinica 6~(5) (2019) 1187--1195.

\bibitem{feng2019low}
K.~Feng, T.~Chaspari, Low-resource language identification from speech using
  transfer learning, in: 2019 IEEE 29th International Workshop on Machine
  Learning for Signal Processing (MLSP), IEEE, 2019, pp. 1--6.

\bibitem{wilkinson2020semi}
N.~Wilkinson, A.~Biswas, E.~Y{\i}lmaz, F.~De~Wet, E.~van~der Westhuizen, T.~R.
  Niesler, Semi-supervised acoustic modelling for five-lingual code-switched
  asr using automatically-segmented soap opera speech, arXiv preprint
  arXiv:2004.06480 (2020).

\bibitem{zelasko2022discovering}
P.~{\.Z}elasko, S.~Feng, L.~M. Vel{\'a}zquez, A.~Abavisani, S.~Bhati,
  O.~Scharenborg, M.~Hasegawa-Johnson, N.~Dehak, Discovering phonetic
  inventories with crosslingual automatic speech recognition, Computer Speech
  \& Language 74 (2022) 101358.

\bibitem{hassan2022improvement}
M.~A. Hassan, A.~Rehmat, M.~U. Ghani~Khan, M.~H. Yousaf, et~al., Improvement in
  automatic speech recognition of south asian accent using transfer learning of
  deepspeech2, Mathematical Problems in Engineering 2022 (2022).

\bibitem{deng2022improving}
K.~Deng, S.~Cao, Y.~Zhang, L.~Ma, G.~Cheng, J.~Xu, P.~Zhang, Improving
  ctc-based speech recognition via knowledge transferring from pre-trained
  language models, in: ICASSP 2022-2022 IEEE International Conference on
  Acoustics, Speech and Signal Processing (ICASSP), IEEE, 2022, pp. 8517--8521.

\bibitem{khurana2022magic}
S.~Khurana, A.~Laurent, J.~Glass, Magic dust for cross-lingual adaptation of
  monolingual wav2vec-2.0, in: ICASSP 2022-2022 IEEE International Conference
  on Acoustics, Speech and Signal Processing (ICASSP), IEEE, 2022, pp.
  6647--6651.

\bibitem{tachbelie2022multilingual}
M.~Y. Tachbelie, S.~T. Abate, T.~Schultz, Multilingual speech recognition for
  globalphone languages, Speech Communication 140 (2022) 71--86.

\bibitem{rolland2022multilingual}
T.~Rolland, A.~Abad, C.~Cucchiarini, H.~Strik, Multilingual transfer learning
  for children automatic speech recognition, in: Proceedings of the Thirteenth
  Language Resources and Evaluation Conference, 2022, pp. 7314--7320.

\bibitem{song2019transfer}
p.~song, Transfer linear subspace learning for cross-corpus speech emotion
  recognition, IEEE Transactions on Affective Computing 10~(2) (2019) 265--275.

\bibitem{liu2018unsupervised}
N.~Liu, Y.~Zong, B.~Zhang, L.~Liu, J.~Chen, G.~Zhao, J.~Zhu, Unsupervised
  cross-corpus speech emotion recognition using domain-adaptive subspace
  learning, in: 2018 IEEE International Conference on Acoustics, Speech and
  Signal Processing (ICASSP), IEEE, 2018, pp. 5144--5148.

\bibitem{liu2021transfer}
N.~Liu, B.~Zhang, B.~Liu, J.~Shi, L.~Yang, Z.~Li, J.~Zhu, Transfer subspace
  learning for unsupervised cross-corpus speech emotion recognition, IEEE
  Access 9 (2021) 95925--95937.

\bibitem{luo2019cross}
H.~Luo, J.~Han, Cross-corpus speech emotion recognition using semi-supervised
  transfer non-negative matrix factorization with adaptation regularization.,
  in: INTERSPEECH, 2019, pp. 3247--3251.

\bibitem{zhang2019transfer}
W.~Zhang, P.~Song, Transfer sparse discriminant subspace learning for
  cross-corpus speech emotion recognition, IEEE/ACM Transactions on Audio,
  Speech, and Language Processing 28 (2019) 307--318.

\bibitem{luo2020nonnegative}
H.~Luo, J.~Han, Nonnegative matrix factorization based transfer subspace
  learning for cross-corpus speech emotion recognition, IEEE/ACM Transactions
  on Audio, Speech, and Language Processing 28 (2020) 2047--2060.

\bibitem{zhang2021cross}
W.~Zhang, P.~Song, D.~Chen, C.~Sheng, W.~Zhang, Cross-corpus speech emotion
  recognition based on joint transfer subspace learning and regression, IEEE
  Transactions on Cognitive and Developmental Systems (2021).

\bibitem{chen2019target}
X.~Chen, X.~Zhou, C.~Lu, Y.~Zong, W.~Zheng, C.~Tang, Target-adapted subspace
  learning for cross-corpus speech emotion recognition, IEICE TRANSACTIONS on
  Information and Systems 102~(12) (2019) 2632--2636.

\bibitem{zhao2021cross}
K.~Zhao, P.~Song, W.~Zhang, W.~Zhang, S.~Li, D.~Chen, W.~Zheng, Cross-corpus
  speech emotion recognition based on sparse subspace transfer learning, in:
  Chinese Conference on Biometric Recognition, Springer, 2021, pp. 466--473.

\bibitem{braunschweiler2021study}
N.~Braunschweiler, R.~Doddipatla, S.~Keizer, S.~Stoyanchev, A study on
  cross-corpus speech emotion recognition and data augmentation, in: 2021 IEEE
  Automatic Speech Recognition and Understanding Workshop (ASRU), 2021, pp.
  24--30.

\bibitem{yi2020adversarial}
J.~Yi, J.~Tao, Y.~Bai, Z.~Tian, C.~Fan, Adversarial transfer learning for
  punctuation restoration, arXiv preprint arXiv:2004.00248 (2020).

\bibitem{li2022sequence}
Q.~Li, H.~Zhu, L.~Luo, G.~Cheng, P.~Zhang, J.~Sun, Y.~Yan, Sequence
  distribution matching for unsupervised domain adaptation in asr, in: 2022
  13th International Symposium on Chinese Spoken Language Processing (ISCSLP),
  IEEE, 2022, pp. 21--25.

\bibitem{zhang2022joint}
K.~Zhang, C.~Gong, W.~Lu, L.~Wang, J.~Wei, D.~Liu, Joint and adversarial
  training with asr for expressive speech synthesis, in: ICASSP 2022-2022 IEEE
  International Conference on Acoustics, Speech and Signal Processing (ICASSP),
  IEEE, 2022, pp. 6322--6326.

\bibitem{boulares2020transfer}
M.~Boulares, T.~Alafif, A.~Barnawi, Transfer learning benchmark for
  cardiovascular disease recognition, IEEE Access 8 (2020) 109475--109491.

\bibitem{takashima2019knowledge}
Y.~Takashima, R.~Takashima, T.~Takiguchi, Y.~Ariki, Knowledge transferability
  between the speech data of persons with dysarthria speaking different
  languages for dysarthric speech recognition, IEEE Access 7 (2019)
  164320--164326.

\bibitem{sertolli2021representation}
B.~Sertolli, Z.~Ren, B.~W. Schuller, N.~Cummins, Representation transfer
  learning from deep end-to-end speech recognition networks for the
  classification of health states from speech, Computer Speech \& Language 68
  (2021) 101204.

\bibitem{Gruzitis2022267}
N.~Gruzitis, R.~Dargis, V.~Lasmanis, G.~Garkaje, D.~Gosko, Adapting automatic
  speech recognition to the radiology domain for a less-resourced language: The
  case of latvian, Lecture Notes in Networks and Systems 333 (2022) 267--276.

\bibitem{hirevs2022convolutional}
M.~Hire{\v{s}}, M.~Gazda, P.~Drot{\'a}r, N.~D. Pah, M.~A. Motin, D.~K. Kumar,
  Convolutional neural network ensemble for parkinson's disease detection from
  voice recordings, Computers in biology and medicine 141 (2022) 105021.

\bibitem{pahar2022covid}
M.~Pahar, M.~Klopper, R.~Warren, T.~Niesler, Covid-19 detection in cough,
  breath and speech using deep transfer learning and bottleneck features,
  Computers in biology and medicine 141 (2022) 105153.

\bibitem{harati2022generalization}
A.~Harati, T.~Rutowski, Y.~Lu, P.~Chlebek, R.~Oliveira, E.~Shriberg, D.~Lin,
  Generalization of deep acoustic and nlp models for large-scale depression
  screening, in: Biomedical Sensing and Analysis: Signal Processing in Medicine
  and Biology, Springer, 2022, pp. 99--132.

\bibitem{rejaibi2022mfcc}
E.~Rejaibi, A.~Komaty, F.~Meriaudeau, S.~Agrebi, A.~Othmani, Mfcc-based
  recurrent neural network for automatic clinical depression recognition and
  assessment from speech, Biomedical Signal Processing and Control 71 (2022)
  103107.

\bibitem{yue2022raw}
Z.~Yue, E.~Loweimi, Z.~Cvetkovic, Raw source and filter modelling for
  dysarthric speech recognition, in: ICASSP 2022-2022 IEEE International
  Conference on Acoustics, Speech and Signal Processing (ICASSP), IEEE, 2022,
  pp. 7377--7381.

\bibitem{almadhor2023e2e}
A.~Almadhor, R.~Irfan, J.~Gao, N.~Saleem, H.~T. Rauf, S.~Kadry, E2e-dasr:
  End-to-end deep learning-based dysarthric automatic speech recognition,
  Expert Systems with Applications 222 (2023) 119797.

\bibitem{hu2021generating}
A.~Hu, D.~Phadnis, S.~R. Shahamiri, Generating synthetic dysarthric speech to
  overcome dysarthria acoustic data scarcity, Journal of Ambient Intelligence
  and Humanized Computing (2021) 1--18.

\bibitem{han2023spatial}
Z.~Han, Y.~Shang, Z.~Shao, J.~Liu, G.~Guo, T.~Liu, H.~Ding, Q.~Hu,
  Spatial-temporal feature network for speech-based depression recognition,
  IEEE Transactions on Cognitive and Developmental Systems (2023).

\bibitem{hu2019adversarial}
S.~Hu, X.~Shang, Z.~Qin, M.~Li, Q.~Wang, C.~Wang, Adversarial examples for
  automatic speech recognition: Attacks and countermeasures, IEEE
  Communications Magazine 57~(10) (2019) 120--126.

\bibitem{sun2018training}
S.~Sun, C.-F. Yeh, M.~Ostendorf, M.-Y. Hwang, L.~Xie, Training augmentation
  with adversarial examples for robust speech recognition, arXiv preprint
  arXiv:1806.02782 (2018).

\bibitem{abdullah2021hear}
H.~Abdullah, M.~S. Rahman, W.~Garcia, K.~Warren, A.~S. Yadav, T.~Shrimpton,
  P.~Traynor, Hear" no evil", see" kenansville"*: Efficient and transferable
  black-box attacks on speech recognition and voice identification systems, in:
  2021 IEEE Symposium on Security and Privacy (SP), IEEE, 2021, pp. 712--729.

\bibitem{schonherr2018adversarial}
L.~Sch{\"o}nherr, K.~Kohls, S.~Zeiler, T.~Holz, D.~Kolossa, Adversarial attacks
  against automatic speech recognition systems via psychoacoustic hiding, arXiv
  preprint arXiv:1808.05665 (2018).

\bibitem{zelasko2021adversarial}
P.~{\.Z}elasko, S.~Joshi, Y.~Shao, J.~Villalba, J.~Trmal, N.~Dehak,
  S.~Khudanpur, Adversarial attacks and defenses for speech recognition
  systems, arXiv preprint arXiv:2103.17122 (2021).

\bibitem{subramanian2020study}
V.~Subramanian, A.~Pankajakshan, E.~Benetos, N.~Xu, S.~McDonald, M.~Sandler, A
  study on the transferability of adversarial attacks in sound event
  classification, in: ICASSP 2020-2020 IEEE International Conference on
  Acoustics, Speech and Signal Processing (ICASSP), IEEE, 2020, pp. 301--305.

\bibitem{carlini2018audio}
N.~Carlini, D.~Wagner, Audio adversarial examples: Targeted attacks on
  speech-to-text, in: 2018 IEEE Security and Privacy Workshops (SPW), IEEE,
  2018, pp. 1--7.

\bibitem{kwon2019selective}
H.~Kwon, Y.~Kim, H.~Yoon, D.~Choi, Selective audio adversarial example in
  evasion attack on speech recognition system, IEEE Transactions on Information
  Forensics and Security 15 (2019) 526--538.

\bibitem{wu2020self}
A.~Wu, C.~Wang, J.~Pino, J.~Gu, Self-supervised representations improve
  end-to-end speech translation, arXiv preprint arXiv:2006.12124 (2020).

\bibitem{zhu2021conwst}
W.~Zhu, H.~Jin, J.~Chen, L.~Luo, J.~Wang, Q.~Lu, A.~Li, Conwst: Non-native
  multi-source knowledge distillation for low resource speech translation, in:
  International Conference on Cognitive Systems and Signal Processing,
  Springer, 2021, pp. 127--141.

\bibitem{azizah2020hierarchical}
K.~Azizah, M.~Adriani, W.~Jatmiko, Hierarchical transfer learning for
  multilingual, multi-speaker, and style transfer dnn-based tts on low-resource
  languages, IEEE Access 8 (2020) 179798--179812.

\bibitem{luo2022physics}
W.~Luo, Z.~Yan, Q.~Song, R.~Tan, Physics-directed data augmentation for deep
  model transfer to specific sensor, ACM Transactions on Sensor Networks 19~(1)
  (2022) 1--30.

\bibitem{s22166292}
M.~Tropea, G.~Fedele, R.~De~Luca, D.~Miriello, F.~De~Rango, Automatic stones
  classification through a cnn-based approach, Sensors 22~(16) (2022).

\bibitem{yoon2023inter}
J.~W. Yoon, B.~J. Woo, S.~Ahn, H.~Lee, N.~S. Kim, Inter-kd: Intermediate
  knowledge distillation for ctc-based automatic speech recognition, in: 2022
  IEEE Spoken Language Technology Workshop (SLT), IEEE, 2023, pp. 280--286.

\bibitem{lee2022knowledge}
M.-H. Lee, J.-H. Chang, Knowledge distillation from language model to acoustic
  model: a hierarchical multi-task learning approach, in: ICASSP 2022-2022 IEEE
  International Conference on Acoustics, Speech and Signal Processing (ICASSP),
  IEEE, 2022, pp. 8392--8396.

\bibitem{chatziagapi2022audio}
A.~Chatziagapi, D.~Sgouropoulos, C.~Karouzos, T.~Melistas, T.~Giannakopoulos,
  A.~Katsamanis, S.~Narayanan, Audio and asr-based filled pause detection, in:
  2022 10th International Conference on Affective Computing and Intelligent
  Interaction (ACII), IEEE, 2022, pp. 1--7.

\bibitem{sahoo2022mic_fuzzynet}
K.~K. Sahoo, R.~Hazra, M.~F. Ijaz, S.~Kim, P.~K. Singh, M.~Mahmud,
  Mic\_fuzzynet: Fuzzy integral based ensemble for automatic classification of
  musical instruments from audio signals, IEEE Access 10 (2022) 100797--100811.

\bibitem{xu2020hybrid}
X.~Xu, Z.~Meng, A hybrid transfer learning model for short-term electric load
  forecasting, Electrical Engineering 102~(3) (2020) 1371--1381.

\bibitem{hu2019heterogeneous}
W.~Hu, Y.~Luo, Z.~Lu, Y.~Wen, Heterogeneous transfer learning for thermal
  comfort modeling, in: Proceedings of the 6th ACM International Conference on
  Systems for Energy-Efficient Buildings, Cities, and Transportation, 2019, pp.
  61--70.

\bibitem{rosenstein2005to}
M.~T. Rosenstein, To transfer or not to transfer, in: NIPS 2005 Workshop on
  Transfer Learning, 2005.

\bibitem{wang2019characterizing}
Z.~Wang, Z.~Dai, B.~P{\'o}czos, J.~Carbonell, Characterizing and avoiding
  negative transfer, in: Proceedings of the IEEE/CVF conference on computer
  vision and pattern recognition, 2019, pp. 11293--11302.

\bibitem{meftah2021hidden}
S.~Meftah, N.~Semmar, Y.~Tamaazousti, H.~Essafi, F.~Sadat, On the hidden
  negative transfer in sequential transfer learning for domain adaptation from
  news to tweets, in: Proceedings of the Second Workshop on Domain Adaptation
  for NLP, 2021, pp. 140--145.

\bibitem{doulaty2015data}
M.~Doulaty, O.~Saz, T.~Hain, Data-selective transfer learning for multi-domain
  speech recognition, arXiv preprint arXiv:1509.02409 (2015).

\bibitem{sousa2014transfer}
R.~Sousa, L.~M. Silva, L.~A. Alexandre, J.~Santos, J.~M. De~S{\'a}, Transfer
  learning: current status, trends and challenges, in: 20th Portuguese
  Conference on Pattern Recognition, RecPad, 2014, pp. 57--58.

\bibitem{delfosse2020deep}
A.~Delfosse, G.~Hebrail, A.~Zerroug, Deep learning applied to nilm: is data
  augmentation worth for energy disaggregation?, in: ECAI 2020, IOS Press,
  2020, pp. 2972--2977.

\bibitem{sayed2022deep}
A.~Sayed, Y.~Himeur, F.~Bensaali, Deep and transfer learning for building
  occupancy detection: A review and comparative analysis, Engineering
  Applications of Artificial Intelligence (2022).

\bibitem{fan2020statistical}
C.~Fan, Y.~Sun, F.~Xiao, J.~Ma, D.~Lee, J.~Wang, Y.~C. Tseng, Statistical
  investigations of transfer learning-based methodology for short-term building
  energy predictions, Applied Energy 262 (2020) 114499.

\bibitem{alghamdi2020detection}
A.~Alghamdi, M.~Hammad, H.~Ugail, A.~Abdel-Raheem, K.~Muhammad, H.~S. Khalifa,
  A.~El-Latif, A.~Ahmed, Detection of myocardial infarction based on novel deep
  transfer learning methods for urban healthcare in smart cities, Multimedia
  tools and applications (2020) 1--22.

\bibitem{das2018evaluation}
B.~Das, B.~Nair, V.~K. Reddy, P.~Venkatesh, Evaluation of multiple linear,
  neural network and penalised regression models for prediction of rice yield
  based on weather parameters for west coast of india, International journal of
  biometeorology 62~(10) (2018) 1809--1822.

\bibitem{himeur2021artificial}
Y.~Himeur, K.~Ghanem, A.~Alsalemi, F.~Bensaali, A.~Amira, Artificial
  intelligence based anomaly detection of energy consumption in buildings: A
  review, current trends and new perspectives, Applied Energy 287 (2021)
  116601.

\bibitem{jha2019enhancing}
D.~Jha, K.~Choudhary, F.~Tavazza, W.-k. Liao, A.~Choudhary, C.~Campbell,
  A.~Agrawal, Enhancing materials property prediction by leveraging
  computational and experimental data using deep transfer learning, Nature
  communications 10~(1) (2019) 1--12.

\bibitem{zhao2017research}
W.~Zhao, Research on the deep learning of the small sample data based on
  transfer learning, in: AIP Conference Proceedings, Vol. 1864, AIP Publishing
  LLC, 2017, p. 020018.

\bibitem{copiaco2023innovative}
A.~Copiaco, Y.~Himeur, A.~Amira, W.~Mansoor, F.~Fadli, S.~Atalla, S.~S. Sohail,
  An innovative deep anomaly detection of building energy consumption using
  energy time-series images, Engineering Applications of Artificial
  Intelligence 119 (2023) 105775.

\bibitem{himeur2022recent}
Y.~Himeur, A.~Alsalemi, F.~Bensaali, A.~Amira, A.~Al-Kababji, Recent trends of
  smart nonintrusive load monitoring in buildings: A review, open challenges,
  and future directions, International Journal of Intelligent Systems 37~(10)
  (2022) 7124--7179.

\bibitem{glorot2011domain}
X.~Glorot, A.~Bordes, Y.~Bengio, Domain adaptation for large-scale sentiment
  classification: A deep learning approach, in: ICML, 2011.

\bibitem{cook2013transfer}
D.~Cook, K.~D. Feuz, N.~C. Krishnan, Transfer learning for activity
  recognition: A survey, Knowledge and information systems 36~(3) (2013)
  537--556.

\bibitem{sullivan2022improving}
P.~Sullivan, T.~Shibano, M.~Abdul-Mageed, Improving automatic speech
  recognition for non-native english with transfer learning and language model
  decoding, arXiv preprint arXiv:2202.05209 (2022).

\bibitem{baevski2020wav2vec}
A.~Baevski, Y.~Zhou, A.~Mohamed, M.~Auli, wav2vec 2.0: A framework for
  self-supervised learning of speech representations, Advances in Neural
  Information Processing Systems 33 (2020) 12449--12460.

\bibitem{xu2021self}
Q.~Xu, A.~Baevski, T.~Likhomanenko, P.~Tomasello, A.~Conneau, R.~Collobert,
  G.~Synnaeve, M.~Auli, Self-training and pre-training are complementary for
  speech recognition, in: ICASSP 2021-2021 IEEE International Conference on
  Acoustics, Speech and Signal Processing (ICASSP), IEEE, 2021, pp. 3030--3034.

\bibitem{lin2021deep}
J.~Lin, J.~Ma, J.~Zhu, H.~Liang, Deep domain adaptation for non-intrusive load
  monitoring based on a knowledge transfer learning network, IEEE Transactions
  on Smart Grid (2021).

\bibitem{zhang2019domain}
Y.~Zhang, J.~Yan, Domain-adversarial transfer learning for robust intrusion
  detection in the smart grid, in: 2019 IEEE International Conference on
  Communications, Control, and Computing Technologies for Smart Grids
  (SmartGridComm), 2019, pp. 1--6.

\bibitem{zhang2020semi}
Y.~Zhang, J.~Yan, Semi-supervised domain-adversarial training for intrusion
  detection against false data injection in the smart grid, in: 2020
  International Joint Conference on Neural Networks (IJCNN), IEEE, 2020, pp.
  1--7.

\bibitem{patricia2014learning}
N.~Patricia, B.~Caputo, Learning to learn, from transfer learning to domain
  adaptation: A unifying perspective, in: Proceedings of the IEEE Conference on
  Computer Vision and Pattern Recognition, 2014, pp. 1442--1449.

\bibitem{paul2018comparison}
A.~Paul, K.~Vogt, F.~Rottensteiner, J.~Ostermann, C.~Heipke, A comparison of
  two strategies for avoiding negative transfer in domain adaptation based on
  logistic regression, International Archives of the Photogrammetry, Remote
  Sensing and Spatial Information Sciences-ISPRS Archives 42 (2018), Nr. 2
  42~(2) (2018) 845--852.

\bibitem{minoofam2021trcla}
S.~A.~H. Minoofam, A.~Bastanfard, M.~R. Keyvanpour, Trcla: A transfer learning
  approach to reduce negative transfer for cellular learning automata, IEEE
  Transactions on Neural Networks and Learning Systems (2021) 1--10.

\bibitem{zhang2020survey}
W.~Zhang, L.~Deng, L.~Zhang, D.~Wu, A survey on negative transfer, arXiv
  preprint arXiv:2009.00909 (2020).

\bibitem{ahmed2020preech}
S.~Ahmed, A.~R. Chowdhury, K.~Fawaz, P.~Ramanathan, Preech: A system for
  $\{$Privacy-Preserving$\}$ speech transcription, in: 29th USENIX Security
  Symposium (USENIX Security 20), 2020, pp. 2703--2720.

\bibitem{ul2020decentralized}
A.~ul~Haque, M.~S. Ghani, T.~Mahmood, Decentralized transfer learning using
  blockchain \& ipfs for deep learning, in: 2020 International Conference on
  Information Networking (ICOIN), IEEE, 2020, pp. 170--177.

\bibitem{wang2021enabling}
X.~Wang, S.~Garg, H.~Lin, M.~J. Piran, J.~Hu, M.~S. Hossain, Enabling secure
  authentication in industrial iot with transfer learning empowered blockchain,
  IEEE Transactions on Industrial Informatics 17~(11) (2021) 7725--7733.

\bibitem{zhang2021federated}
P.~Zhang, H.~Sun, J.~Situ, C.~Jiang, D.~Xie, Federated transfer learning for
  iiot devices with low computing power based on blockchain and edge computing,
  Ieee Access 9 (2021) 98630--98638.

\bibitem{maurya2021federated}
S.~Maurya, S.~Joseph, A.~Asokan, A.~A. Algethami, M.~Hamdi, H.~T. Rauf,
  Federated transfer learning for authentication and privacy preservation using
  novel supportive twin delayed ddpg (s-td3) algorithm for iiot, Sensors
  21~(23) (2021) 7793.

\bibitem{arefeen2021transjury}
M.~A. Arefeen, S.~Tabassum~Nimi, M.~Y. Sarwar~Uddin, Y.~Lee, Transjury: Towards
  explainable transfer learning through selection of layers from deep neural
  networks, in: 2021 IEEE International Conference on Big Data (Big Data),
  2021, pp. 978--984.

\bibitem{kirchhof2021chances}
M.~Kirchhof, L.~Schmid, C.~Reining, M.~t. Hompel, M.~Pauly, Chances of
  interpretable transfer learning for human activity recognition in
  warehousing, in: International Conference on Computational Logistics,
  Springer, 2021, pp. 163--177.

\bibitem{incahuanaco2022coffeese}
F.~Incahuanaco-Quispe, E.~Hinojosa-Cardenas, D.~A. Pilares-Figueroa, C.~A.
  Beltr{\'a}n-Casta{\~n}{\'o}n, Coffeese: Interpretable transfer learning
  method for estimating the severity of coffee rust, in: Annual International
  Conference on Information Management and Big Data, Springer, 2022, pp.
  340--355.

\bibitem{ramakrishnan2016towards}
R.~Ramakrishnan, J.~Shah, Towards interpretable explanations for transfer
  learning in sequential tasks (2016).

\bibitem{kim2019the}
D.~Kim, W.~Lim, M.~Hong, H.~Kim, The structure of deep neural network for
  interpretable transfer learning, in: 2019 IEEE International Conference on
  Big Data and Smart Computing (BigComp), 2019, pp. 1--4.

\bibitem{lee2021interpretable}
S.~Lee, B.~C. Song, Interpretable embedding procedure knowledge transfer via
  stacked principal component analysis and graph neural network, in:
  Proceedings of the AAAI Conference on Artificial Intelligence, Vol.~35, 2021,
  pp. 8297--8305.

\bibitem{carr2021interpretable}
T.~Carr, J.~Sanderson, D.~Broadway, S.~Sami, Interpretable staged transfer
  learning improves oct classification and clinical explanation of retinal
  diseases from small sample sizes, Investigative Ophthalmology \& Visual
  Science 62~(8) (2021) 2119--2119.

\bibitem{zhao2014online}
P.~Zhao, S.~C. Hoi, J.~Wang, B.~Li, Online transfer learning, Artificial
  intelligence 216 (2014) 76--102.

\bibitem{wu2017online}
Q.~Wu, H.~Wu, X.~Zhou, M.~Tan, Y.~Xu, Y.~Yan, T.~Hao, Online transfer learning
  with multiple homogeneous or heterogeneous sources, IEEE Transactions on
  Knowledge and Data Engineering 29~(7) (2017) 1494--1507.

\bibitem{wu2019online}
H.~Wu, Y.~Yan, Y.~Ye, H.~Min, M.~K. Ng, Q.~Wu, Online heterogeneous transfer
  learning by knowledge transition, ACM Transactions on Intelligent Systems and
  Technology (TIST) 10~(3) (2019) 1--19.

\bibitem{alasbahi2022an}
R.~Alasbahi, X.~Zheng, An online transfer learning framework with extreme
  learning machine for automated credit scoring, IEEE Access 10 (2022)
  46697--46716.

\bibitem{qiu2021easytransfer}
M.~Qiu, P.~Li, C.~Wang, H.~Pan, A.~Wang, C.~Chen, X.~Jia, Y.~Li, J.~Huang,
  D.~Cai, et~al., Easytransfer: A simple and scalable deep transfer learning
  platform for nlp applications, in: Proceedings of the 30th ACM International
  Conference on Information \& Knowledge Management, 2021, pp. 4075--4084.

\bibitem{gruetzemacher2022deep}
R.~Gruetzemacher, D.~Paradice, Deep transfer learning \& beyond: Transformer
  language models in information systems research, ACM Computing Surveys (CSUR)
  54~(10s) (2022) 1--35.

\bibitem{kasneci2023chatgpt}
E.~Kasneci, K.~Se{\ss}ler, S.~K{\"u}chemann, M.~Bannert, D.~Dementieva,
  F.~Fischer, U.~Gasser, G.~Groh, S.~G{\"u}nnemann, E.~H{\"u}llermeier, et~al.,
  Chatgpt for good? on opportunities and challenges of large language models
  for education, Learning and Individual Differences 103 (2023) 102274.

\bibitem{hu2023advancing}
M.~Hu, S.~Pan, Y.~Li, X.~Yang, Advancing medical imaging with language models:
  A journey from n-grams to chatgpt, arXiv preprint arXiv:2304.04920 (2023).

\bibitem{sohail2023future}
S.~S. Sohail, F.~Farhat, Y.~Himeur, M.~Nadeem, D.~{\O}. Madsen, Y.~Singh,
  S.~Atalla, W.~Mansoor, The future of gpt: A taxonomy of existing chatgpt
  research, current challenges, and possible future directions, Current
  Challenges, and Possible Future Directions (April 8, 2023) (2023).

\bibitem{wu2023brief}
T.~Wu, S.~He, J.~Liu, S.~Sun, K.~Liu, Q.-L. Han, Y.~Tang, A brief overview of
  chatgpt: The history, status quo and potential future development, IEEE/CAA
  Journal of Automatica Sinica 10~(5) (2023) 1122--1136.

\bibitem{strzelecki2023use}
A.~Strzelecki, To use or not to use chatgpt in higher education? a study of
  students’ acceptance and use of technology, Interactive Learning
  Environments (2023) 1--14.

\bibitem{laskar2023systematic}
M.~T.~R. Laskar, M.~S. Bari, M.~Rahman, M.~A.~H. Bhuiyan, S.~Joty, J.~X. Huang,
  A systematic study and comprehensive evaluation of chatgpt on benchmark
  datasets, arXiv preprint arXiv:2305.18486 (2023).

\bibitem{cheng2023potential}
K.~Cheng, Z.~Sun, Y.~He, S.~Gu, H.~Wu, The potential impact of chatgpt/gpt-4 on
  surgery: will it topple the profession of surgeons?, International Journal of
  Surgery (2023) 10--1097.

\bibitem{sohail2023using}
S.~S. Sohail, D.~{\O}. Madsen, Y.~Himeur, M.~Ashraf, Using chatgpt to navigate
  ambivalent and contradictory research findings on artificial intelligence,
  Available at SSRN 4413913 (2023).

\end{thebibliography}

\end{document}